\documentclass[11pt]{article}

\usepackage{amsthm,amsmath,amssymb,bbm}
\usepackage{natbib}
\usepackage{multirow}
\usepackage{subfigure}
\usepackage{makecell}
\usepackage{booktabs}
\usepackage{array}
\usepackage{tabularx}
\usepackage{tabulary}
\usepackage{caption}
\usepackage{booktabs}
\usepackage{url}
\usepackage{algorithm}
\usepackage{algorithmic}
\usepackage{bm}
\usepackage{wrapfig}
\usepackage{lipsum}
\usepackage{mathrsfs}
\usepackage{dsfont}
\usepackage{titling}
\usepackage{txfonts}

\usepackage{xr}
\usepackage[usenames,dvipsnames,svgnames,table]{xcolor}
\usepackage[colorlinks,
linkcolor=blue,
anchorcolor=blue,
urlcolor  = blue,
citecolor=blue
]{hyperref}

\newcommand{\sgn}{\mathop{\mathrm{sign}}}

\usepackage{smile}

\newcommand{\nn}{\nonumber}

\def\##1\#{\begin{align}#1\end{align}}
\def\$#1\${\begin{align*}#1\end{align*}}

\def\sn{\sum_{i=1}^n}
\def\Sb{\mathbf{S}}

\newcommand{\BB}{\mathbb{B}}

\newcommand{\bzero}{{\mathbf 0}}

\usepackage{relsize}
\newcommand{\T}{{\mathsmaller {\rm T}}}

\newcommand{\wt}{\widetilde}

\newcommand{\bfsym}[1]{\ensuremath{\boldsymbol{#1}}}
       \def \bbeta    {\bfsym{\beta}}
       \def \bdelta   {\bfsym{\delta}}

\newcommand{\Rom}[1]{\text{\uppercase\expandafter{\romannumeral #1\relax}}}

\usepackage{geometry}
\usepackage{enumitem}

\newcommand{\cc}{{\rm c}}

\newcommand{\ora}{{\rm ora}}
\numberwithin{equation}{section}

\begin{document}

\title{High-Dimensional Quantile Regression: \\Convolution Smoothing and Concave Regularization}

\author{Kean Ming Tan\thanks{Department of Statistics, University of Michigan, Ann Arbor, Michigan 48109, USA. E-mail:\href{mailto:keanming@umich.edu}{\textsf{keanming@umich.edu}}.},~~Lan Wang\thanks{Miami Herbert Business School, University of Miami, Coral Gables, FL 33146, USA. E-mail:\href{mailto:lanwang@mbs.miami.edu}{\textsf{lanwang@mbs.miami.edu}}.}~~and~Wen-Xin Zhou\thanks{Department of Mathematics, University of California, San Diego, La Jolla, CA 92093, USA. E-mail:\href{mailto:wez243@ucsd.edu}{\textsf{wez243@ucsd.edu}}.} }

\date{}
\maketitle

\vspace{-0.5in}

\begin{abstract}

$\ell_1$-penalized  quantile regression is widely used for analyzing high-dimensional data with heterogeneity.
It is now recognized that the $\ell_1$-penalty introduces non-negligible estimation bias, while a proper use of concave regularization may lead to estimators with refined convergence rates and oracle properties as the signal strengthens.
Although folded concave penalized $M$-estimation with strongly convex loss functions have been well studied, the extant literature on quantile regression is relatively silent. The main difficulty is that the quantile loss is piecewise linear: it is non-smooth and has curvature concentrated at a single point. To overcome the lack of smoothness and strong convexity, we propose and study a convolution-type smoothed quantile regression with iteratively reweighted $\ell_1$-regularization. The resulting smoothed empirical loss  is twice continuously differentiable and (provably) locally strongly convex with high probability. We show that the iteratively reweighted $\ell_1$-penalized smoothed quantile regression estimator, after a few iterations, achieves the optimal  rate of convergence, and moreover, the oracle rate and the strong oracle property under an almost necessary and sufficient minimum signal strength condition. Extensive numerical studies corroborate our theoretical results.
\end{abstract}


\section{Introduction}
\label{sec:1}
Massive complex datasets bring challenges to data analysis due to the presence of outliers and heterogeneity.
Consider regression of a scalar response $y$ on a $p$-dimensional predictor $\bx \in \mathbb{R}^p$.
The least squares method focuses on the conditional mean of the outcome given the predictor.
Despite its popularity in the statistical and econometric literature, it is sensitive to outliers and fails to capture heterogeneity in the set of important features.  Moreover, in many applications, the scientific question of interest may not be fully addressed by inferring the conditional mean. Since the seminal work of \cite{KB1978}, quantile regression (QR) has gained increasing attention by offering a set of complementary methods designed to explore data features invisible to the inveiglements of least squares methods. Quantile regression is robust to data heterogeneity and outliers, and also offers unique insights into the entire conditional distribution of the outcome given the predictor.
We refer to \citet{K2005} and \citet{KCHP2017} for an overview of quantile regression theory, methods and applications.

In the high-dimensional setting in which the number of features, $p$, exceeds the number of observations, $n$, it is often the case that only a small subset of a large pool of features  influences the conditional distribution of the outcome.
To perform estimation and variable selection simultaneously, the standard approach is to minimize the empirical loss plus a penalty on the model complexity.
The $\ell_1$-penalty is arguably the most commonly used penalty function that induces sparsity \citep{Tibs1996}.
Least squares methods with $\ell_1$-regularization have been  extensively studied in the past two decades. 
Because of the extremely long list of relevant literature, we refer the reader to the monographs \citet{BV2011}, \citet{HTW2015}, \citet{W2019}, \citet{FLZZ2020}, and the references therein.
In the context of quantile regression, \citet{BC2011} provided a comprehensive analysis of the $\ell_1$-penalized quantile regression as well as post-penalized QR estimator.
Since then, the literature on high-dimensional quantile regression has grown rapidly, and we refer to Chapter 15 of  \citet{KCHP2017} for an overview.

It is now a consensus that the $\ell_1$-penalty induces non-negligible bias \citep{FL2001,Z2006, ZZ2012}, due to which the selected model tends to include spurious variables unless stringent conditions are imposed on the design matrix, such as the strong irrepresentable condition \citep{ZY2006,MB2006}.
To reduce the bias induced by the $\ell_1$-penalty when the signal is sufficiently strong, various concave penalty functions have been designed \citep{FL2001,ZMCP2010,Z2010}.
For concave penalized $M$-estimation with convex and locally strongly convex losses, a large body of literature has shown that there exists a local solution that possesses the oracle property, i.e., a solution that is as efficient as the oracle estimator obtained by assuming the true active set is known {\it a priori}, under certain minimum signal strength condition, also known as the beta-min condition. We refer the reader to \cite{FL2001}, \cite{ZL2008}, \cite{KCO2008}, \cite{Z2010}, \cite{FL2011}, \cite{ZZ2012}, \cite{KK2012}, \cite{LW2015}, and \cite{L2017} for more details.

Comparably, quantile regression with concave regularization is much less understood theoretically primarily due to the challenges in analyzing the piecewise linear quantile loss and the concave penalty simultaneously.
Let $\bbeta^* \in \RR^p$ be the  $s$-sparse underlying parameter vector with support $\cS = \{ 1\leq j\leq p: \beta^*_j \neq 0 \}$, and define the minimum signal strength $\| \bbeta^*_{\cS} \|_{\min} = \min_{j\in \cS} |\beta^*_j|$.
Under a beta-min condition $\| \bbeta^*_{\cS} \|_{\min} \gg  n^{-1/2}\max\{ s, \sqrt{\log(p)  } \}$,  \citet{WWL2012} showed that the oracle QR estimator belongs to the set of local minima of the non-convex penalized quantile objective function with probability approaching one.
From a different angle, \citet{FXZ2014} proved that the oracle QR estimator can be obtained via the one-step local linear approximation (LLA) algorithm \citep{ZL2008} under a beta-min condition $\| \bbeta^*_{\cS} \|_{\min} \gtrsim \sqrt{s\log(p)/n}$, that is, the minimal non-zero coefficient is of order $\sqrt{s\log(p)/n}$ in magnitude.
We refer to Chapter 16 of \citet{KCHP2017} for an overview of the existing results on non-convex regularized quantile regression.  
Existing work on folded concave penalized QR either impose stringent signal strength assumptions or only establish theoretical guarantees for some local optimum which, due to non-convexity, is not necessarily the solution obtained by any practical algorithm. In other words, there is no guarantee that the solution obtained from a given algorithm will satisfy the desired statistical properties, leaving a gap between theory and practice.

A natural way to resolve the non-differentiability issue is to smooth the piecewise linear quantile loss using a kernel.
The idea of kernel smoothing was first considered by \citet{H1998} in the context of bootstrap inference for median regression.  \citet{H1998} showed that the estimator obtained from the smoothed quantile loss is asymptotically equivalent to that of the standard quantile regression estimator.
This motivates a series of work on smoothed quantile regression when the number of features is fixed \citep{WH2006,WMY2015,GK2016}. However, smoothing the piecewise linear loss directly yields a non-convex function for which global minimum is not guaranteed. This poses even more challenges in the high-dimensional setting.

\begin{figure}[!htp]
\centering
\includegraphics[scale=0.5]{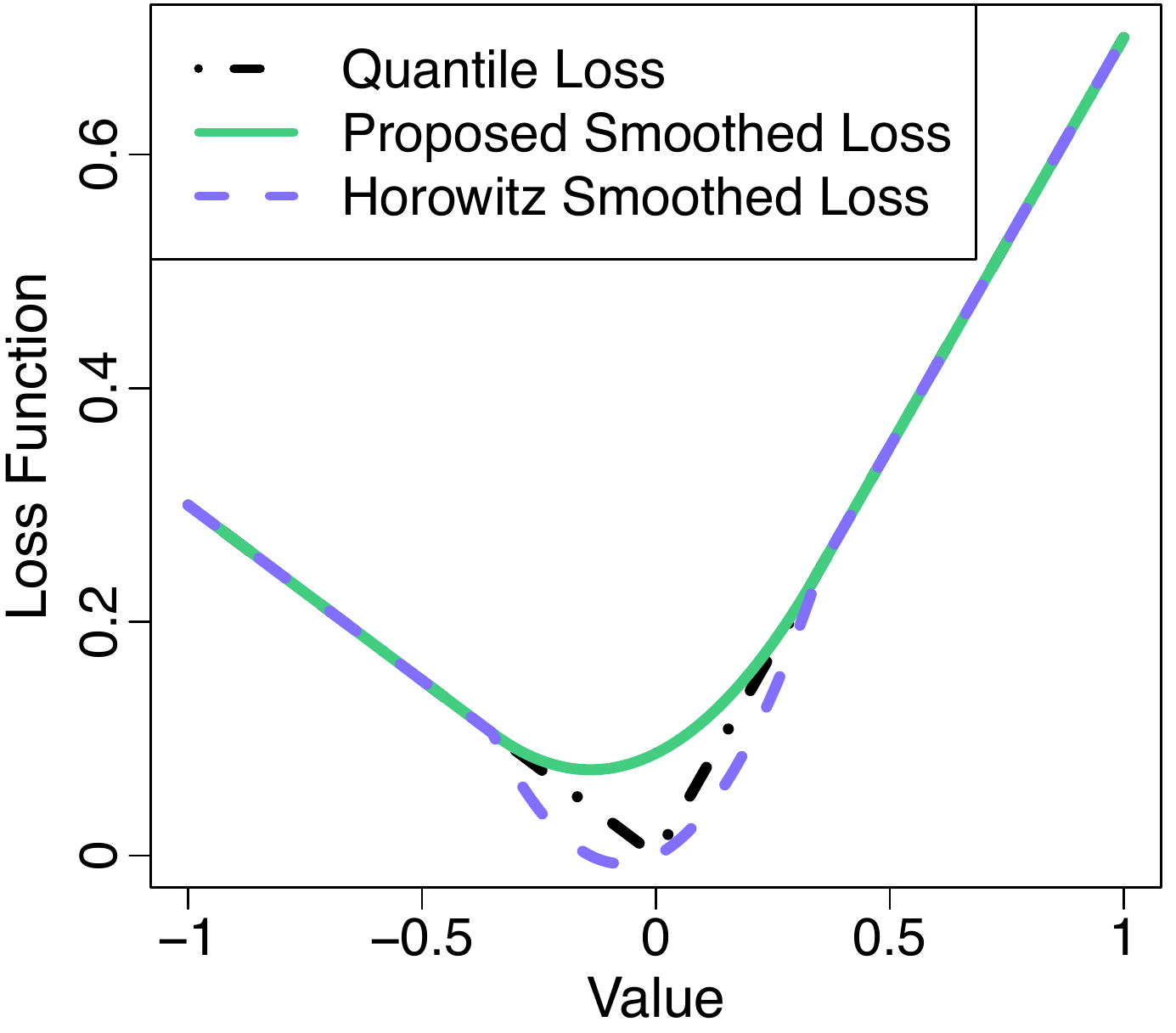}
\caption{Plots of a standard quantile loss, Horowitz's smoothed quantile loss \citep{H1998}, and a convolution-type smoothed quantile loss.}
\label{fig:lossfunction}
\end{figure}

In this paper, we propose and study a new method for quantile regression in high-dimensional sparse models, which is based on convolution smoothing and iteratively reweighted $\ell_1$-penalization.
To deal with non-smoothness, we smooth the piecewise linear quantile loss via convolution. The idea is to smooth the subgradient of the quantile loss, and then integrate it to obtain a  smoothed  loss function that is also convex.
See Figure~\ref{fig:lossfunction} for a visualization of Horowitz's and convolution smoothing methods.
 \citet{FGH2019} developed the traditional asymptotic theory for convolution smoothing  in the context of linear quantile regression when the sample size $n$ tends to infinity while $p$ is kept fixed. For high-dimensional sparse models, we extend the one-step LLA algorithm proposed by \cite{ZL2008}, and propose a multi-step, iterative procedure which solves a weighted $\ell_1$-penalized smoothed quantile objective function at each iteration. This multi-step procedure consists of a sequence of convex programs, which is similar to the multi-stage convex relaxation method for sparse regularization \citep{Z2010,FLSZ2018}. Computationally, for different smoothing kernels, typified by the uniform and Gaussian kernels, we propose efficient algorithms to minimize the weighted $\ell_1$-penalized smoothed quantile objective function at each stage. Comparing with existing methods for fitting high-dimensional quantile regression, the proposed gradient-based algorithms are more scalable to large-scale problems with either large sample size or high dimensionality.

Since the proposed multi-step procedure delivers a sequence of solutions iteratively, to understand how these estimators evolve statistically, we provide a delicate analysis of the estimator at each stage whose overall estimation error consists of three components: shrinkage bias, oracle rate, and smoothing bias.
The theoretical analysis in \citet{Z2010} and \citet{FLSZ2018} is primarily suited for the quadratic case, although the method applies to more general loss functions. In this work, we  aim at establishing theoretical underpinnings of why and how convolution smoothing and iteratively reweighted $\ell_1$-penalization help with achieving oracle properties for quantile regression. 

In particular, we show that the solution for the first iteration, i.e., the $\ell_1$-penalized smoothed quantile regression, is near minimax optimal, and coincide with those of existing results for $\ell_1$-penalized QR estimator.
Moreover, our analysis reveals that the multi-step, iterative algorithm refines the statistical rate in a sequential manner: every relaxation step shrinks the estimation error from the previous step by a $\delta$-fraction for some predetermined $\delta \in (0,1)$. All the results are non-asymptotic with explicit errors depending on $(s,p, n)$, including the deterministic smoothing bias and stochastic statistical errors.
With a minimal requirement on the signal strength---$\| \bbeta^*_{\cS} \|_{\min} \gtrsim \sqrt{\log(p)/n}$, we show that after as many as $\ell \gtrsim \lceil \log(\max\{ \log(p) , s \}) \rceil$ iterations, the multi-step algorithm will deliver an estimator that achieves the oracle rate of convergence as well as the strong oracle property. The latter implies variable selection consistency as a byproduct. To our knowledge, these are the first statistical characterizations of computationally feasible concave regularized quantile regression estimators.

The rest of the paper is organized as follows.  In Section~\ref{sec2}, we describe the convolution-type smoothing approach for quantile regression, followed by an iteratively reweighted $\ell_1$-penalized procedure for fitting high-dimensional sparse models. At each stage, the problem boils down to minimizing a weighted $\ell_1$-penalized smoothed quantile objective function, for which  we propose efficient and scalable algorithms in Section~\ref{sec:algorithm} with a particular focus on uniform and Gaussian kernels. In Section~\ref{sec:theory}, we provide theoretical guarantees for the sequence of estimators obtained by the multi-step method, including estimation error bounds (in high probability) and strong oracle property. A numerical demonstration of the proposed method on simulated data and a real data application are provided in Sections~\ref{sec:numerical} and \ref{sec:real}, respectively. The proofs of all theoretical results are given in the online supplementary material.  The Python code that implements the proposed iteratively reweighted regularized quantile regression procedure is available at \href{https://github.com/WenxinZhou/conquer}{https://github.com/WenxinZhou/conquer}.

\medskip
\noindent \textbf{Notation:}
For every integer $k\geq 1$, we use $\RR^k$ to denote the the $k$-dimensional Euclidean space, and write $[k]=\{1,\ldots, k\}$. The inner product of any two vectors $\bu=(u_1, \ldots, u_k)^\T, \bv=(v_1, \ldots ,v_k)^\T \in \RR^k$ is defined by $\bu^\T \bv = \langle \bu, \bv \rangle= \sum_{i=1}^k u_i v_i$.
Moreover, let $\bu \circ \bv = (u_1 v_1, \ldots, u_k v_k )^\T$ denote the Hadamard product of $\bu$ and $\bv$.
For a subset $\cS \subseteq [k]$ with cardinality $|\cS|$, we write $\bu_{\cS} \in \RR^{| \cS |}$ as the subvector of $\bu$ that consists of the entries of $\bu$ indexed by $\cS$.
We use $\| \cdot \|_p$ $(1\leq q \leq \infty)$ to denote the $\ell_q$-norm in $\RR^k$: $\| \bu \|_q = ( \sum_{i=1}^k | u_i |^q )^{1/q}$ and $\| \bu \|_\infty = \max_{1\leq i\leq k} |u_i|$. For $k\geq 2$, $\mathbb{S}^{k-1} = \{ \bu \in \RR^k : \| \bu \|_2 = 1 \}$ denotes the unit sphere in $\RR^k$.
For any function $f:\RR \mapsto \RR$ and vector $\bu = (u_1,\ldots, u_k)^\T \in \RR^k$, we write $f(\bu) = ( f(u_1), \ldots, f(u_k) )^\T \in \RR^k$.

Throughout this paper, we use bold uppercase letters to represent matrices. For $k\geq 2$, $\Ib_k$ represents an $k\times k$ identity matrix. For any $k\times k$ symmetric, positive semidefinite matrix $\Ab \in \RR^{k\times k}$, we use $\gamma(\Ab)  \in \RR^k$ to denote its vector of eigenvalues, ordered as $\gamma_1(\Ab) \geq \cdots \geq \gamma_p(\Ab) \geq 0$, and let $\| \Ab \|_2 = \gamma_1(\Ab)$ be the operator norm of $\Ab$.
Moreover, let  $\| \cdot \|_{\Ab}$ denote  the vector norm induced by $\Ab$: $\| \bu \|_{\Ab} = \| \Ab^{1/2} \bu \|_2$ for $\bu \in \RR^k$.
For any two real numbers $u$ and $v$, we write $u\vee v = \max(u,v)$ and $u \wedge v = \min(u,v)$. For two sequences of non-negative numbers $\{ a_n \}_{n\geq 1}$ and $\{ b_n \}_{n\geq 1}$, $a_n \lesssim b_n$ indicates that there exists a constant $C>0$ independent of $n$ such that $a_n \geq Cb_n$; $a_n \gtrsim b_n$ is equivalent to $b_n \lesssim a_n$; $a_n \asymp b_n$ is equivalent to $a_n \lesssim b_n$ and $b_n \lesssim a_n$. For two numbers $C_1$ and $C_2$, we write $C_2=C_2(C_1)$ if $C_2$ depends only on $C_1$.

\section{Sparse quantile regression: convolution smoothing and iterative regularization}
\label{sec2}
\subsection{Penalized quantile regression}
\label{subsec:problem}
We consider a scalar response variable $y\in\RR$ and a $p$-dimensional feature vector $\bx = (x_1, \ldots, x_p)^\T \in \RR^p$ such that the $\tau$-th conditional quantile of $y$ given $\bx$ is modeled as $F^{-1}_{y| \bx }(\tau |  \bx ) =   \bx^\T   \bbeta^*  $ for some $0<\tau<1$, where $\bbeta^*  = (\beta^*_1, \ldots, \beta^*_p)^\T \in \RR^p$.
Let $\{ (y_i, \bx_i)\}_{i=1}^n$ be a random sample from $(y,\bx)$.
The preceding model assumption is equivalent to
\#
	y_i =     \bx_i^\T \bbeta^*    + \varepsilon_i ~~\mbox{ and }~~ \PP(\varepsilon_i \leq 0 \, |\, \bx_i ) = \tau .   \label{model}
\#
Throughout the paper, we set $x_1 \equiv 1$ so that $\beta^*_1$ denotes the intercept.
To avoid notational clutter, the dependence of $\bbeta^*$ and $\varepsilon_i$ on $\tau$ will be assumed without displaying.

Given a random sample $\{ (y_i, \bx_i) \}_{i=1}^n$, a penalized QR estimator is generally defined as either the global optimum or one of the local optima to the optimization problem
\# \label{l1.qr}
  \underset{\bbeta = (\beta_1, \ldots, \beta_p )^\T \in \RR^{p}}{\mathrm{minimize}}~ \Biggl\{ \underbrace{  \frac{1}{n} \sn \rho_\tau( y_i -    \bx_i^\T  \bbeta  ) }_{  =:  \hat Q(\bbeta) }  + \sum_{j=1}^p q_{\lambda} ( |\beta_j| ) \Biggr\},
\#	
where $ \rho_\tau(u) = u \{ \tau -  \mathbbm{1}(u<0)\}$ is the $\tau$-quantile function, also referred to as the check function, and $q_{\lambda}(\cdot): [0, \infty) \to [0, \infty)$ is a sparsity-inducing penalty function parametrized by $\lambda>0$.

Due to convexity, the $\ell_1$-penalized method for which $q_\lambda(t) = \lambda t$ ($t\geq 0$) has dominated the literature on high-dimensional statistics. Work in the context of quantile regression include that of \cite{WLJ2007}, \cite{BC2011}, \cite{BFW2011}, \cite{W2013}, and \cite{ZPH2015}, \citet{SB2017}, among others. 
Various algorithms can be employed to solve the resulting $\ell_1$-penalized problem \citep{BJMO2011,B2010,KCHP2017,Gu2018}. To alleviate the non-negligible bias induced by the $\ell_1$ penalty, folded concave penalties have been used in, for example, \cite{WWL2012} and \cite{FXZ2014}, leading to non-convex optimization problems. Together, the non-differentiable quantile loss and the non-convex penalty bring fundamental statistical and computational challenges.  

Statistical theory of non-convex regularized quantile regression is relatively underdeveloped.  
Most of the existing results are developed either under stringent minimum signal strength conditions, or for the hypothetical global optimum (or one of the local optima). 
Motivated from the algorithmic approaches developed by \cite{ZL2008} and \cite{FLSZ2018}, we consider a multi-step iterative method that solves a sequence of convex problems, which bypasses the computational issues from solving the non-convex problem \eqref{l1.qr} directly. Theoretically, a major difficulty is that the quantile loss is  piecewise linear, so that its ``curvature energy" is concentrated in a single point. This is in contrast to many popular loss functions considered in the statistical literature, such as the squared, logistic, or Huber loss, which are at least locally strongly convex. Therefore, a proper smoothing scheme that creates smoothness and local strong convexity is the key to the success of the proposed framework.

\subsection{Convolution-type smoothing approach}
\label{sec:smoothing}
Let $F_{\varepsilon|\bx}(\cdot)$ be the conditional distribution of $\varepsilon$ given $\bx$. The population quantile loss can then be written as
\[
Q(\bbeta) = \EE_{\bx} \Biggl\{ \int_{-\infty}^\infty \rho_\tau(u-\langle\bx, \bbeta -\bbeta^* \rangle) \,{\rm d}  F_{\varepsilon|\bx} ( u ) \Biggr\},
\]
where $\EE_{\bx}(\cdot)$ is the expectation taken with respect to $\bx$.
Provided that the conditional distribution $F_{\varepsilon |\bx}(\cdot)$ is sufficiently smooth, $Q(\bbeta)$ is twice differentiable and strongly convex  in a neighborhood of $\bbeta^*$. For every $\bbeta \in \RR^p$, let $\hat F(\cdot;\bbeta)$ be the empirical cumulative distribution function (ECDF)  of the residuals $\{r_i(\bbeta) := y_i -  \bx_i^\T \bbeta \}_{i=1}^n$, i.e., $\hat F(u;\bbeta) = (1/n) \sn \mathbbm{1}\{ r_i(\bbeta)  \leq u \}$ for any $u \in \RR$.
Then, the empirical quantile loss $\hat Q(\cdot)$ in  \eqref{l1.qr} can be expressed as
\#
\hat Q(\bbeta) = \int_{-\infty}^\infty \rho_\tau(u) \,{\rm d} \hat F(u;\bbeta). \label{empirical.qr}
\#
Since the ECDF $\hat F(\cdot; \bbeta)$ is discontinuous, the standard empirical quantile loss $\hat Q(\cdot)$ has the same degree of smoothness as $\rho_\tau(\cdot)$. This motivates \cite{FGH2019} to use a kernel CDF estimator.
Given the residuals $r_i(\bbeta) = y_i - \bx_i^\T \bbeta$ and a smoothing parameter/bandwidth $h=h_n >0$, let $\hat F_h(\cdot;\bbeta)$ be the distribution function of the classical Rosenblatt--Parzen kernel density estimator:
\#
	\hat F_h(u;\bbeta) = \int_{-\infty}^u \hat f_h(t;\bbeta) \,{\rm d}   t~~\mbox{ with }~~ \hat f_h(t;\bbeta) = \frac{1}{n} \sn K_h \big(  t-r_i(\bbeta) \big) , \nn
\#
where $K:\mathbb R \to \mathbb [0,\infty)$ is a symmetric, non-negative kernel that integrates to one, and $K_h(u) := (1/h) K(u/h)$ for $u\in \RR$. Replacing $\hat F(u ; \bbeta)$ in \eqref{empirical.qr} with its kernel-smoothed counterpart $\hat F_h(u;\bbeta)$ yields the following smoothed empirical quantile loss
\#
	\hat Q_h(\bbeta) :=   \int_{-\infty}^\infty \rho_\tau(u) \,{\rm d} \hat F_h(u;\bbeta) = \frac{1}{nh} \sn \int_{-\infty}^\infty \rho_\tau(u) K \bigg( \frac{u + \bx_i^\T \bbeta  - y_i  }{h}  \bigg) \,{\rm d} u . \label{smooth.qloss}
\#
Define the integrated kernel function $\bar{K}: \RR \to [0,1]$ as $\bar{K}(u ) =  \int_{-\infty}^u K(t) \,{\rm d} t$.
As will be shown in Section~\ref{sec:theory:bias}, the smoothed empirical quantile objective function $ \hat Q_h  (\bbeta) $ is twice continuously differentiable with gradient $\nabla  \hat Q_h(\bbeta)  =  (1/n) \sn  \{ \bar{K}  (  -r_i(\bbeta)/h  ) - \tau  \} \bx_i $ and Hessian matrix $\nabla^2  \hat Q_h(\bbeta)  =  (1/n)\sn K_h(-r_i(\bbeta)) \bx_i \bx_i^\T$.
Moreover, we will show that the smoothed objective function $\hat Q_h(\cdot)$ is strongly convex in a cone local neighborhood of $\bbeta^*$ with high probability; see Proposition~\ref{prop:RSC}.

\begin{remark} \label{rmk:kernel}
For a given kernel function $K(\cdot)$ and bandwidth $h>0$, the smoothed quantile loss $\hat Q_h(\cdot)$ defined in \eqref{smooth.qloss} can be equivalently written as $\hat Q_h(\bbeta) = (1/n) \sn \ell_h(y_i - \bx_i^\T \bbeta)$, where
\#
	\ell_h(u) =  (\rho_\tau * K_h )(u) = \int_{-\infty}^{\infty} \rho_\tau(v) K_h(v- u ) \, {\rm d} v ,  \ \ u\in \RR . \label{convolution.loss}
\#	
Here $*$ denotes the convolution operator. To better understand this smoothing mechanism, we compute the smoothed loss $\ell_h=\rho_\tau * K_h$ explicitly for several widely used kernel functions. Recall that $\rho_\tau(u) = |u|/2 + (\tau - 1/2) u$.

\begin{enumerate}
\item[(i)] (Uniform kernel) For the uniform kernel $K(u) = (1/2)\mathbbm{1} (|u|\leq 1)$, which is the density function of the uniform distribution on $[-1, 1]$, the resulting smoothed loss takes the form $\ell_h(u) =  (h/2) U(u/h) + (\tau - 1/2) u$, where $U(u) = (u^2/2 + 1/2) \mathbbm{1} (|u|\leq 1) +   |u|  \mathbbm{1} (|u|>1)$ is a Huber-type loss. Convolution plays a role of random smoothing in the sense that $ \ell_h(u)  = (1/2) \EE (|Z_u|) + (\tau - 1/2)u$, where for every $u\in \RR$, $Z_u$ denotes a random variable uniformly distributed between $u-h$ and $u+h$.

\item[(ii)] (Gaussian kernel) For the Gaussian kernel $K(u) = \phi(u)$, the density function of a standard normal distribution, the resulting smoothed loss is $ \ell_{ h}(u)  = (1/2) \EE (|G_u|) + (\tau - 1/2)u$, where $G_u \sim N(u, h^2)$. Note that $|G_u|$ follows a folded normal distribution \citep{LNN1961} with mean  $\EE |G_u| =  (2/\pi)^{1/2}  h e^{-u^2/(2h^2)} + u \{ 1 - 2 \Phi(- u/h )  \}$. Hence, the smoothed loss can be written as $ \ell_{ h}(u) = (h/2) G(u/h) + (\tau - 1/2 ) u$, where $G(u) =(2/\pi)^{1/2} e^{-u^2/2 } + u \{ 1- 2\Phi(-u) \}$.

\item[(iii)] (Laplacian kernel) In the case of the Laplacian kernel $K(u) =  e^{-|u|}/2$, we have $\ell_h(u) =\rho_\tau(u)  + h e^{-|u|/h}/2$.

\item[(iv)] (Logistic kernel) In the case of the logistic kernel $K(u) = e^{-u}/(1+e^{-u})^2$, the resulting smoothed loss is $\ell_h(u) =  \tau u  + h \log(1+e^{-u/h})$.

\item[(v)] (Epanechnikov kernel) For the Epanechnikov kernel $K(u) = (3 / 4) (1 - u^2)  \mathbbm{1}(|u|\leq 1)$, the resulting smoothed loss is $ \ell_{  h}(u) = (h/2) E(u/h) +(\tau - 1/2)u$, where $E(u) = (3u^2/4 - u^4/8+ 3/8) \mathbbm{1}(|u|\leq 1) + |u| \mathbbm{1}(|u|>1)$.


\end{enumerate}
\end{remark}

\subsection{Iteratively reweighted $\ell_1$-penalized method}
 \label{sec:multi-stage}

 Let $\{ (y_i , \bx_i) \}_{i=1}^n$ be independent data vectors from the conditional quantile model \eqref{model} with a sparse target parameter $\bbeta^* \in \RR^p$. Extending the one-step LLA algorithm proposed by \cite{ZL2008}, we consider a multi-step, iteratively regularized method as follows. Let $q_\lambda(\cdot)$ be a prespecified penalty function that is differentiable almost everywhere. 
Starting at iteration 0 with an initial estimator  $\hat \bbeta^{(0)}$, for $\ell=1, 2,\ldots$, we iteratively update the previous estimator $\hat \bbeta^{(\ell-1)}$ by solving
\#
\hat \bbeta^{(\ell )}  = (\hat \beta_1^{(\ell)}, \ldots,\hat \beta_p^{(\ell)})^\T \in  \argmin_{\bbeta = (  \beta_1, \ldots,\beta_p)^\T  } \Bigg\{  \hat Q_h(\bbeta) + \sum_{j=1}^p q'_\lambda( |\hat \beta_j^{(\ell-1)} | )  | \beta_j |  \Bigg\},\label{weighted.lasso-lad}
\#
where $q_{\lambda}' (\cdot)$ is the first-order derivative of $q_\lambda(\cdot)$, and $\hat Q_h(\cdot)$ is the convolution smoothed quantile objective function defined in \eqref{smooth.qloss}. To avoid notational clutter, we suppress the dependence of $\{ \hat \bbeta^{(\ell)} = \hat \bbeta_h^{(\ell)}(\tau, \lambda) \}_{\ell\geq 0}$ on the quantile index $\tau$, bandwidth $h$, and penalty level $\lambda$.

The penalty function $q_\lambda(\cdot)$, or its derivative to be exact, plays the role of producing sparse solutions. We consider a class of penalty functions that satisfies the following conditions.
\begin{enumerate}
\item[(A1)]  The penalty function $q_\lambda$ is of the form $q_\lambda(t) = \lambda^2 q(t/\lambda)$ for $t\geq 0$, where $q: [0, \infty) \mapsto [0, \infty)$ satisfies:
(i) $q$ is non-decreasing on $[0,\infty)$ with $q(0)=0$; (ii) $q(\cdot)$ is differentiable almost everywhere on $(0,\infty)$, $0\leq q'(t) \leq 1$ and $\lim_{t \downarrow 0 } q'(t) = 1$; (iii) $q'(t_1) \leq q'(t_2)$ for all $t_1 \geq t_2 \geq 0$.
\end{enumerate}
Examples of penalties that satisfy Condition (A1) include:
\begin{enumerate}
\item  $\ell_1$-penalty: $q(t) = |t|$. In this case, $q'(t) = 1$ for all $t>0$. Therefore, $\hat \bbeta^{(1)}$ defined in \eqref{weighted.lasso-lad} with $\ell = 1$ is the $\ell_1$-penalized SQR estimator, and the procedure stops after the first step.  

\item  Smoothly clipped absolute deviation (SCAD) penalty \citep{FL2001}: The function $q(\cdot)$ is defined through its derivative  $q'(t) = \mathbbm{1} (t\leq 1) +  \frac{(a-t)_+}{a-1} \mathbbm{1} (t>1)$ for $t\geq 0$ and some $a>2$, and $q(0)=0$. \cite{FL2001} suggested $a=3.7$ by a Bayesian argument. 

\item  Minimax concave penalty (MCP) \citep{ZMCP2010}: The function $q(\cdot)$ is defined through its derivative $q'(t) = (1-t/a)_+$ for $t\geq 0$ and some $a\geq 1$, and $q(0)=0$.

\item Capped-$\ell_1$ penalty \citep{Z2010}: $q(t ) = \min(a/2, t)$ and $q'(t) = \mathbbm{1}(t\leq a/2)$ for $t\geq 0$ and some $a\geq 1$.
\end{enumerate}

If we start the multi-step procedure using any penalty $q_\lambda$ that satisfies Condition (A1) and a trivial initialization $\hat \bbeta^{(0)} = \textbf{0}$, then $q_{\lambda}' (|\hat{\beta}_j^{(0)}|)  = q_\lambda'(0)= \lambda$ for $j=1,\ldots,p$, and hence the first step is essentially computing an $\ell_1$-penalized smoothed QR estimator.
 At each subsequent iteration,  the subproblem \eqref{weighted.lasso-lad} can be expressed as a weighted $\ell_1$-penalized smoothed quantile loss minimization:
\#
	   \underset{\bbeta\in \RR^p}{\mathrm{minimize}}~   \bigl\{ \hat  Q_h(\bbeta) +  \| \blambda   \circ  \bbeta \|_1 \bigr\} ,   \label{general.lasso}
\#
where $\blambda = (\lambda_1, \ldots, \lambda_p )^\T$ is a $p$-vector of regularization parameters with $\lambda_j \geq 0$, and $\circ$ denotes the Hadamard product.
We summarize this iteratively reweighted $\ell_1$-penalized method in Algorithm~\ref{alg:multistage}.

\begin{algorithm}[!htp]
\small
\caption{Iteratively Reweighted $\ell_1$-Penalized Smoothed QR.}
\label{alg:multistage}
   \textbf{Input:} Data vectors $\{(y_i, \bx_i)\}_{i=1}^n$, quantile index $\tau\in (0,1)$, bandwidth $h >0$, and an initial estimator $\hat{\bbeta}^{(0)} \in \RR^p$. \\
For $\ell=1,2,\ldots$, repeat 
\begin{enumerate}
\item Set $\lambda_j^{(\ell-1)} =  q'_\lambda(|\hat{\beta}_j^{(\ell-1)}|)$ for $j=1,\ldots,p$;
\item Compute
\begin{equation}
\label{eq:weightedl1kq}
\hat{\bbeta}^{(\ell)} \in \underset{\bbeta \in \RR^{p}}{\argmin} ~  \bigl\{ \hat Q_h  (\bbeta) +  \|\blambda^{(\ell-1)}  \circ \bbeta \|_1 \bigr\} ;
\end{equation}
\end{enumerate}
 until convergence.
\end{algorithm}

In Section~\ref{sec:theory}, we will establish non-asymptotic statistical theory for the sequence of estimators $\{ \hat \bbeta^{(\ell)}\}_{\ell\geq 0}$ initialized with $\hat \bbeta^{(0) } = \textbf{0}$ when the penalty $q_\lambda(t) = \lambda^2 q(t/\lambda)$ obeys Condition (A1). 
In order to reduce the (regularization) bias when the signal is sufficiently strong,  we are particularly interested in the concave penality $q(\cdot)$, which not only satisfies Condition (A1) but also has a redescending derivative, i.e., $q'(t) = 0$ for all sufficiently large $t$. 

Another widely applicable idea for bias reduction is adaptive Lasso \citep{Z2006}, which is a one-step procedure that solves, in the context of quantile regression,
\#
 \wt \bbeta \in \argmin_{\bbeta \in \RR^p}  \Bigg\{    \hat \cQ(\bbeta) + \lambda  \sum_{j=1}^p w( |\wt \beta_j^{(0)} | ) |\beta_j | \Bigg\} ,
\#
where $\wt \bbeta^{(0)} = (\wt \beta^{(0)}_1, \ldots, \wt \beta^{(0)}_p)^\T$ is an initial estimator of $\bbeta^*$, say the $\ell_1$-QR (or QR-Lasso) estimator \citep{BC2011}, and $w(t) := t^{-\gamma}$ for $t> 0$ and some $\gamma>0$. Note that the weight function $\lambda w(\cdot)$ for adaptive Lasso is quite different from $q_\lambda'(\cdot) = \lambda q'(\cdot/\lambda)$ in \eqref{weighted.lasso-lad}. 
As discussed in \cite{FL2008}, an advantage of the concave penalty, such as SCAD and MCP, is that zero is not an absorbing state: once a coefficient is shrunk to zero, it will remain zero throughout the remaining iterations.
 As a result, any true positive that is left out by the initial Lasso estimator will be missed in the second stage as well.  The aforementioned is an important phenomenon which was empirically verified by \cite{FLSZ2018}.

\begin{remark}
In practice, it is common to leave a subset of parameters, such as the intercept and coefficients which correspond to features that are already viewed relevant, unpenalized throughout the multi-step procedure \eqref{weighted.lasso-lad}. Given a predetermined index set $\cR \subseteq [p]$, we can modify Algorithm~\ref{alg:multistage} by taking $\blambda^{(\ell)}= (\lambda^{(\ell)}_1 ,\ldots, \lambda^{(\ell)}_p)^\T$ ($\ell \geq 0$) to be $\lambda^{(\ell)}_j=0$ for $j\in \cR$ and $\lambda^{(\ell)}_j = q'_\lambda(|\hat{\beta}_j^{(\ell)}|)$ for $j\notin \cR$. Theoretically, we will study the sequence of estimates $\{ \hat \bbeta^{(\ell)}\}_{\ell \geq 1}$ obtained from Algorithm~\ref{alg:multistage} because a special treatment of leaving parameters indexed by $\cR$ unpenalized only makes things more convoluted and does not bring new insights from a theoretical viewpoint.
\end{remark}

\section{Algorithm}
\label{sec:algorithm}
As discussed in Section~\ref{sec:multi-stage}, the multi-step convex relaxation method leads to a sequence of iteratively reweighted $\ell_1$-penalized problems. Computationally, it suffices to develop efficient algorithms for solving the convex  problem \eqref{eq:weightedl1kq}. For several commonly used kernels,  explicit forms of  the smoothed check loss functions are given in Remark~\ref{rmk:kernel}. In the following sections, we present specialized algorithms for two representative kernel functions: the uniform kernel and the Gaussian kernel.

\subsection{A coordinate descent algorithm for uniform kernel}
\label{subsec:uniform kernel}
First we describe a coordinate descent algorithm for solving \eqref{eq:weightedl1kq} with the uniform kernel, i.e., $K(u) = 1/2$ for $|u|\le 1$. The coordinate descent algorithm is an iterative method that minimizes the objective function with respect to one variable at a time while fixing the other variables.
To implement the algorithm, we calculate the partial derivative of the loss function in~\eqref{eq:weightedl1kq} with respect to each variable, and derive the corresponding update for each variable while keeping the others fixed.

The gradient of the loss function in~\eqref{eq:weightedl1kq} involves $\bar{K}(\cdot)$. For the uniform kernel, we have
\begin{equation*}
\bar{K}\biggl(\frac{\bx_i^\T \bbeta -y_i}{h}\biggr)
= \begin{cases}
1 \qquad\qquad\qquad~~ \mathrm{if~}{\bx_i^\T \bbeta -y_i} \ge h,\\
\frac{1}{2} \Bigl(\frac{\bx_i^\T \bbeta -y_i}{h} + 1 \Bigr)\quad \mathrm{if~}|\bx_i^\T \bbeta -y_i |\le h,\\
0 \qquad\qquad\qquad~~  \mathrm{if~}{\bx_i^\T \bbeta -y_i} \le -h.
\end{cases}
\end{equation*}
Let $C_1=\{i:\bx_i^\T \bbeta -y_i \leq -h \}$, $C_2=\{i:|\bx_i^\T \bbeta -y_i |\le h \}$, and $C_3=\{i:\bx_i^\T \bbeta -y_i \geq h \}$.
Then, the first-order optimality condition of minimizing $\beta_j \to \hat Q_h  (\bbeta) +  \|\blambda^{(\ell-1)}  \circ \bbeta_- \|_1 $ can be written as
\[
-\tau \sum_{i=1}^n x_{ij} +\frac{1}{2}\sum_{i \in C_2} x_{ij}+\sum_{i \in C_3} x_{ij}
+ \frac{1}{2h}\sum_{i\in C_2} ( \bx_i^\T \bbeta -y_i )x_{ij} + n \lambda_j^{(\ell-1)} \hat{z}_j=0,
\]
where $\hat{z}_j \in \partial |\hat \beta_j |$ is the subgradient.
This leads to the following closed-form solution for $\hat \beta_j$:
\[
\hat{\beta}_j = S\left\{ \frac{2h\tau \sum_{i=1}^nx_{ij} -2h \sum_{i\in C_3}^n x_{ij}-h \sum_{i\in C_2}^n x_{ij}+\sum_{i\in C_2}x_{ij} (y_i - \langle \bx_{i,-j}, \bbeta_{-j} \rangle)}{\sum_{i\in C_2}x_{ij}^2}, \frac{2n h\lambda_j^{(\ell-1)}}{\sum_{i\in C_2}x_{ij}^2}\right\},
\]
where $S(a,b) = \mathrm{sign}(a) \max (|a|-b,0) $ denotes the soft-thresholding operator.
Therefore, a solution of \eqref{eq:weightedl1kq} can be obtained by iteratively updating each $\hat{\beta}_j$ until convergence.
The details are summarized in Algorithm~\ref{Alg:general}.
\begin{algorithm}[!htp]
\small
\caption{ Coordinate Descent Algorithm for Solving \eqref{eq:weightedl1kq} with Uniform Kernel.}
\label{Alg:general}
\textbf{Input} quantile level $\tau$, smoothing parameter $h$, regularization parameter $\blambda^{(\ell-1)}$, and   convergence criterion $\epsilon$.\\
\textbf{Initialization}  $\hat{\bbeta}^{(0)}=\boldsymbol{0}$.\\
\textbf{Iterate} the following until the stopping criterion $\|\hat{\bbeta}^{(t)}-\hat{\bbeta}^{(t-1)}\|_2 \le \epsilon$ is met, where $\hat{\bbeta}^{(t)}$ is the value of $\bbeta$ obtained at the $t$th iteration.  That is, for each $j=1,\ldots,p$:
\begin{enumerate}
\item Set $C_1=\{i:\bx_i^\T \bbeta -y_i \geq h \}$, $C_2=\{i:|\bx_i^\T \bbeta -y_i |\le h \}$, and $C_3=\{i:\bx_i^\T \bbeta -y_i \le -h \}$, where we use $\bbeta$ to denote the updated solution at the current iteration.
\item Set
\[
\hat{\beta}_{j}^{(t)} = S\left\{ \frac{2h\tau \sum_{i=1}^nx_{ij} -2h \sum_{i\in C_3}^n x_{ij}-h \sum_{i\in C_2}^n x_{ij}+\sum_{i\in C_2}x_{ij} (y_i - \langle \bx_{i,-j}, \bbeta_{-j} \rangle)}{\sum_{i\in C_2}x_{ij}^2}, \frac{2n h\lambda_j^{(\ell-1)}}{\sum_{i\in C_2}x_{ij}^2}\right\},
\]
where $S(a,b) = \mathrm{sign}(a) \max (|a|-b,0) $ is the soft-thresholding operator.
\end{enumerate}
\textbf{Output} the estimated parameter $\hat{\bbeta}^{(t)}$.
\end{algorithm}

Compared to the existing algorithms for solving $\ell_1$-regularized quantile regression, Algorithm~\ref{Alg:general} is computationally efficient especially for large-scale problems.
The computational complexity is similar to that of the coordinate descent algorithm for Lasso.

\subsection{An alternating direction method of multiplier algorithm for Gaussian kernel}

Next we consider the case of smoothing via  the Gaussian kernel function. In this case, we have
\[
\bar{K}\bigg(\frac{\bx_i^\T \bbeta -y_i}{h}\bigg)
=\Phi \bigg(  \frac{\bx_i^\T \bbeta -y_i}{h} \bigg),
\]
where $\Phi(\cdot)$ is the cumulative distribution function of the standard normal distribution.
The coordinate descent approach in the previous section can no longer be employed, at least trivially, to solve~\eqref{eq:weightedl1kq} since there is no closed-form solution of minimizing $\beta_j \to \hat Q_h  (\bbeta) +  \|\blambda^{(\ell-1)}  \circ \bbeta_- \|_1 $ with the Gaussian kernel. To address this issue, we introduce an alternating direction method of multiplier (ADMM) algorithm to solve~\eqref{eq:weightedl1kq} by decoupling terms that are difficult to optimize jointly.
A similar approach has been considered in  \citet{Gu2018} for solving standard quantile regression with $\ell_1$-regularization.
Let $\br = (r_1,\ldots, r_n)^\intercal$ with $r_i  = y_i- \langle \bx_i,\bbeta \rangle$.  Optimization problem~\eqref{eq:weightedl1kq} can then be rewritten as
 \#
\underset{\bbeta \in \RR^{p},\br \in \RR^{n}}{\mathrm{minimize}} ~  &\bigl\{ \hat Q_h  (\br) +  \|\blambda^{(\ell-1)}  \circ \bbeta_- \|_1 \bigr\},\nn\\
\mathrm{subject~to}~ & \br = \yb-\Xb \bbeta.\label{opt:weighted2}
\#
The augmented Lagrangian for~\eqref{opt:weighted2} is
\#
\cL_{\rho}(\bbeta,\br,\boldsymbol{\eta}) = \hat Q_h  (\br)+  \|\blambda^{(\ell-1)}  \circ \bbeta_- \|_1 + \langle  \boldsymbol{\eta}, \br - \yb+\Xb\bbeta\rangle +\frac{\rho}{2}\|\br -\yb+\Xb\bbeta \|_2^2,
\#
 where $\boldsymbol{\eta}$ is the Lagrange multiplier and $\rho$ is a tuning parameter for the ADMM algorithm.
Updates for the ADMM   can be derived by minimizing each parameter while keeping the others fixed.  We summarize the details in Algorithm~\ref{alg:admm}.

\begin{algorithm}[!htp]
\small
\caption{ADMM Algorithm for Solving \eqref{eq:weightedl1kq} with Gaussian Kernel.}
\label{alg:admm}
\textbf{Input}  quantile parameter $\tau$, smoothing parameter $h$, regularization parameter $\blambda^{(\ell-1)}$, and the convergence criterion $\epsilon$.\\
\textbf{Initialize} the primal variables $\hat{\bbeta}^{(0)}= \hat{\br}^{(0)}=\boldsymbol{0}$ and the dual variable $\hat{\boldsymbol{\eta}}^{(0)} = \boldsymbol{0}$.\\
\textbf{Iterate} the following until the stopping criterion $\|\hat{\bbeta}^{(t)}-\hat{\bbeta}^{(t-1)}\|_2 \le \epsilon$ is met:
\begin{enumerate}
\item Update $\bbeta$ as
\[
 \hat{\bbeta}^{(t)} = \underset{\bbeta\in \RR^{p}}{\mathrm{argmin}}~ \left\{\frac{\rho}{2} \left\|\yb -\hat{\br}^{(t-1)}-\frac{1}{\sqrt{\rho}} \hat{\boldsymbol{\eta}}^{(t-1)}   -\Xb\bbeta  \right\|_2^2+ \|\blambda^{(\ell-1)}  \circ \bbeta_- \|_1 \right\}.
\]
\item Iterate the following until convergence: for each $i=1,\ldots,n$, update $r_i$ by solving
\[
 \tau- \Phi\left(\frac{-r_i}{h}\right) + \hat{\eta}_{i}^{(t-1)} + \rho \bigl(r_i - y_i + \langle \bx_i,\hat{\bbeta}^{(t)} \rangle\bigr)= 0.
 \]
\item Update $\boldsymbol{\eta}$ as
\[
    \hat{\boldsymbol{\eta}}^{(t)} =\hat{\boldsymbol{\eta}}^{(t-1)} +\rho\bigl(\hat{\br}^{(t)}-\yb + \Xb \hat{\bbeta}^{(t)}\bigr) .
\]
\end{enumerate}
\textbf{Output} the estimated parameter $\hat{\bbeta}^{(t)}$.
\end{algorithm}

 The updates for $\bbeta$ involves solving a Lasso regression problem for which efficient software is available. Alternatively, one can also linearize the loss function as in \citet{Gu2018} to obtain a closed-form solution.
 The updates for $\br$ can be obtained using coordinate descent algorithm by updating each coordinate of $\br$ using standard numerical methods such as the bisection method.
See Algorithm~\ref{alg:admm} for details. 

\section{Statistical theory}
\label{sec:theory}
In this section, we provide a comprehensive   analysis of the sequence of regularized quatile regression estimators $\{ \hat \bbeta^{(\ell)} \}_{\ell \geq 1}$ obtained by solving \eqref{weighted.lasso-lad} iteratively, initialized with $\hat \bbeta^{(0)} = \textbf{0}$. 
For simplicity, we restrict our attention to a fixed quantile level $\tau \in (0,1)$ of interest.
We first characterize the (deterministic) bias induced  by convolution smoothing described in Section~\ref{sec:theory:bias}.
In Section~\ref{sec:theory:lasso}, we provide  high probability bounds (under $\ell_1$- and $\ell_2$-errors) for the one-step estimator $\hat \bbeta^{(1)}$, i.e., the $\ell_1$-penalized smoothed QR estimator ($\ell_1$-SQR) which is of independent interest. With a flexible choice of the bandwidth $h$, these error bounds for $\hat \bbeta^{(1)}$ are near-minimax optimal \citep{Wang2019}, and coincide with those of the $\ell_1$-QR estimator \cite{BC2011}.
In Section~\ref{sec:theory:concave}, we analyze $\hat \bbeta^{(\ell)}$ ($\ell\geq 2$) whose overall estimation error consists of three parts: shrinkage bias, oracle rate, and smoothing bias. 
Our analysis reveals that the multi-step iterative algorithm refines the statistical rate in a sequential manner: every relaxation step shrinks the estimation error from the previous step by a $\delta$-fraction for some $\delta \in (0,1)$.  %
Under a necessary beta-min condition, we show that  the multi-step estimator $\hat \bbeta^{(\ell)}$ with $\ell \gtrsim \log\{\log (p)\}$ achieves the oracle rate of convergence, i.e., it shares the convergence rate of the oracle estimator that has access to the true active set. Under a sub-Gaussian condition on the feature vector and a stronger sample size requirement, we further show in Section~\ref{sec:oracle} that the multi-step estimator $\hat{\bbeta}^{(\ell)}$ with $\ell \gtrsim \log(s)$ coincides with the oracle estimator with high probability, and hence achieves variable selection consistency. Throughout, we   use the notation ``$\lesssim$" to indicate ``$\leq$" up to constants that are independent of $(s,p,n)$.

\subsection{Smoothing bias}
\label{sec:theory:bias}
To begin with, note that the smoothed quantile objective $\hat Q_h(\cdot)$ defined in \eqref{smooth.qloss} can be written as
 \[
  \hat Q_h(\bbeta) = (1-\tau) \int_{-\infty}^0 \hat F_h(u;\bbeta)  \,{\rm d} u + \tau \int_0^\infty \{ 1-\hat F_h(u;\bbeta) \} \,{\rm d}  u.
  \]
Recall the integrated kernel function $\bar{K}(u ) =  \int_{-\infty}^u K(t) \,{\rm d} t$, which is non-decreasing and takes values in $[0,1]$. With $r_i(\bbeta) = y_i - \bx_i^\T \bbeta$,  the gradient vector and Hessian matrix of $\hat Q_h(\bbeta)$ are, respectively,
\#
	\nabla  \hat Q_h(\bbeta)  = \frac{1}{n} \sn  \bigl\{ \bar{K}\bigl(  -r_i(\bbeta)/h \bigr) - \tau \bigr\} \bx_i  ~~\mbox{ and }~~
\nabla^2  \hat Q_h(\bbeta)  = \frac{1}{n} \sn K_h(-r_i(\bbeta)) \bx_i \bx_i^\T . \label{smooth.Hessian}
\#

To examine the bias induced by smoothing, define the expected smoothed loss function $Q_h (\bbeta) = \EE \{\hat Q_h(\bbeta)\}$, $\bbeta \in \RR^p$, and the pseudo parameter
\#
	 \bbeta^*_h  = (\beta^*_{h,1} ,\ldots, \beta^*_{h,p} )^\T \in \argmin_{\bbeta \in \RR^p } Q_h(\bbeta),  \label{pseudo.parameter}
\#
which is the population minimizer of the smoothed quantile loss and varies with $h$.
In general, $\bbeta^*_h$ differs from $\bbeta^*$ -- the unknown parameter vector in model \eqref{model}. The latter is identified as the unique minimizer of the population quantile objective $Q(\bbeta) := \EE \{\hat Q(\bbeta)\}$.
However, as the smoothed quantile loss $\ell_h(\cdot)$ in~\eqref{convolution.loss}  approximates the quantile loss $\rho_\tau(\cdot)$ as $h=h_n \to 0$, $\bbeta^*_h$ is expected to converge to $\bbeta^*$, and  we refer to $\| \bbeta^*_h - \bbeta^* \|_2$ as the approximation error or bias due to smoothing.

The following result provides upper bounds of the smoothing bias under mild conditions on the random covariates $\bx\in \RR^p$, the conditional density of $\varepsilon$ given $\bx$, and the kernel function. Throughout Section~\ref{sec:theory}, we assume that the second moment $\bSigma  = (\sigma_{jk})_{1\leq j, k\leq p} = \EE(  \bx    \bx^\T )$ of $\bx = (x_1, \ldots, x_p)^\T$ (with $x_1\equiv 1$) exists and is positive definite. Moreover, let $\gamma_1 = \gamma_1(\bSigma) \geq 1$, $\gamma_p = \gamma_p(\bSigma) \in (0, 1]$, and  $\sigma_{\bx}^2= \max_{1\leq j\leq p} \sigma_{jj}$.

\begin{enumerate}
\item[(B1)] The conditional density of $\varepsilon$ given $\bx$, denoted by $f_{\varepsilon|\bx}$, satisfies  $ f_l \leq f_{\varepsilon |\bx}(0) \leq f_u $ almost surely (over $\bx$) for some $f_u \geq f_l>0$.
Moreover, there exists a constant $l_0 >0$ such that $|f_{\varepsilon|\bx}(u)- f_{\varepsilon|\bx}(v)| \leq l_0 |u-v|$ for all $u,v\in \RR$ almost surely (over $\bx$).

\item[(B2)] The kernel function $K: \RR \to  [0,\infty)$ is symmetric around zero, and satisfies $\int_{-\infty}^\infty K(u) \, {\rm d} u=1$ and $\int_{-\infty}^\infty u^2 K(u) \, {\rm d} u <\infty$. For $\ell=1, 2, \ldots$, let $\kappa_\ell =  \int_{-\infty}^\infty |u|^\ell K(u) \, {\rm d}u$ be the $\ell$-th absolute moment of $K(\cdot)$.
\end{enumerate}

\begin{proposition} \label{prop:bias}
Assume that Conditions~(B1) and (B2) hold, and $\mu_3 := \sup_{\bu \in \mathbb S^{p-1}} \EE|\bz^\T \bu|^3<\infty$ with $\bz = \bSigma^{-1/2} \bx$.
Provided $0<h <  f_l / ( c_0 l_0)$, $\bbeta^*_h$ is the unique minimizer of $\bbeta \mapsto Q_h(\bbeta)$ and satisfies
\#
	      \| \bbeta^*_h - \bbeta^* \|_{\bSigma} \leq  c_0 l_0   f_l^{-1}  h^2 , \label{bias.ubd}
\#
where $c_0  = (\mu_3+\kappa_2)/2+\kappa_1$. In addition, assume $\kappa_3  <\infty$ and $f_{\varepsilon | \bx}$ has an $l_1$-Lipschitz continuous derivative almost everywhere for some $l_1>0$. Then
\# \label{bias.leading}
	 \biggl\|   \bSigma^{-1} \Jb( \bbeta^*_h - \bbeta^* ) +   \frac{1}{2}  \kappa_2 h^2 \cdot \bSigma^{-1} \EE   \bigl\{  f_{\varepsilon | \bx}'(0)\bx \bigr\}    \biggr\|_{\bSigma} \leq C  h^3,
\#
where $\Jb = \EE   \{  f_{\varepsilon|\bx}(0) \cdot  \bx \bx^\T  \}$, and $C >0$ depends only on $(f_l,  l_0, l_1, \mu_3 )$ and the kernel $K$.
\end{proposition}

Proposition~\ref{prop:bias} is a non-asymptotic version of Theorem~1 in \cite{FGH2019}, and explicitly captures the dependence of the bias on several model-based quantities.
Note that the $p\times p$ matrix $\Jb = \EE \{  f_{\varepsilon|\bx}(0) \cdot \bx \bx^\T  \}$ is the Hessian of the population quantile objective $Q(\cdot)$ evaluated at $\bbeta^*$, i.e., $\Jb = \nabla^2 Q(\bbeta^*)$. Under Condition~(B1), $f_l \gamma_p(\bSigma) \leq \gamma_p(\Jb) \leq \gamma_1(\Jb) \leq f_u  \gamma_1(\bSigma)$.
An interesting implication of Proposition~\ref{prop:bias}  is that, when both $f_{\varepsilon|\bx}(0)$ and $f'_{\varepsilon|\bx}(0)$ are independent of $\bx$ (i.e., $f_{\varepsilon|\bx}(0)=f_{\varepsilon }(0)$ and $f'_{\varepsilon|\bx}(0)=f'_{\varepsilon }(0)$),  the bias decomposition bound \eqref{bias.leading} simplifies to
\#
 \Bigg\|  f_{\varepsilon }(0) ( \bbeta_h^* - \bbeta^* ) + 0.5 f'_\varepsilon(0) \kappa_2 h^2 \begin{bmatrix}
1  \\
 \textbf{0}_{p-1}
\end{bmatrix} \Bigg\|_{\bSigma} \leq C  h^3. \nn
\#
In other words, the smoothing bias is concentrated primarily on the intercept. To some extent, this observation further certifies the benefit of smoothing  in variable selection of which the main focus is on the slope coefficients rather than the intercept.

\subsection{$\ell_1$-penalized smoothed quantile regression}
\label{sec:theory:lasso}

Given a bandwidth $h>0$ and a regularization parameter $\lambda>0$, let $\hat \bbeta_h = \hat \bbeta_h(\tau,\lambda)$ be the $\ell_1$-penalized SQR ($\ell_1$-SQR) estimator, defined as the solution to the following convex optimization problem:
\#
 \min_{\bbeta \in \RR^p} ~  \bigl\{ \hat Q_h  (\bbeta) + \lambda \| \bbeta \|_1 \bigr\}.
\label{lasso.sqr}
\#
In this section, we characterize the estimation error of $\hat \bbeta_h \in \RR^{p}$ under $\ell_2$- and $\ell_1$-norms. First we impose a moment condition on the (random) covariate vector $\bx = (x_1, \ldots, x_p)^\T \in \RR^p$ with $x_1 \equiv 1$. Without loss of generality, assume $\mu_j = \EE(x_j) = 0$ for $2\leq j\leq p$; otherwise, consider a change of variable $(\beta_1, \beta_2, \ldots, \beta_p)^\T \mapsto (\beta_1 + \sum_{j=2}^p \mu_j \beta_j, \beta_2, \ldots, \beta_p)^\T$ so that the obtained results apply to model $F_{y|\bx}^{-1}(\tau) = \beta^\flat_0 +    \sum_{j=2}^p (x_j-\mu_j) \beta^*_j$, where $ \beta^\flat_0 = \beta^*_0 + \sum_{j=2}^p \mu_j \beta^*_j$.

\begin{enumerate}
\item[(B3)] $\bSigma=\EE(\bx\bx^\T)$ is positive definite and $\bz = \bSigma^{-1/2} \bx \in\RR^p$ is sub-exponential: there exist   constants $\upsilon_0  ,c_0  \geq 1$ such that $ \PP(   | \bz^\T \bu | \geq \upsilon_0  \| \bu \|_2 \cdot  t ) \leq c_0 e^{-t }$ for  all $\bu \in \RR^{p }$ and $t\geq 0$.  For convenience, we assume $c_0=1$, and write $\sigma_{\bx}^2=\max_{1\leq j\leq p} \EE(x_j^2)$.
\end{enumerate}

Moreover,
for $r , l>0$, define the (rescaled) $\ell_2$-ball and $\ell_1$-cone as
\#
\BB_{\bSigma}(r) &= \{ \bdelta \in \RR^p : \|    \bdelta  \|_{\bSigma} \leq  r \} ~~\mbox{ and }~~  \CC_{\bSigma}(l)  = \bigl\{ \bdelta \in \RR^p : \| \bdelta \|_1 \leq l \|\bdelta \|_{\bSigma} \bigr\}    . \label{def:cone}
\#
Our theoretical analysis of the $\ell_1$-SQR estimator depends crucially on the following ``good" event, which is related to the local restricted strong convexity (RSC) of the empirical smoothed quantile loss function. We refer the reader to  \cite{NRWY2012} and \cite{LW2015} for detailed discussions of the restricted strong convexity for regularized $M$-estimation in high dimensions.

\begin{definition}(Local Restricted Strong Convexity) 
Given radius parameters $r, l>0$ and a curvature parameter $\kappa >0$, define the event
\#
 \cE_{{\rm rsc}}(r, l , \kappa) = \left\{ \frac{ \langle \nabla \hat Q_h(\bbeta) -   \nabla \hat Q_h(\bbeta^*) , \bbeta - \bbeta^* \rangle}{\| \bbeta - \bbeta^* \|_{\bSigma}^2}  \geq \kappa  ~\mbox{ for all }~ \bbeta \in \bbeta^* + \BB_{\bSigma}(r) \cap \CC_{\bSigma}(l)   \right\} . \label{event.rsc}
\#
\end{definition}

Our first result shows that, with suitably chosen $(r, l, \kappa)$, the event $ \cE_{{\rm rsc}}(r, l , \kappa)$ occurs with high probability. 
In order for the local RSC condition to hold, the radius parameter $r$ has to be of the same order as, or possibly smaller than the bandwidth $h$.

\begin{proposition} \label{prop:RSC}
Assume Conditions~(B1)--(B3) hold, and $\kappa_l = \min_{|u|\leq 1} K(u) >0$.  Moreover, let $(r, l, h)$ and $n$ satisfy 
\#
20 \upsilon_0^2  \,r \leq   h \leq  f_l/(2 l_0)   ~~\mbox{ and }~~   n \geq C \sigma_{\bx}^2  f_u   f_l^{-2} ( l/r)^2 h  \log(2p)    \label{RSC.scaling0}
\#
for a sufficiently large constant $C$. Then, the local RSC event $ \cE_{{\rm rsc}}(r, l , \kappa)$ with $\kappa = ( \kappa_l f_l)/2$ occurs with probability at least $1-(2p)^{-1}$.
\end{proposition}

\begin{remark}
We do not  claim that the values of the constants appearing in Proposition~\ref{prop:RSC} are optimal.
They result from non-asymptotic probabilistic bounds which reflect worst-case scenarios.
The condition $\min_{|u|\leq 1} K(u) >0$ is only for theoretical and notational convenience.
If the kernel $K(\cdot)$ is compactly supported on $[-1,1]$, we may rescale it to obtain $K_a(u) =(1/a)K(u/a)$ for some $a>1$. Then, $K_a(\cdot)$ is supported on $[-a,a]$ with $\min_{|u|\leq 1} K(u) >0$. For example,
\begin{itemize}
\item[(i)] (Gaussian kernel) if $K(u)= (2\pi)^{-1/2} e^{-u^2/2}$ is the Gaussian kernel, we have $\kappa_l = (2\pi e)^{-1/2} \approx 0.242$ and $\kappa_2 = 1$;

\item[(ii)] (Uniform kernel) if $K(u)= (1/2) \mathbbm{1}(|u|\leq 1)$  is the uniform kernel, we may consider its rescaled version  $K_{3/2}(u) = (1/3)\mathbbm{1}(|u|\leq 3/2)$. In this case, $\kappa_l= 1/3$ and $\kappa_2=3/4$.
\end{itemize}
Throughout, we view $(\kappa_l, \kappa_2)$ as absolute constants.
\end{remark}

\begin{theorem}
\label{thm:lasso-qr}
Under the conditional quantile model \eqref{model} with $\bbeta^* \in \RR^p$ being $s$-sparse, assume Conditions~(B1)--(B3) hold with $\kappa_l = \min_{|u|\leq 1} K(u)>0$.  
Then, the $\ell_1$-SQR estimator $\hat \bbeta = \hat \bbeta_h$ with $\lambda \asymp \sigma_{\bx} \sqrt{\tau(1-\tau)  \log (p)/n}$ satisfies the bounds
\#
 	\| \hat \bbeta   - \bbeta^* \|_2  \leq   C_1 f_l^{-1} s^{1/2} \lambda     ~~\mbox{ and }~~	 \| \hat \bbeta - \bbeta^* \|_1 \leq C_2   f_l^{-1}    s\lambda     \label{lasso.qr.bounds}
\#
with probability at least $1-  p^{-1}$, provided that  the bandwidth satisfies
$$
	\max\Bigg(\frac{ \sigma_{\bx}}{  f_l }\sqrt{\frac{s\log p}{n}}  , \frac{ \sigma_{\bx}^2 f_u }{ f_l^2 } \frac{s \log p}{n} \Bigg) \lesssim h   \leq \min\big\{  f_l/(2l_0) , (s^{1/2} \lambda)^{1/2} \big\} ,
$$
where the constants $C_1, C_2>0$ depend only on $(l_0, \upsilon_0, \gamma_p, \kappa_l, \kappa_2)$.
\end{theorem}

The above theorem shows that with a proper yet flexible choice of the bandwidth, the $\ell_1$-penalized smoothed QR estimator achieves the same rate of convergence as the $\ell_1$-QR estimator under both $\ell_1$- and $\ell_2$-errors \citep{BC2011}. Technically, we assume the random feature vector is sub-exponential, which is arguably the weakest moment condition in high-dimensional regression analysis under random design \citep{W2019}.
This preliminary result is of independent interest, and more importantly, it paves the way for further analysis of smoothed quantile regression with iteratively reweighted $\ell_1$-regularization.

\subsection{Concave regularization and oracle rate of convergence}
\label{sec:theory:concave}

In this section, we derive rates of convergence for the solution path $\{ \hat \bbeta^{(\ell)}\}_{\ell=1,2,\ldots}$ of the multi-step iterative algorithm defined in \eqref{weighted.lasso-lad}. Starting from $\hat \bbeta^{(0)}=\textbf{0}$, we note that  $\hat \bbeta^{(1)}$ is exactly the $\ell_1$-SQR estimator studied in the previous section; see Theorem~\ref{thm:lasso-qr}.
For subsequent $\hat{\bbeta}^{(\ell)}$'s, we first state the result as a deterministic claim in Theorem~\ref{thm:LLA}, but conditioned on some ``good" event  regarding the local RSC property and the gradient of $\hat Q_h(\cdot)$ at $\bbeta^*$.
Under Condition (B3) on the random covariate vector,  probabilistic claims enter in certifying that this ``good" event holds  with high probability with a suitable choice of $\lambda$ and $h$; see Theorem~\ref{thm:ncvx}.

Recall the event $\cE_{{\rm rsc}}(r, l, \kappa)$ defined in \eqref{event.rsc} on which a local RSC property of the smoothed quantile objective $\hat Q_h(\cdot)$ holds, where $\kappa$ is a curvature parameter.
Moreover, define
\#
 \bw^*_h = \bw_h(\bbeta^*)  \in \RR^p ~~\mbox{ and }~~ b_h^* = \| \bSigma^{-1/2}   \nabla Q_h(\bbeta^*)  \|_2 , \label{def:wb}
\#
where $\bw_h(\bbeta)= \nabla \hat Q_h(\bbeta) - \nabla Q_h(\bbeta)$ is the centered score function, and $b_h^* \geq 0$ quantifies the bias induced by smoothing. For the standard quantile loss, we have $ \nabla Q(\bbeta^*)=\textbf{0}$. Under Conditions~(B1) and (B2), examine the proof of Proposition~\ref{prop:bias} yields $b_h^* \leq  l_0 \kappa_2 h^2/2$, that is, the smoothing bias has magnitude of the order $h^2$.
To refine the statistical rate obtained in Theorem~\ref{thm:lasso-qr}, which is near-minimax optimal for estimating sparse targets, we need an additional beta-min condition on $\| \bbeta^*_{\cS} \|_{\min} = \min_{j \in \cS} |\beta^*_{j}| $, where $\cS = \{ 1 \leq j\leq p: \beta^*_j \neq 0 \}  $ is the active set of $\bbeta^*$. 
For a deterministic analysis, we first derive the contraction property of the solution path $\{ \hat \bbeta^{(\ell)} \}_{\ell \geq 1}$  conditioned on some ``good" event.

\begin{theorem} \label{thm:LLA}
Given $\kappa>0$ and a penalty function $q(\cdot)$ satisfying (A1), assume that  there exists some constant $\alpha_0>0$ such that
\#
  \frac{\alpha_0 }{\sqrt{1+ \{ q'(\alpha_0)/2 \}^2}}>  \frac{1}{\kappa \gamma_p}  ~~\mbox{ and }~~ q'(\alpha_0) >0   . \label{alpha0.constraint}
\#
Let the penalty level $\lambda$ and bandwidth $h$ satisfy $b_h^* \leq (s/\gamma_p)^{1/2} \lambda$.
Moreover, define $r_{{\rm opt}} = \gamma_p^{1/2} \alpha_0 c    s^{1/2} \lambda$ and $l=  \{  (2 + \frac{2}{q'(\alpha_0)}) (c^2+1)^{1/2}  + \frac{2}{q'(\alpha_0)}  \} (s/\gamma_p)^{1/2}$, where the constant $c>0$ is defined through the equation
\#
  0.5 q'(\alpha_0) (c^2+1)^{1/2} + 2 =    \alpha_0 \kappa \gamma_p  \cdot c. \label{eqn.c}
\#
Then, for any $r\geq r_{{\rm opt}}$, conditioned on the event $\cE_{{\rm rsc}} (r ,l,\kappa)  \cap \{  \| \bw_h^* \|_\infty \leq 0.5 q'(\alpha_0) \lambda \}$, the sequence of solutions $\{\hat \bbeta^{(\ell)}\}_{\ell \geq 1}$ to programs \eqref{weighted.lasso-lad} satisfies
\#
\| \hat \bbeta^{(\ell)} -\bbeta^* \|_{\bSigma} \leq \delta \cdot \| \hat \bbeta^{(\ell-1)} -\bbeta^* \|_{\bSigma} +  \underbrace{  \kappa^{-1} \gamma_p^{-1/2}  \bigl\{     \|  q_\lambda'(  (  |\bbeta^*_{ \cS}| -   \alpha_0 \lambda )_+ ) \|_2  +   \|   \bw_{h, \cS  }^*  \|_2    \bigr\} }_{=: r_{{\rm ora}}}  + \,\kappa^{-1} b^*_h ,  \label{contraction.inequality}
\#
where $\delta = \sqrt{1+ \{ q'(\alpha_0)/2 \}^2}/( \alpha_0  \kappa \gamma_p)  \in (0,1)$ and $u_+ = \max(u, 0)$.
In addition, 
\#
\| \hat \bbeta^{(\ell)} -\bbeta^* \|_{\bSigma}  & \leq  \delta^{\ell-1}  r_{{\rm opt}} + (1-\delta)^{-1} \big( r_{{\rm ora}}  + \kappa^{-1} b_h^* \big)  ~\mbox{ for any } \ell \geq 2  . \label{sequential.bound}
\# 
\end{theorem}

Theorem~\ref{thm:LLA} reveals how iteratively reweighted $\ell_1$-penalization refines the statistical rate in a sequential manner:
every relaxation step shrinks the estimation error from the previous step by a $\delta$-fraction. 
The error term that does not vary with reweighted penalization consists of
\#
	\underbrace{ \big\|  q_\lambda'\big(  ( |\bbeta^*_{ \cS}| -   \alpha_0 \lambda )_+ \big) \big\|_2}_{{\rm shrinkage~bias}}, \quad
	\underbrace{\big\|   \bw_{h, \cS  }^* \big\|_2}_{{\rm oracle~rate}}, ~~~~~\mbox{ and }~
	\underbrace{b_h^*}_{{\rm smoothing~bias}} . \nn
\#
The first term $ \|  q_\lambda'  ( (|\bbeta^*_{ \cS}| -   \alpha_0 \lambda )_+ ) \|_2$  is known as the shrinkage bias induced by the folded-concave penalty function \citep{FLSZ2018}. For the $\ell_1$-norm penalty, i.e., $q_\lambda(t) = \lambda |t|$ and $q_\lambda'(t) = \lambda \sgn(t)$, the  shrinkage bias can be as large as $s^{1/2} \lambda$.
Without any prior knowledge on the signal strength, we have $\|  q_\lambda'  ( (|\bbeta^*_{ \cS}| -   \alpha_0 \lambda )_+ ) \|_2 \leq \|  q_\lambda'  ( \bzero_{\cS} ) \|_2 = s^{1/2} \lambda $ for any penalty $q_\lambda$ satisfying Condition~(A1).
Assume $q_\lambda(t)=\lambda^2 q(t/\lambda)$ is a concave penalty defined on $\RR^+$ with $\alpha_* := \inf\{ \alpha>0: q'(\alpha) =0 \}<\infty$. Given a regularization parameter $\lambda>0$, consider the decomposition $\cS = \cS_0 \cup \cS_1$, where
\$
\cS_0  = \big\{  j \in \cS : |\beta_j | < (\alpha_0+ \alpha_*)\lambda \big\} ~~\mbox{ and }~~\cS_1  = \big\{  j \in \cS : |\beta_j | \geq  (\alpha_0+ \alpha_*)\lambda \big\} 
\$
have cardinalities $s_0$ and $s_1$, respectively. The shrinkage bias term can then be bounded by 
\$
\|  q_\lambda'  ( (|\bbeta^*_{ \cS}| -   \alpha_0 \lambda )_+ ) \|_2 \leq  \|  q_\lambda'  ( \bzero_{\cS_0} ) \|_2   = s_0^{1/2} \lambda .
\$
Under the beta-min condition $\| \bbeta^*_{ \cS} \|_{\min}\geq (\alpha_0+ \alpha_*)\lambda$, the shrinkage bias vanishes, and hence the final rate of convergence is determined by  $\| \bw^*_{h, \cS} \|_2$ and $b_h^*$. 
As previously noted, the latter is the smoothing bias term, and satisfies $b_h^* \leq l_0 \kappa_2 h^2/2$.

The terminology ``oracle" stems from the ``oracle estimator", defined as the QR estimator that knows in advance the true subset of the important features. For a better comparison, we define the oracle smoothed QR estimator as
\#
	\hat \bbeta^{\ora} = \argmin_{ \bbeta  \in \RR^p :  \bbeta_{ \cS^{\cc}} = \textbf{0}  }  \hat Q_h(\bbeta) =  \argmin_{ \bbeta \in \RR^p   :  \bbeta_{  \cS^{\cc}} = \textbf{0}  }\frac{1}{n} \sn \ell_h(y_i -    \bx_{i,  \cS}^\T  \bbeta_{ \cS}   ) , \label{def:oracle}
\#
where $\ell_h(\cdot)$ is the smoothed quantile loss given in \eqref{convolution.loss}.
As we will show in Section~\ref{sec:oracle},  the oracle SQR estimator $\hat \bbeta^{\ora}$ satisfies the bound
$$
	\| \hat \bbeta^{\ora} - \bbeta^* \|_2 \lesssim  \|  \bw^*_{h, \cS}  \|_2 + h^2
$$
with high probability, and $\|   \bw_{h, \cS  }^*  \|_2$ is of order $\sqrt{s/n}$.

Theorem~\ref{thm:LLA} is a deterministic result. Probabilistic claims enter in certifying that the local RSC condition holds with high probability (see Proposition~\ref{prop:RSC}), and in verifying that the ``good" event $\{ \| \bw^*_h \|_\infty \leq 0.5 q'(\alpha_0) \lambda \}$ occurs with high probability with a specified choice of $\lambda$.
The following theorem states, under a necessary beta-min condition, the iteratively reweighted $\ell_1$-penalized SQR (IRW-$\ell_1$-SQR) estimator $\hat \bbeta^{(\ell)}$, after a few iterations, achieves the estimation error of the oracle that knows the sparsity pattern of $\bbeta^*$.

\begin{theorem} \label{thm:ncvx}
In addition to Conditions (A1), (B1)--(B3), assume there exist $\alpha_1 > \alpha_0 >0$ such that
\#
	q'(\alpha_0)>0, \quad \frac{ \alpha_0 }{\sqrt{4 +\{ q'(\alpha_0\}^2 }}>    ( \kappa_l f_l \gamma_p)^{-1}  ~~\mbox{ and }~~ q'(\alpha_1 )=0  ,   \label{oracle.rate.cond}
\#  
where $\kappa_l = \min_{|u|\leq 1} K(u)>0$.  Moreover, let the regularization parameter $\lambda$ and bandwidth $h$ satisfy $\lambda \asymp \sigma_{\bx} \sqrt{\tau(1-\tau) \log(p)/n}$ and 
\$
	\max\Bigg(\frac{ \sigma_{\bx}}{ f_l }\sqrt{\frac{s\log p}{n}}  , \frac{ \sigma_{\bx}^2 f_u }{  f_l^2} \frac{s \log p}{n} \Bigg) \lesssim h  \lesssim (s^{1/2}\lambda)^{1/2} . 
\$
For any $t\geq 0$, under the beta-min condition $\| \bbeta_{\cS}^* \|_{\min} \geq (\alpha_0 + \alpha_1) \lambda$ and scaling $n\gtrsim \max\{ s\log(p), s+ t\}$, the IRW-$\ell_1$-SQR estimator $\hat \bbeta^{(\ell)}$ with $\ell \gtrsim \lceil \log\{\log (p)\}/\log(1/\delta) \rceil$ satisfies the bounds
\#
\|\hat{\bbeta}^{(\ell)} -\bbeta^* \|_{2} \lesssim  f_l^{-1} \Bigg( \sqrt{\frac{s+t}{n}} + h^2  \Bigg) ~~\mbox{ and }~~ \|\hat{\bbeta}^{(\ell)} -\bbeta^* \|_1  \lesssim  f_l^{-1} s^{1/2}  \Biggl( \sqrt{\frac{s+t}{n}} +   h^2 \Biggr)   \label{weak.oracle.rate}
\#
with probability at least $1- p^{-1} - e^{-t}$, where $\delta = \sqrt{4 + \{ q'(\alpha_0) \}^2}/(\alpha_0 \kappa_l f_l  \gamma_p ) \in (0,1)$.
\end{theorem}


\begin{remark}[Oracle rate of convergence and high-dimensional scaling]
The conclusion of Theorem~\ref{thm:ncvx} is referred to as the weak oracle property: the IRW-$\ell_1$-SQR estimator achieves the convergence rate of the oracle $\hat \bbeta^{{\rm ora}}$ when the support set $\cS$ were known {\it a priori}.
Starting from $\hat \bbeta^{(0)} = \textbf{0}$, the one-step estimator $\hat \bbeta^{(1)}$ ($\ell_1$-SQR) has an estimation error (under $\ell_2$-norm) of order $\sqrt{ s \cdot \log(p) /n}$ (see Theorem~\ref{thm:lasso-qr}). 
Under an almost necessary and sufficient beta-min condition---$\| \bbeta^*_{\cS} \|_{\min} \gtrsim \sqrt{\log(p)/n}$, a refined near-oracle statistical rate $\sqrt{s/n} + h^2$ can be attained by a multi-step iterative procedure, which solves a sequence of convex programs. Here, $\sqrt{s/n}$ is referred to as the oracle rate, and the $h^2$-term quantifies the smoothing bias (Proposition~\ref{prop:bias}). In order to certify the local RSC property of the smoothed objective function, the bandwidth should have magnitude at least of the order $\sqrt{s \log(p)/n}$. If we choose a bandwidth $h\asymp \sqrt{s\log(p)/n}$, the $\ell_2$-error of the multi-step  estimator will be of order $\sqrt{s/n} + s \log(p)/n$ under the high-dimensional scaling $n\gtrsim s \log(p)$. Intuitively, the main reason for having an extra term $s\log(p)/n$ is that even if the underlying vector $\bbeta^*$ is $s$-sparse, the population parameter $\bbeta^*_h \in \RR^p$ corresponding to the smoothed objective function (see \eqref{pseudo.parameter}) may be denser.  As a result, there is a statistical price to pay for smoothing.
\end{remark} 

\begin{remark}[Minimum signal strength and oracle rate]
In a linear regression model $y = \bx^\T \bbeta^*+ \varepsilon$ with a Gaussian error $\varepsilon \sim N(0, \sigma^2)$, consider the parameter space $\Omega_{s, a} = \{ \bbeta \in \RR^p: \| \bbeta \|_0 \leq s, \min_{j : \beta_j \neq 0} |\beta_j| \geq a\}$ for $a>0$. 
Assuming that the design matrix $\mathbb{X} = (\bx_1, \ldots, \bx_n)^\T \in \RR^{n\times p}$ satisfies a restricted isometry property and has normalized columns (each column has an $\ell_2$-norm equal to $\sqrt{n}$), \cite{N2019} derived the following sharp lower bounds for the minimax risk $\psi(s,a) := \inf_{\hat \bbeta} \sup_{\bbeta^* \in \Omega_{s,a}} \EE \| \hat \bbeta - \bbeta^* \|_2^2$: for any $\epsilon\in (0,1)$,
\$
\psi(s,a)  \geq \{ 1 + o(1) \} \frac{2\sigma^2 s \log(ep/s)}{n} ~\mbox{ for any } a \leq (1-\epsilon) \sigma \sqrt{\frac{2\log(ep/s)}{n}}
\$
and 
\$
\psi(s,a) \geq  \{ 1 + o(1) \} \frac{ \sigma^2 s }{n} ~\mbox{ for any } a \geq (1 + \epsilon) \sigma \sqrt{\frac{2\log(ep/s)}{n}} ,
\$
where the limit corresponds to $s/p \to 0$ and $s\log(ep/s) /n \to 0$. The minimax rate $2\sigma^2 s \log(ep/s)/n$ can be attained by both Lasso and Slope \citep{BLT2018},  while the oracle rate $\sigma^2 s/n$ can only be achieved when the magnitude of the minimum signal is of order $\sigma \sqrt{\log(p/s)/n}$. For estimating an $s$-sparse vector $\bbeta^* \in \RR^p$ in the conditional quantile model \eqref{model}, \cite{Wang2019} proved the lower bound $\sqrt{s\log(p/s) /n}$ for the minimax estimation error under $\ell_2$-norm. In order to achieve the refined oracle rate, \cite{FXZ2014} required a stronger beta-min condition, i.e., $\|\bbeta^*_{\cS}\|_{\min }\gtrsim \sqrt{s\log(p)/n}$, and a stringent independence assumption between $\varepsilon$ and $\bx$ in the conditional quantile model \eqref{model}. The beta-min condition imposed in Theorems~\ref{thm:LLA} and \ref{thm:ncvx} is almost necessary and sufficient, and is the weakest possible up to constant factors.
\end{remark}

\subsection{Strong oracle property}
\label{sec:oracle}
In this section, we establish the strong oracle property for the multi-step estimator $\hat \bbeta^{(\ell)}$ when $\ell$ is sufficiently large, i.e., $\hat \bbeta^{(\ell)}$ equals the oracle estimator $\hat \bbeta^\ora$ with high probability \citep{FL2011}.
To this end, we define a similar local RSC event to $\cE_{{\rm rsc}}(r, l, \kappa)$ given in \eqref{event.rsc}. Recall that $\cS \subseteq [p]$ is the support of $\bbeta^*$. Given radius parameters $r, l>0$ and a curvature parameter $\kappa>0$, define
\#
 \cG_{{\rm rsc}} (r, l ,\kappa) = \left\{  \frac{\langle \hat Q_h(\bbeta_1) - \nabla \hat Q_h(\bbeta_2) , \bbeta_1 - \bbeta_2  \rangle }{ \| \bbeta_1 - \bbeta_2 \|_{\bSigma}^2 }  \geq \kappa  ~\mbox{ for all } (\bbeta_1 , \bbeta_2 ) \in \Lambda(r, l)  \right\}  , \label{event.rsc2}
\#
where $\Lambda(r, l)  := \{ (\bbeta_1 , \bbeta_2 ): \bbeta_1 \in \bbeta_2 + \BB_{\bSigma}(r) \cap \CC_{\bSigma}(l), \bbeta_2 \in \bbeta^* + \BB_{\bSigma}(r/2) , {\rm supp} (\bbeta_2) \subseteq \cS  \}$. Similarly to \eqref{def:wb}, we  define the oracle score 
\#
	\bw_h^\ora = \nabla  \hat Q_h( \hat \bbeta^\ora)  \in \RR^p ,  \label{def:oracle.score}
\#
where $\hat \bbeta^\ora$ is defined in \eqref{def:oracle}. By the optimality of $\hat \bbeta^\ora$, we have  
$ \bw^\ora_{h, \cS} = (-1/n) \sn \ell_h'(y_i - \bx_{i,\cS}^\T  \hat \bbeta^\ora_{\cS} ) \bx_{i,\cS} =  \textbf{0}_s$. Like Theorem~\ref{thm:LLA}, the following result is also deterministic given the stated conditioning.

\begin{theorem} \label{thm:strong.oracle1}
Assume Condition~(A1) holds, and for some predetermined $\delta \in (0,1)$ and $\kappa>0$, there exist constants $\alpha_1 > \alpha_0 >0$ such that 
\#
	q'(\alpha_0 ) >0 , \quad  \frac{\alpha_0}{\sqrt{1+ \{q'(\alpha_0) /2 \}^2}} >  \frac{1}{\delta \kappa \gamma_p} ~~\mbox{ and }~~ q'(\alpha_1) = 0.  \label{oracle.cond}
\#
Moreover, let $r\geq  \gamma_p^{1/2}    \alpha_0 c_1 s^{1/2} \lambda $ and $l= \{ 2 + \frac{2}{q'(\alpha_0)} \}(c_1^2+1)^{1/2} (s/\gamma_p)^{1/2}$, where $c_1>0$ is a constant determined by 
\#
    0.5 q'(\alpha_0)  (c_1^2+1)^{1/2} + 1=  \alpha_0  \kappa \gamma_p c_1 .  \label{eqn.c1}
\#
Assume the beta-min condition $\| \bbeta^*_{\cS} \|_{\min} \geq (\alpha_0+\alpha_1) \lambda$ holds. Then, conditioned on the event 
\#
 \big\{ \| \bw^{\ora}_h \|_\infty \leq 0.5 q'(\alpha_0) \lambda \big\} & \cap  \big\{ \| \hat \bbeta^{\ora} - \bbeta^* \|_{\bSigma} \leq r /2 \big\}  \cap 
 \cG_{{\rm rsc}}(r, l, \kappa) \nn \\
 & \cap \left\{ \| \hat \bbeta^{\ora} - \bbeta^* \|_\infty \leq  \left[ \alpha_0- \frac{\sqrt{1+ \{q'(\alpha_0) /2 \}^2}}{\delta \kappa \gamma_p } \right] \lambda   \right\} ,  \label{oracle.events}
\#
the strong oracle property holds: $\hat \bbeta^{(\ell)} = \hat \bbeta^\ora$ provided $\ell \geq \lceil  \log(s^{1/2}/\delta) / \log(1/\delta)\rceil$.
\end{theorem}

Our next goal is is to control the probability of the events in \eqref{oracle.events}.  To this end, we need the following statistical properties of the oracle estimator $\hat \bbeta^{\ora}$, including a deviation bound and a non-asymptotic  Kiefer-Bahadur representation  that are of independent interest. The latter requires a slightly stronger moment condition on the random feature.

\begin{enumerate}
\item[(B1$'$)] In addition to Condition~(B1), assume $\sup_{u\in \RR}|f_{\varepsilon|\bx}(u)| \leq f_u  <\infty$ almost surely over $\bx$. 

\item[(B2$'$)] In addition to Condition~(B2), assume $\sup_{u\in \RR} K(u) \leq \kappa_u$ for some $ \kappa_u \in (0, 1]$.

\item[(B3$'$)] The (random) covariate vector $\bx = \bSigma^{1/2} \bz \in \RR^p$ is sub-Gaussian: there exists some  $\upsilon_1  \geq 1$ such that $\PP(   | \bz^\T \bu | \geq \upsilon_1 \| \bu \|_2 \cdot  t  )  \leq 2 e^{-t^2/2 }$ for  all $\bu \in \RR^{p }$ and $t\geq 0$.
\end{enumerate}

Note that the oracle $\hat \bbeta^{\ora} \in \RR^p$ with $\hat \bbeta^{\ora}_{\cS^{\cc}}=\textbf{0}$ is essentially an unpenalized smoothed QR estimator in the low-dimensional regime ``$s\ll n$''. We refer to \cite{FGH2019} for a comprehensive asymptotic analysis when $s$ is fixed, and \cite{HPTZ2020} for a finite sample theory when $s$ is allowed to grow with $n$. 
This paper concerns the case where both $s$ (intrinsic dimension) and $p$ (ambient dimension) can grow with sample size $n$. We therefore summarize the estimation bound and Bahadur representation for $\hat \bbeta^{\ora}_{\cS}$ by \cite{HPTZ2020} in the following proposition. Let 
\#
	\Sb = \EE (\bx_{\cS} \bx_{\cS}^\T) ~~\mbox{ and }~~ \Db  = \EE \{ f_{\varepsilon | \bx}(0) \cdot \bx_{\cS} \bx_{\cS}^\T \}  \label{def.submatrix}
\#
be, respectively, the $s\times s$ sub-matrices of $\bSigma$ and $\Jb$ indexed by the true support $\cS \subseteq [p]$.

\begin{proposition} \label{thm:oracle.smoothqr}
Assume Conditions~(B1$'$)--(B3$'$) hold.  For any $t\geq 0$, suppose the sample size $n$ and the bandwidth $h=h_n$ are such that $n\gtrsim s + t$ and $\sqrt{(s+t)/n }\lesssim h \lesssim 1$. Then, the oracle estimator $\hat \bbeta^\ora$ defined in \eqref{def:oracle} satisfies
\#
 \|  \hat \bbeta^\ora - \bbeta^*  \|_{\bSigma}  =  \| (\hat \bbeta^\ora - \bbeta^*)_{  \cS} \|_{\Sb} \lesssim  f_l^{-1}\Bigg( \sqrt{ \frac{s+t }{n}}  + h^2 \Bigg) 	 \label{oracle.concentration}
\#
with probability at least $1-2e^{-t}$. Moreover,
\#
	 \bigg\|  \Db  (\hat \bbeta^\ora - \bbeta^*)_{\cS} +  \frac{1}{n}  \sn  \bigl\{   \bar K(-\varepsilon_i/h) - \tau  \bigr\} \bx_{i, \cS}  \biggr\|_{\Sb^{-1}}  \lesssim   \frac{ s +t}{  h^{1/2} n   } +   h  \sqrt{\frac{s+t }{n}}  + h^3 \label{oracle.bahadur}
\#
with probability at least $1- 3e^{-t}$.
\end{proposition}

Finally, with the above preparations, we are able to establish the strong oracle property of $ \hat \bbeta^{(\ell)}$  when $\ell$ is sufficiently large.

\begin{theorem} \label{thm:strong.oracle2}
Assume Conditions (B1$'$)--(B3$'$) and (A1) hold with $\kappa_l = \min_{|u|\leq 1} K(u)>0$ and 
\# \label{irrepresentable.cond}
	  \max_{j \in  \cS^\cc}   \|  \Jb_{j \cS}  (\Jb_{\cS \cS})^{-1}   \|_1 \leq A_0 .
\#
for some $A_0\geq 1$. For a prespecified $\delta \in (0,1)$, suppose there exist constants $\alpha_1>\alpha_0$ satisfying \eqref{oracle.cond} with $\kappa= \kappa_l f_l/2$, and the beta-min condition $\| \bbeta_{\cS}^* \|_{\min} \geq (\alpha_0  + \alpha_1) \lambda$.
Choose the bandwidth $h$ and penalty level $\lambda$ as $h \asymp \{ \log(p)/n \}^{1/4}$ and $\lambda \asymp \sqrt{\log (p)/n}$. Then, with probability at least $1-2p^{-1}-5n^{-1}$, $\hat \bbeta^{(\ell)} = \hat \bbeta^\ora$ for all $\ell \geq \lceil  \log(s^{1/2}/\delta) / \log(1/\delta)\rceil$, provided that the sparsity $s$ and ambient dimension $p$ obey the growth condition $\max\{ s^2 \log(p) , s^{8/3}/(\log p) \} \lesssim n$.
\end{theorem}

As stated in Theorem~\ref{thm:strong.oracle2}, in addition to the beta-min condition $\| \bbeta^*_{\cS} \|_{\min} \gtrsim \sqrt{\log(p)/n}$, we need an extra assumption \eqref{irrepresentable.cond} to establish the strong oracle property. Informally speaking, if we regress every spurious (density-weighted) feature $f_{\varepsilon | \bx}(0) \cdot x_{j}$ ($j\in \cS^{\cc}$) on the important (density-weighted) features $f_{\varepsilon | \bx}(0) \cdot x_{\cS}$,  \eqref{irrepresentable.cond} requires the $\ell_1$-norm of the resulting regression coefficient vector to be bounded by $A_0$.
It is worth noting that assumption \eqref{irrepresentable.cond} is much weaker than the irrepresentable condition, which is sufficient and nearly necessary for model consistency of the Lasso \citep{ZY2006,MB2006, L2021}  in the conditional mean model. A population version of  the irrepresentable condition is that, for some $\alpha \in (0,1)$, $\max_{j\in \cS^{\cc}} \| \bSigma_{j \cS} (\bSigma_{\cS \cS} )^{-1}  \|_1 \leq \alpha$.

For conditional mean regression with heavy-tailed errors, \cite{L2017} established the strong oracle property for any local stationary point of the folded concave penalized optimization problem \eqref{l1.qr} subject to an $\ell_1$-ball constraint, when the loss function is twice differentiable. The required growth condition on $(s, p)$ is $\max\{   s\log(p) , s^2 \} \lesssim n$; see Theorem~2 in  \cite{L2017}. For sparse quantile regression, our result requires a slightly stronger scaling   $\max\{ s^2 \log(p) , s^{8/3}/(\log p) \} \lesssim n$ due to the non-smoothness of the quantile loss.
Intuitively, the strong oracle property is related to the second-order accuracy and efficiency: the oracle estimator is asymptotically normal provided that the sparsity $s$ does not grow too fast with the sample size.  For Huber's $M$-estimator, \cite{HS2000} proved the asymptotic normality for its linear functionals under the scaling $s^2 \log(s) = o(n)$; while in the context of quantile regression, the same asymptotic results usually hold under stronger growth conditions due to both non-linearity and non-smoothness of the problem, such as $s^3 (\log n)^2 = o(n)$ \citep{W1989,HS2000} and $s^{8/3} = o(n)$ \citep{HPTZ2020}. To some extent, this explains why the high-dimensional scaling in our Theorem~\ref{thm:strong.oracle2} is slightly stronger than those needed for regularized $M$-estimators with smooth loss functions.

\section{Numerical study}
\label{sec:numerical}
We perform numerical studies to assess the performance of the proposed  regularized quantile regression method using $\ell_1$ and SCAD penalties.
The SCAD penalty \citep{FL2001} is defined through its derivative that takes the form $q'_\lambda(t) =\lambda  \mathbbm{1}(t\leq \lambda) + (a-1)^{-1} (a \lambda - t)_+ \mathbbm{1} (t>\lambda)$ for $t\geq 0$, where we pick $a=3.7$ as suggested in \citet{FL2001}, although it may not be the optimal value for quantile regression.
We use uniform and Gaussian kernels to smooth the quantile loss, and then employ the multi-stage convex relaxation method described in Algorithm~\ref{alg:multistage} with $\ell=3$ iterations. We will show later in this section that for moderately large $p$,
 $\ell=3$ iterations is often sufficient and that more iterations will lead to little to no improvement in terms of estimation accuracy.

We compare our proposal---iteratively reweighted $\ell_1$-penalized smoothed quantile regression, with the standard Lasso implemented by the \texttt{R} packageg \texttt{glmnet}, and both $\ell_1$- and folded concave penalized  quantile regressions implemented by the \texttt{R} package \texttt{FHDQR} \citep{Gu2018}.
As a benchmark, we also compute the oracle estimator by fitting unpenalized quantile regression using the important covariates.  The regularization parameter $\lambda$ for Lasso and penalized QR is selected via five-fold cross-validation; for the latter, we use the check loss to define the validation error.
Specifically, we choose the $\lambda$ value that yields the minimum cross-validation error under the $\ell_2$-loss and check loss for Lasso and penalized QR, respectively.
The proposed method involves a smoothing parameter $h$,  which can also be tuned via cross-validation in practice.
Recall that convolution smoothing facilitates optimization through a balanced trade-off between statistical accuracy and computational complexity. Our numerical experiments show that the results are rather insensitive to the choice of the bandwidth provide that it is in a reasonable range (neither too small nor too large). The default value of $h$ is set to be $\max\{ 0.05, \sqrt{\tau(1-\tau)} \{\log (p)/n\}^{1/4} \}$. We note that this particular choice of $h$ is by no means optimal numerically.

For all the numerical experiments,  we generate synthetic data $\{ ( y_i , \bx_i) \}_{i=1}^n$ from a linear model $y_i =  \bx_i^\T \bbeta^*  + \varepsilon_i$ with $\bbeta^*=(1.8,0,1.6,0,1.4,0,1.2,0,1,0,-1,0,-1.2,0,-1.4,0,-1.6,0,-1.8,\mathbf{0}_{p-19})^{\T}$, and $\bx_i \sim N_p(\mathbf{0},\bSigma)$ with $\bSigma = (0.7^{|j-k|})_{1\leq j, k \leq p}$.
The random error follows one of the following four distributions: (i) standard normal distribution $N(0,1)$;  (ii) $t$-distribution with $1.5$ degrees of freedom; (iii) standard Cauchy distribution; and (iv) a mixture of normal distributions -- $0.7N(0,1) + 0.3N(0,25)$. 

To evaluate the performance across different methods, we report the true and false positive rates (TPR and FPR), defined as the proportion of  correctly estimated nonzeros and the proportion of falsely estimated nonzeros, respectively.
We also report the sum of squared errors (SSE), i.e., $\| \hat \bbeta -\bbeta^* \|_2^2$.
Results for four different noise distributions under moderate ($n=500, p=400$) and high-dimensional settings ($n=500,p=1000$), averaged over 100 replications, are displayed in Tables~\ref{tablegaussian}--\ref{tablemixture}.

Under the Gaussian random noise, we see from Table~\ref{tablegaussian} that all methods have similar TPR and FPR.  The Lasso has the lowest SSE compared to QR-Lasso and SQR-Lasso, which coincides with the fact that quantile regression does lose some efficiency in a normal model.
For both standard and smoothed quantile regressions,  iteratively reweighted regularization with the SCAD penalty considerably reduces the estimation error,  is proximate to the oracle procedure.  Similar results hold when the minimax concave penalty is used.
This supports our theoretical results on SQR that concave regularization improves the estimation error from $\sqrt{s\log (p)/n}$ to the near-oracle rate $\sqrt{\{s+\log(p)\}/n}$.
Among all regularized quantile regression methods,  the proposed procedure---iteratively reweighted $\ell_1$-penalized SQR with either uniform or Gaussian kernel smoothing---has the best overall performance.


\begin{table}[!htp]
\footnotesize
{\scriptsize
\begin{center}
\caption{Numerical comparisons  under Gaussian model.  The empirical average (and standard error) of the true and false positive rates (TPR and FPR) as  well as the sum of squared errors (SSE),  over 100 simulations, are reported.  
}
\begin{tabular}{l | c    c c|c cccc}
  \hline
  & \multicolumn{3}{c}{Moderate Dimension ($n=500$, $p=400$)}& \multicolumn{3}{c}{High Dimension ($n=500$, $p=1000$)} \\\hline
Methods&TPR&FPR & Error&TPR&FPR &Error\\
\hline
Lasso   & 1 (0)  & 0.067 (0.003)  & 0.147 (0.006) & 1 (0) & 0.033 (0.001)& 0.167 (0.006)   \\
{SCAD}   & 1 (0) & 0.055 (0.003) & 0.062 (0.012)  & 1 (0) &0.026 (0.001)& 0.051 (0.003)  \\
{QR-Lasso}   & 1 (0)  & 0.119 (0.006) & 0.240 (0.009)  &1 (0) & 0.068 (0.003)& 0.284 (0.009)   \\
{QR-SCAD}   & 1 (0) & 0.112 (0.006)& 0.183 (0.014)  &1 (0)  & 0.069 (0.004)& 0.161 (0.010)   \\
{SQR-Lasso (uniform)}  &1 (0) & 0.066 (0.003)  & 0.224 (0.013)   &1 (0)& 0.036 (0.002)& 0.234 (0.007)   \\
{SQR-SCAD (uniform)}   & 1 (0) &0.057 (0.004) & 0.129 (0.011) & 1 (0)& 0.032 (0.002)& 0.116 (0.008)   \\
{SQR-Lasso (Gaussian)}   & 1 (0)  & 0.072 (0.004) & 0.191 (0.007)  &1 (0)& 0.034 (0.002)& 0.223 (0.007)   \\
{SQR-SCAD (Gaussian)}   &  1 (0)  & 0.056 (0.003) & 0.131 (0.010)    &1 (0)& 0.028 (0.002)& 0.108 (0.007)   \\
{Oracle}   & 1 (0)  & 0 (0) & 0.049 (0.003)  &1 (0)& 0 (0) &  0.053 (0.003)\\
\hline
\end{tabular}
\label{tablegaussian}
\end{center}
}
\end{table}

Next, we examine the performance of  different methods when outliers are present.  From Table~\ref{tablet} we see that the Lasso has the highest SSE with TPR merely above 0.5 in both moderate- and high-dimensional settings.  In contrast,  regularized quantile regression methods have high TPR while maintain low FPR.  The FPR and SSE for SQR are further reduced by a visible margin when the SCAD penalty is used.   This corroborates our main message that high-dimensional quantile regression significantly benefits from smoothing and non-convex regularization. 
Similar results can be found in Table~\ref{tablecauchy} and \ref{tablemixture} for Cauchy  and a mixture normal error distributions. 

\begin{table}[!htp]
\footnotesize
{\scriptsize
\begin{center}
\caption{Numerical comparisons under $t_{1.5}$ model.} 
\begin{tabular}{l | c    c c|c cccc}
  \hline
  & \multicolumn{3}{c}{Moderate Dimension ($n=500$, $p=400$)}& \multicolumn{3}{c}{High Dimension ($n=500$, $p=1000$)} \\\hline
Methods&TPR&FPR &Error &TPR&FPR &Error\\
\hline
Lasso   & 0.908 (0.016)  & 0.052 (0.002)  & 4.615 (0.401) & 0.854 (0.022) & 0.023 (0.001)   &5.668 (0.524)   \\
{SCAD}   & 0.842 (0.020)  & 0.044 (0.002) & 7.138 (0.739 &0.790 (0.024)& 0.019 (0.001) & 8.253 (0.762)   \\
{QR-Lasso}   & 1  (0) & 0.112 (0.005)& 0.417 (0.015)  & 1  (0) & 0.065 (0.003)& 0.541 (0.021)   \\
{QR-SCAD}   &1  (0)  &0.103 (0.005) & 0.346 (0.024) & 1 (0)& 0.062 (0.003)& 0.362 (0.022)   \\
{SQR-Lasso (uniform)}   & 0.999 (0.001) & 0.067 (0.004) & 0.387 (0.032)  &1 (0) & 0.032 (0.002)& 0.433 (0.017)    \\
{SQR-SCAD (uniform)}   & 0.999 (0.001) & 0.055 (0.004) & 0.266 (0.028)  & 1 (0)& 0.028 (0.002)& 0.230 (0.017)   \\
{SQR-Lasso (Gaussian)}   &1  (0) & 0.066 (0.003)& 0.332 (0.012)   & 1 (0)& 0.030 (0.001)& 0.420 (0.017)   \\
{SQR-SCAD (Gaussian)}   & 1  (0) & 0.048 (0.003)& 0.238 (0.018)    &1 (0)& 0.024 (0.001)& 0.220 (0.015)   \\
{Oracle}   & 1 (0)  & 0 (0) & 0.065 (0.004)  &1 (0)& 0 (0)& 0.074 (0.004)   \\
\hline
\end{tabular}
\label{tablet}
\end{center}
}
\end{table}

\begin{table}[!htp]
{\scriptsize
\begin{center}
\caption{Numerical comparisons under Cauchy model.}
\begin{tabular}{l | c    c c|c cccc}
  \hline
  & \multicolumn{3}{c}{Moderate Dimension ($n=500$, $p=400$)}& \multicolumn{3}{c}{High Dimension ($n=500$, $p=1000$)} \\\hline
Methods&TPR&FPR &Error &TPR&FPR &Error\\
\hline
Lasso   &0.344 (0.032) & 0.021 (0.003)& 16.799 (0.522)  & 0.305 (0.033) & 0.009 (0.001)& 17.479 (0.953)   \\
{SCAD}   & 0.297 (0.028) & 0.020 (0.002)& 20.382 (0.860)& 0.272 (0.029)& 0.009 (0.001)& 19.526 (0.871)  \\
{QR-Lasso}   & 1 (0)  & 0.118 (0.004) & 0.546 (0.022)  & 1 (0)& 0.060 (0.002)& 0.709 (0.025)   \\
{QR-SCAD}   & 1 (0) & 0.112 (0.005) & 0.585 (0.047) & 1 (0)& 0.058 (0.002)& 0.473 (0.034)   \\
{SQR-Lasso (uniform)}   & 0.990 (0.004)  & 0.054 (0.002) & 0.628 (0.070)  &0.999 (0.010)& 0.030 (0.002)& 0.588 (0.042)   \\
{SQR-SCAD (uniform)}   & 0.992 (0.004)  & 0.045 (0.003) & 0.391 (0.047)  & 0.998 (0.002)& 0.026 (0.001)& 0.308 (0.031)   \\
{SQR-Lasso (Gaussian)}   & 1 (0) & 0.058 (0.002) & 0.434 (0.017)  &   1 (0) & 0.028 (0.001)& 0.533 (0.019)\\
{SQR-SCAD (Gaussian)}   & 1 (0) & 0.042 (0.002) & 0.298 (0.021)  &  1 (0) & 0.022 (0.001)& 0.276 (0.021) \\
{Oracle}   & 1 (0) & 0 (0) & 0.076 (0.004) &1 (0)& 0 (0)& 0.080 (0.004) \\
\hline
\end{tabular}
\label{tablecauchy}
\end{center}
}
\end{table}

\begin{table}[!htp]
{\scriptsize
\begin{center}
\caption{Numerical comparisons under mixture normal model.}
\begin{tabular}{l | c    c c|c cccc}
  \hline
  & \multicolumn{3}{c}{Moderate Dimension ($n=500$, $p=400$)}& \multicolumn{3}{c}{High Dimension ($n=500$, $p=1000$)} \\\hline
Methods&TPR&FPR &Error &TPR&FPR &Error\\
\hline
Lasso   & 0.999 (0.001) & 0.062 (0.003)  & 1.253 (0.058) & 1 (0) & 0.030 (0.001)& 1.346 (0.047)   \\
{SCAD}   & 0.996 (0.002) & 0.048 (0.002)  & 0.606 (0.063)  & 0.995 (0.002)& 0.025 (0.001)& 0.746 (0.070)  \\
{QR-Lasso}   & 1 (0) & 0.126 (0.005)& 0.507 (0.019)  &1 (0) & 0.059 (0.002)& 0.559 (0.017)   \\
{QR-SCAD}   &  1 (0) & 0.121 (0.006)& 0.546 (0.041) &1 (0)  & 0.057 (0.002)& 0.361 (0.020)   \\
{SQR-Lasso (uniform)}   & 0.999 (0.001)  & 0.070 (0.004)& 0.496 (0.040) &1 (0)& 0.030 (0.002)& 0.462 (0.013)   \\
{SQR-SCAD (uniform)}   & 1 (0)  & 0.060 (0.004)& 0.366 (0.029)  & 1 (0)& 0.026 (0.002)& 0.244 (0.016)   \\
{SQR-Lasso (Gaussian)}   &   1 (0) & 0.072 (0.003) & 0.405 (0.015)  &1 (0)& 0.029 (0.001) & 0.443 (0.013)   \\
{SQR-SCAD (Gaussian)}   &  1 (0) & 0.054 (0.003) & 0.346 (0.024)  &1 (0)& 0.024 (0.001) & 0.242 (0.015)  \\
{Oracle}   & 1 (0) & 0 (0) & 0.087 (0.005)  &1 (0)& 0 (0) &  0.086 (0.004)\\
\hline
\end{tabular}
\label{tablemixture}
\end{center}
}
\end{table}

\begin{figure}[!t]
\centering
\includegraphics[scale=0.39]{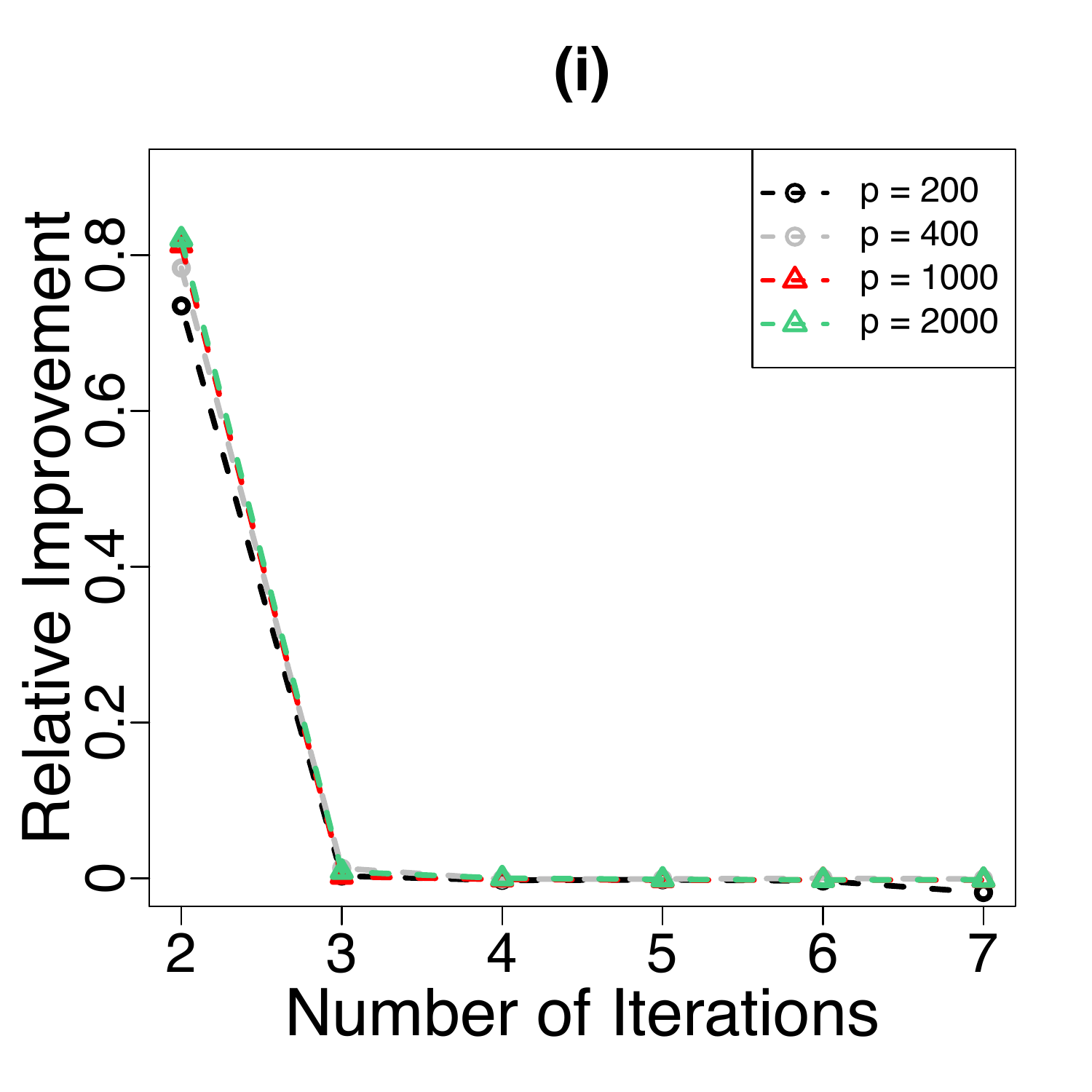}~~~
\includegraphics[scale=0.39]{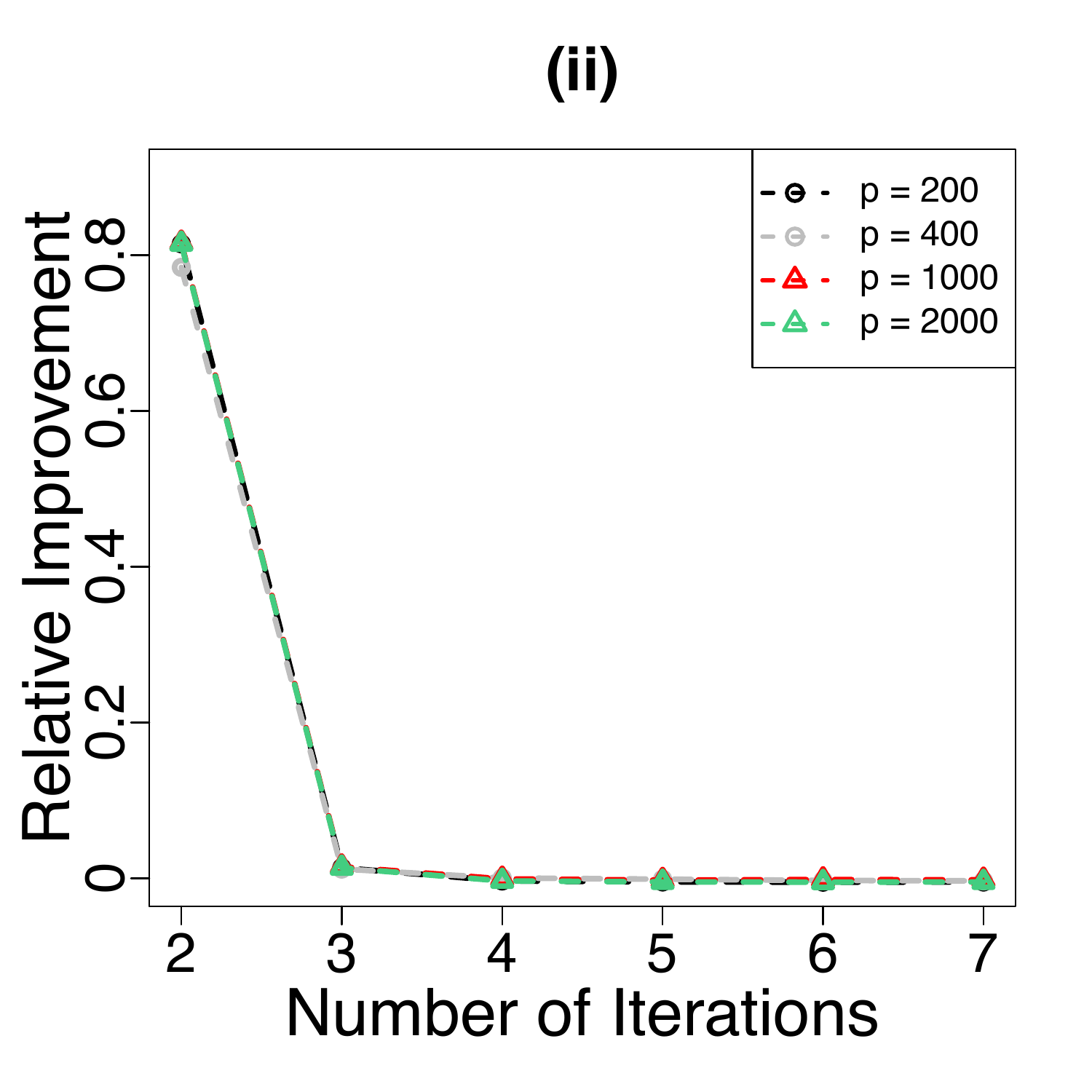}~~~
\includegraphics[scale=0.39]{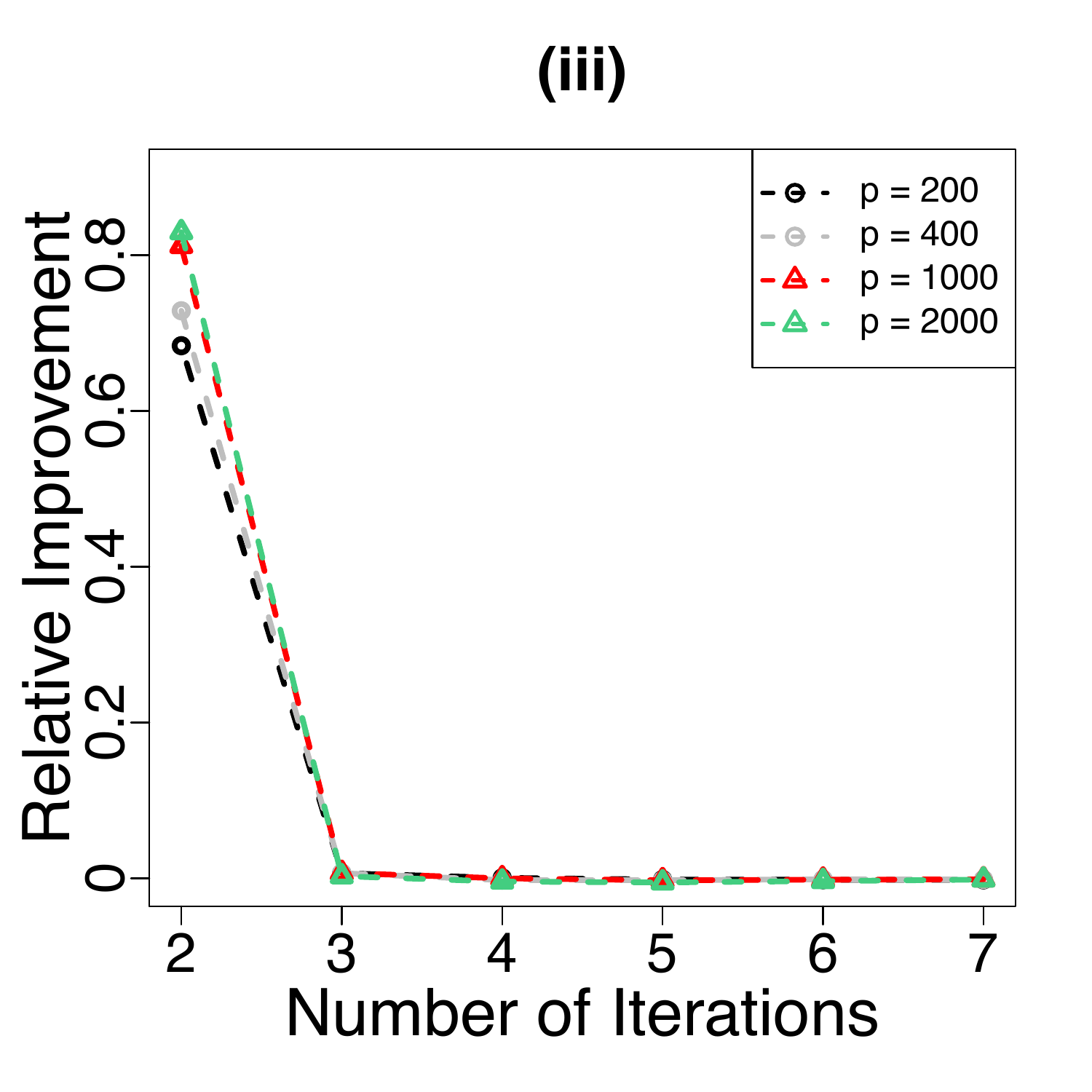}
\caption{Plots of relative improvement defined in~\eqref{eq:relativeimp} versus number of iterations when $n=500$ and $p \in \{200,400,1000,2000\}$.  The three panels correspond to models with different noise distributions: (i) $t$ distribution with 1.5 degrees of freedom; (ii) standard Cauchy distribution; and (iii) a mixture normal distribution.
}
\label{fig:cvline}
\end{figure}

Lastly, we assess more closely the effects of iteratively reweighted $\ell_1$-regularization; see Algorithm~\ref{alg:multistage}.
We keep the above model settings and focus on three different noise distributions: (i) $t$ distribution with 1.5 degrees of freedom; (ii) standard Cauchy distribution; and (iii) a mixture normal distribution.
For simplicity, we set the tuning parameter $\lambda= 0.5 \sqrt{\log (p)/n}$.   
We run Algorithm~\ref{alg:multistage} with uniform kernel and stop after 7 iterations.
Starting with $\hat \bbeta^{(0)} = \textbf{0}$, recall that $\hat{\bbeta}^{(1)}$ is the SQR-Lasso estimator.
To quantify the relative performance of the solution path, at $\ell$th iteration, we define the relative improvement of $\hat \bbeta^{(\ell)}$ with respect to $\hat{\bbeta}^{(\ell - 1)}$ as
\begin{equation}
\label{eq:relativeimp}
\frac{\|\hat{\bbeta}^{(\ell-1)}-\bbeta^*  \|_2^2-\|\hat{\bbeta}^{(\ell)}-\bbeta^*  \|_2^2}{\|\hat{\bbeta}^{(1) }-\bbeta^*  \|_2^2} ,  \ \ \ell \geq 2.
\end{equation}
The relative improvement is a value between zero and one. A value close to zero indicates that there is little  improvement in estimation error and vice versa.
The results for $n=500$ and $p\in \{200,400,1000,2000\}$, averaged over 100 replications, are summarized in Figure~\ref{fig:cvline}. We see  that running an additional iteration $(\ell=2)$ leads to the most significant improvement. 
The estimator, after $\ell=3$ iterations, can still be improved under the $t$ and Cauchy models.
In all the $(n,p)$ settings considered, running $\ell \geq 4$ iterations only shows marginal improvement, suggesting that the multi-step procedure with $\ell=3$ is sufficient for moderate-scale datasets.

\section{An application to gene expression data}
\label{sec:real}
We apply the proposed method to an expression quantitative trait locus (eQTL) dataset previously analyzed in \cite{Scheetz2006}, \cite{KCO2008} and \cite{WWL2012}.
The dataset was collected on a study that used eQTL mapping in laboratory rats to investigate and identify genetic variation in the mammalian eye that is relevant to human eye disease \citep{Scheetz2006}.
 Following \citet{WWL2012}, we study the association between gene TRIM32, which was found to be associated with human eye disease, and the other expressions at other probes.
 The data consists of expression values of 31,042 probe sets on 120 rats.  After some data pre-processing steps as described in \citet{WWL2012}, the number of probes are reduced to 18,958.
We further select the top 500 probes that have the highest absolute correlation with the expression of the response.
We apply the proposed method using the uniform kernel and SCAD penalty, with regularization parameter selected by ten-fold cross-validation.
For comparisons, we also implement the $\ell_1$- and concave regularized quantile regression methods, denoted by QR-Lasso and QR-SCAD, using the \texttt{R} package \texttt{FHDQR}.

 Similar to \citet{WWL2012}, we conduct 50 random partitions of the data by randomly selecting the expression values for 80 rats as the training data and the remaining 40 rats as the testing data.
 The selected model size and prediction error (under quantile loss), averaged over 50 random partitions, are reported in Table~\ref{table:realdata}.   We observe from Table~\ref{table:realdata} that the SQR has consistently lower prediction errors than the standard QR across all three quantile levels considered.
The prediction error is also  improved for SQR when the SCAD penalty is used.
In contrary, QR-SCAD exhibits no improvement over QR-Lasso in prediction accuracy, which is in line with the observation in \cite{WWL2012}.
One explanation may be that the lack of smoothness and strong convexity of the quantile loss overshadows the bias-reducing property of the concave penalty. These results suggest that high-dimensional quantile regression considerably benefits from smoothing and concave regularization in terms of model selection ability, prediction accuracy and computational feasibility.

\begin{table}[!htp]
\begin{center}
\caption{The average selected model size and prediction error (under quantile loss), with standard errors in the parenthesis, over 50 random partitions. }
\begin{tabular}{l |      c c}
  \hline
Methods & Model Size & Prediction Error\\
\hline
QR-Lasso ($\tau=0.3$)      & 38.28 (3.192) &  0.225 (0.005) \\
QR-SCAD ($\tau=0.3$)    & 34.66 (3.291) &  0.241 (0.006)  \\
SQR-Lasso ($\tau=0.3$)     & 45.28 (1.866)  & 0.118 (0.003) \\
SQR-SCAD ($\tau=0.3$)    & 31.32 (1.827)  &0.106 (0.003) \\
\hline
QR-Lasso ($\tau=0.5$)      & 33.76 (1.985) & 0.222 (0.003)  \\
QR-SCAD ($\tau=0.5$)    & 30.28 (2.114)  & 0.236 (0.004)  \\
SQR-Lasso ($\tau=0.5$)     & 36.76 (1.533)  &0.142 (0.003)  \\
SQR-SCAD ($\tau=0.5$)    & 29.58 (2.006) & 0.132 (0.003) \\
\hline
QR-Lasso ($\tau=0.7$)      &  29.66 (1.669) & 0.195 (0.003)  \\
QR-SCAD ($\tau=0.7$)    & 24.22 (1.942) & 0.205 (0.003)   \\
SQR-Lasso ($\tau=0.7$)    & 41.44 (2.262)  & 0.124 (0.003) \\
SQR-SCAD ($\tau=0.7$)     & 27.52 (2.269) & 0.116 (0.004) \\
\hline
\end{tabular}
\label{table:realdata}
\end{center}
\end{table}

\section{Discussions}
\label{sec:discussion}

In this paper we introduced a  class of penalized convolution smoothed methods for fitting sparse quantile regression models in high dimensions.  Convolution smoothing turns the non-differentiable check loss into a twice-differentiable and convex surrogate, and the resulting empirical loss is proven to be locally strongly convex (with high probability). To reduce the $\ell_1$-regularization bias as the signal strengthens, we considered a multi-step, iterative procedure which solves a weighted $\ell_1$-penalized smoothed quantile objective function at each iteration. Statistically, we established the oracle-like performance of the output of this procedure, such as the oracle convergence rate and variable selection consistency, under an almost necessary and sufficient minimum signal strength condition. From a computational perspective, together convolution smoothing and convex relaxation enable the use of gradient-based algorithms that are much more scalable to large-scale datasets. In summary, through convolution smoothing with a suitably chosen bandwidth, we aim to seek a better trade-off between statistical accuracy and computational precision for high-dimensional quantile regression. The proposed procedures will be implemented in the R package \texttt{conquer}, available at
$$
\href{https://cran.r-project.org/web/packages/conquer/index.html}{\texttt{https://cran.r-project.org/web/packages/conquer/index.html}}.
$$  
The Python code is also publicly accessible at \href{https://github.com/WenxinZhou/conquer}{https://github.com/WenxinZhou/conquer}, with an option to perform post-selection-inference (via bootstrap).

There are several avenues for future work. When the parameter of interest arises in a matrix form, the low-rankness is often used to capture its low intrinsic dimension.  This falls into the general category of ill-posed inverse problems, where the number of observations/measurements is much smaller than the ambient dimension of the model. 
See \cite{CRPW2012} for a general framework to convert notions of simplicity into convex penalty functions, resulting in convex optimization solutions to linear, underdetermined inverse problems.
The idea of concave penalization can also be applied to low-rank matrix recovery problems. In essence, one can use a concave function to penalize the vector of singular values of matrix $\bTheta \in \RR^{p_1\times p_2}$. We refer to \cite{WZG2017} for a unified computational and statistical framework for non-convex low-rank matrix estimation when the Frobenius norm is used as the data-fitting measure.
We conjecture that the proposed multi-step reweighted convex penalization approach and convolution smoothing will lead to oracle statistical guarantees and fast computational methods for quantile matrix regression and quantile matrix completion problems \citep{BCPW2019}. We leave this as future work.


\appendix

\section{Regularized Smoothed Quantile Regression under Independence}
\label{sec:zero.bias}

In Section~\ref{sec:smoothing}, we discussed the bias induced by smoothing. Recall that 	$\bbeta^*_h  \in \argmin_{\bbeta \in \RR^p}  Q_h(\bbeta)$
is the population mimimizer under the smoothed quantile objective, where $Q_h(\bbeta) = \EE \{ \hat Q_h(\bbeta)\}$. Proposition~\ref{prop:bias}  shows that under a Lipschitz condition on the conditional density $f_{\varepsilon | \bx}(\cdot)$, the smoothing bias $\| \bbeta^*_h - \bbeta^* \|_2$ is of the order $h^2$. The assumed sparsity of $\bbeta^* \in \RR^p$, however, is not  necessarily inherited by $\bbeta^*_h$. Therefore, there is a statistical price to be paid by not having a sparse $\bbeta^*_h$ after smoothing. This results in stronger growth conditions on $(s, p)$ in Theorem~\ref{thm:ncvx} and Theorem~\ref{thm:strong.oracle2}.  Specifically, we have shown that with suitably chosen penalty level $\lambda$ and bandwidth $h$,
\#
	\| \hat \bbeta^{(\ell)} - \bbeta^* \|_2  \lesssim \sqrt{\frac{s+ \log(p)}{n}}  ~~\mbox{ (weak oracle property)} \nn
\#
with probability at least $1-C_1 p^{-1}$ as long as $\ell \gtrsim \log(\log p)$ and $n\gtrsim s^2 \log(p)$.  In addition,
\#
  \hat \bbeta^{(\ell)} = \hat \bbeta^{\ora} ~~\mbox{ (strong oracle property)} \nn
\#
with probability at least $1-C_2 (p^{-1} + n^{-1})$ as long as $\ell \gtrsim \log(s)$ and $n\gtrsim \max   \{ s^{8/3} , s^2 \log(p)  \}$.

In the following, we show that under a stronger independence assumption between the random feature vector $\bx$ and error variable $\varepsilon$, smoothing only introduces bias on the intercept, and therefore $\bbeta^*_h$ preserves the sparsity of the true parameter $\bbeta^*$. This observation guarantees that we pay almost no price for estimating $\bbeta^*_- := (\beta^*_2, \ldots, \beta^*_p)^\T$ with the use of convolution smoothing. It is worth noticing that such an independence assumption is typically stringent in the context of quantile regression, whose main feature is the ability to capture heterogeneity in the set of important predictors at different quantile levels of the response distribution.  The results of this section complement the existing theory for composite quantile regression in high dimensions \citep{ZY2008,BFW2011}.

\begin{enumerate}
\item[(B1$^*$)] $\varepsilon \in \RR$ and $\bx\in \RR^p$ are independent, and the function $m(\alpha) :=  \EE \bigl\{  \ell_h(\varepsilon- \alpha)   \bigr\}$ has a unique minimizer $b_h$, where $\ell_h(\cdot)$ is defined in \eqref{convolution.loss}. The density function of $\varepsilon$, denoted by $f_{\varepsilon}(\cdot)$, satisfies $f_\varepsilon(0)>0$ and $|f_{\varepsilon }(u)- f_{\varepsilon }(v)| \leq l_0 |u-v|$ for all $u,v\in \RR$ and some $l_0>0$.
\end{enumerate}

\begin{proposition} \label{prop:unique}
Assume Conditions~(B1$^*$) and (B2) hold. Then, $\bbeta^*_h$ is the unique minimizer of $Q_h(\cdot)$ and satisfies $\bbeta^*_h = (\beta^*_1 + b_h, \bbeta_-^{* \T})^\T$. Provided that $0<h < f_0/(c_1 l_0)$, we have
\#
	 |b_h | \leq   l_0  c_1 \kappa_2^{1/2}   f_0^{-1} h^2 ,  \label{inter.bias.ubd1}
\#
where $c_1 = \kappa_1 + \kappa_2^{1/2}$ and $f_0 = f_\varepsilon(0)$.
\end{proposition}

\begin{remark}
According to \eqref{smooth.Hessian}, the first and second derivatives of $\alpha \mapsto m(\alpha )$ are
\#
	m'(\alpha )  =    \int_{-\infty}^\infty  K(u) F_{\varepsilon }(\alpha - hu)   \, {\rm d} u - \tau ~~\mbox{ and }~~ m''(\alpha ) =   \EE  \bigl \{ K_h(\alpha - \varepsilon )   \bigr\} =  \int_{-\infty}^\infty K(u)  f_{\varepsilon }(\alpha- hu ) \, {\rm d} u , \nn
\#
where $F_{\varepsilon}$ and $f_{\varepsilon}$ denote, respectively, the distribution and density functions of $\varepsilon$. Moreover, note that $\lim_{\alpha \to \infty} m'(\alpha ) = 1-\tau$ and $\lim_{\alpha \to - \infty} m'(\alpha ) = -\tau$.  Provided that $F_{ \varepsilon}(\cdot)$ is strictly increasing, there exits a unique $b_h $ for which $m'(b_h )=0$. In other words, $b_h$ is the unique minimizer of $m(\cdot)$.
\end{remark}

Under independence, our first result is on the weak oracle property, which is in parallel with Theorem~\ref{thm:ncvx}.

\begin{theorem} \label{thm:new.weak.oracle}
Assume that Conditions (A1) and (B1$^*$), (B2) and (B3) hold, and there exist $\alpha_1>\alpha_0>0$ satisfying \eqref{oracle.rate.cond} with $f_l$ replaced by $f_0 = f_\varepsilon(0)$.
Let the penalty level $\lambda$ and bandwidth $h$ satisfy $\lambda \asymp \sigma_{\bx}\sqrt{\log(p)/n}$ and $\sigma_{\bx} f_0^{-1}\sqrt{s\log(p)/n} \lesssim h \lesssim f_0 $. Under the beta-min condition $\|\bbeta^*_{\cS} \|_{\min} \geq (\alpha_0 +\alpha_1)\lambda$ and sample size requirement $n\gtrsim s\log(p) + t$, the multi-step estimator $\hat \bbeta^{(\ell)}$ with $\ell \gtrsim \lceil \log(\log p) / \log(1/\delta) \rceil$ satisfies, for any $t\geq 0$,
\$
 \| \hat \bbeta^{(\ell)}  - \bbeta_h^* \|_2 \lesssim f_0^{-1}\sqrt{\frac{s+t}{n}} ~~\mbox{ and }~~
  \| \hat \bbeta^{(\ell)}  - \bbeta_h^* \|_1 \lesssim f_0^{-1} s^{1/2}\sqrt{\frac{s+t}{n}} 
\$
with probability at least $1- p^{-1} - e^{-t}$, where $\delta  =\sqrt{4+ \{ q'(\alpha_0)\}^2}/(\alpha_0 \kappa_l f_0 \gamma_p)$.
\end{theorem}

To establish the strong oracle property, we first refine Proposition~\ref{thm:oracle.smoothqr} on the oracle estimator.

\begin{proposition} \label{thm:new.oracle}
Assume Conditions~(B1$^*$) and (B1$'$)--(B3$'$) hold. For any $t\geq 0$, et the sample size $n$ and the bandwidth $h=h_n$ be such that $n\gtrsim s + t$ and $\sqrt{(s+t)/n }\lesssim h \lesssim 1$. Then, the oracle estimator $\hat \bbeta^\ora$ satisfies
\#
  \|   \Sb^{1/2} (\hat \bbeta^\ora - \bbeta_h^*)_{  \cS} \|_2 \lesssim  f_0^{-1} \sqrt{\frac{s+t }{n}}   	 \label{new.oracle.concentration}
\#
with probability at least $1-2 e^{-t}$, where $ \Sb := \bSigma_{\cS \cS}= \EE(\bx_{\cS} \bx_{\cS}^\T )$. Moreover,
\#
	 \bigg\| \Sb^{-1/2}  \Db_h  (\hat \bbeta^\ora - \bbeta_h^*)_{  \cS} + \Sb^{-1/2} \frac{1}{n}  \sn  \bigl\{   \bar K((b_h - \varepsilon_i)/h) - \tau  \bigr\} \bx_{i,  \cS}  \biggr\|_2  \lesssim   \frac{ s +t}{  h^{1/2} n   }   \label{new.oracle.bahadur}
\#
with probability at least $1- 3e^{-t}$, where $\Db_h = m''(b_h)\cdot\Sb$ with $m(\cdot)$ and $b_h$ defined in Condition~(B1$^*$).
\end{proposition}

Finally, Theorem~\ref{thm:new.strong.oracle} below relaxes the sample size scaling required for the strong oracle property given in Theorem~\ref{thm:strong.oracle2}.

\begin{theorem}  \label{thm:new.strong.oracle}
Assume that Conditions (A1), (B1$^*$) and (B1$'$)--(B3$'$) hold, and 
\# \label{irrepresentable.cond2}
	  \max_{j \in  \cS^\cc}   \|  \bSigma_{j \cS} (\bSigma_{\cS \cS})^{-1}   \|_1 \leq A_1 
\#
for some $A_1\geq 1$. For a prespecified $\delta \in (0,1)$, suppose there exist constants $\alpha_1>\alpha_0$ satisfying \eqref{oracle.cond} with $\kappa= \kappa_l f_\varepsilon(0)/2$, and the beta-min condition $\| \bbeta_{\cS}^* \|_{\min} \geq (\alpha_0  + \alpha_1) \lambda$ with the penalty level $\lambda \asymp \sqrt{\log(p)/n}$.   
Then, with probability at least $1-2p^{-1}-5n^{-1}$, $\hat \bbeta^{(\ell)} = \hat \bbeta^\ora$ for all $\ell \geq \lceil  \log(s^{1/2}/\delta) / \log(1/\delta)\rceil$, provided that the bandwidth $h$ and sample size $n$ are subject to
$$
 \max\Bigg\{  \sqrt{\frac{s\log(p)}{n}} , \frac{s^2}{n\log(p)} \Bigg\} \lesssim h \lesssim 1 .
$$
\end{theorem}

\section{Proof of Proposition~\ref{prop:bias} in Section~\ref{sec:theory:bias}}
\label{appendix:bias}
We derive an upper bound for $\| \bbeta^*_h - \bbeta^* \|_{\bSigma}$ via a localized analysis exploited by \cite{FLSZ2018}. Define the local vicinity $\Theta_h = \bbeta^* + \BB_{\bSigma}(   h)$ of $\bbeta^*$. To begin with, it is unclear that whether $\bbeta^*_h \in \argmin_{\bbeta \in \RR^p} Q_h(\bbeta)$ falls into this local region. Instead,  we consider an intermediate vector $\bbeta^\dagger_h  = (1-\eta) \bbeta^* + \eta \bbeta^*_h$, where $\eta = \sup\{ u \in [0,1] : \bbeta^* + u( \bbeta^*_h - \bbeta^* )  \in \Theta_h \}$, which is the large value between 0 and 1 such that the corresponding convex combination of $\bbeta^*$ and $\bbeta^*_h$ falls into $\Theta_h$.
If $\bbeta^*_h \notin \Theta_h $, then $\eta \in (0,1)$ and $\bbeta^\dagger_h$ falls onto the boundary of $\Theta_h$, i.e., $\| \bbeta^\dagger_h - \bbeta^* \|_{\bSigma } =   h $; otherwise if $\bbeta^*_h \in \Theta_h $, $\eta=1$ and hence $\bbeta^\dagger_h = \bbeta^*_h$.

By the convexity of $\bbeta\mapsto Q_h(\bbeta)$, the optimality of $\bbeta^*_h$, and Lemma~F.2 in the supplementary material of \cite{FLSZ2018}, we obtain that
\#
	0 & \leq  \langle \nabla Q_h( \bbeta^\dagger_h )  - \nabla Q_h(\bbeta^*) ,  \bbeta^\dagger_h  - \bbeta^* \rangle \nn \\
	& \leq \eta  \cdot  \langle \nabla Q_h(\bbeta^*_h )  - \nabla Q_h(\bbeta^*) , \bbeta^*_h - \bbeta^* \rangle  =  \langle    - \nabla Q_h(\bbeta^*) , \bbeta^\dagger_h - \bbeta^* \rangle . \label{pop.foc}
\#
Applying the mean value theorem for vector-valued functions yields
\#
 \nabla Q_h( \bbeta^\dagger_h )  - \nabla Q_h(\bbeta^*)  =   \int_0^1   \nabla^2 Q_h\big( (1-t)   \bbeta^* + t \bbeta^\dagger_h  \big)  \, {\rm d} t  \,  \big(  \bbeta^\dagger_h  - \bbeta^* \bigr) , \label{pop.grad.diff}
\# 
where  $\nabla^2 Q_h( \bbeta  )  =  \EE \bigl\{ K_h( \langle \bx , \bbeta -\bbeta^*\rangle - \varepsilon ) \bx \bx^\T \bigr\}$ for $\bbeta \in \RR^p$. With $\bdelta := \bbeta - \bbeta^*$, note that
\#
	& \EE \bigl\{ K_h(   \bx^\T \bdelta  - \varepsilon ) | \bx \bigr\}  = \frac{1}{h} \int_{-\infty}^\infty K\bigg(  \frac{  \bx^\T \bdelta - t}{h} \bigg) f_{\varepsilon | \bx} (t) \, {\rm d} t =  \int_{-\infty}^\infty K(u) f_{\varepsilon | \bx} (  \bx^\T \bdelta - hu ) \, {\rm d}u . \nn
\#
By the Lipschitz continuity of $f_{\varepsilon|\bx}(\cdot)$,
\#
	\EE  \bigl\{ K_h(   \bx^\T \bdelta  - \varepsilon ) | \bx \bigr\}  =  f_{\varepsilon | \bx }(0 ) + R_h(\bdelta)  \label{Kh.mean.ubd1}
\#
with $R_h(\bdelta)$ satisfying $|R_h(\bdelta)|  \leq   l_0  \bigl( |  \bx^\T \bdelta | + \kappa_1 h \bigr)$.
Substituting \eqref{Kh.mean.ubd1} into \eqref{pop.foc} and \eqref{pop.grad.diff} yields
\#
& 	\langle \nabla Q_h( \bbeta^\dagger_h   )  - \nabla Q_h(\bbeta^*) ,  \bbeta^\dagger_h  - \bbeta^* \rangle \nn \\
& \geq  \|  \bbeta^\dagger_h   - \bbeta^*  \|_{\Jb}^2    -    \frac{l_0}{2}  \EE |  \langle \bx,  \bbeta^\dagger_h   - \bbeta^* \rangle  |^3 - l_0 \kappa_1 h \cdot \|  \bbeta^\dagger_h   - \bbeta^* \|_{\bSigma}^2  \nn \\
& \geq  \|   \bbeta^\dagger_h   - \bbeta^*  \|_{\Jb}^2    -    \frac{l_0}{2} \mu_3 \cdot   \| \bbeta^\dagger_h   - \bbeta^* \|^3_{\bSigma}  - l_0 \kappa_1 h \cdot \| \bbeta^\dagger_h  - \bbeta^* \|_{\bSigma}^2 , \label{pop.foc.lbd}
\#
where $\Jb= \EE \{ f_{\varepsilon | \bx} (0) \bx \bx^\T \}$.
 
On the other hand, we have
\#
		& \langle    - \nabla Q_h(\bbeta^*) , \bbeta^\dagger_h   - \bbeta^* \rangle \leq \| \bSigma^{-1/2} \nabla Q_h(\bbeta^*) \|_2 \cdot\| \bbeta^\dagger_h  - \bbeta^* \|_{\bSigma} , \nn
\#
where $\nabla Q_h(\bbeta^*) = \EE \{ \bar K(-\varepsilon /h) - \tau \} \bx $. Using integration by parts and a Taylor series expansion yields
\#
   \EE \bigl\{  \bar K(-\varepsilon /h) | \bx \bigr\} &  = \int_{-\infty}^\infty \bar K(-t/h) \, {\rm d} F_{\varepsilon | \bx}(t) \nn \\
& = -\frac{1}{h}\int_{-\infty}^\infty  K(-t/h) F_{\varepsilon | \bx}(t)  \, {\rm d}t = \int_{-\infty}^\infty  K(u) F_{\varepsilon | \bx}(-h u)  \, {\rm d}u  \nn \\
& = \tau    + \int_{-\infty}^\infty  K(u) \int_{0}^{-hu} \bigl\{  f_{\varepsilon | \bx} (t) - f_{\varepsilon | \bx} (0) \bigr\} \, {\rm d} t  \, {\rm d}u ,\nn
\#
from which it follows that $| \EE  \bar K(-\varepsilon/h) - \tau | \leq \frac{l_0}{2} \kappa_2   h^2$. 
Consequently,
\#
 &  \| \bSigma^{-1/2} \nabla Q_h(\bbeta^*) \|_2 = \sup_{\bu \in \mathbb S^{p-1}}  \EE  \bigl\{ \bar K(-\varepsilon/h) -\tau \bigr\} \langle \bSigma^{-1/2} \bx, \bu \rangle \leq \frac{l_0}{2} \kappa_2 h^2 .  \label{score.mean.ubd}
\#
Putting together the pieces, we conclude that
\#
 \langle    - \nabla Q_h(\bbeta^*) , \bbeta^\dagger_h    - \bbeta^* \rangle \leq  \frac{l_0}{2} \kappa_2 h^2 \cdot \|  \bbeta^\dagger_h - \bbeta^* \|_{\bSigma} . \label{pop.foc.ubd}
\#

Recall that $f_{\varepsilon |\bx}(0) \geq f_l >0$ almost surely and $ \bbeta^\dagger_h  \in \Theta_h$. Together, \eqref{pop.foc}, \eqref{pop.foc.lbd} and \eqref{pop.foc.ubd} imply
\#
  f_l \cdot  \|  \bbeta^\dagger_h  - \bbeta^* \|_{\bSigma}^2   \leq    \big(  0.5 \mu_3 + 0.5  \kappa_2  +  \kappa_1   \big)   l_0  h^2 \cdot \| \bbeta^\dagger_h  - \bbeta^* \|_{\bSigma} . \nn 
\#
Canceling $\|  \bbeta^\dagger_h  - \bbeta^* \|_{\bSigma}$ on both sides gives
\#
   \|  \bbeta^\dagger_h   - \bbeta^* \|_{\bSigma}  \leq    \underbrace{  \big(  0.5 \mu_3 + 0.5  \kappa_2  +  \kappa_1   \big)  }_{= : c_0}  \frac{l_0 h }{f_l}    h  = \frac{c_0 l_0 h}{f_l}    h   . \nn 
\#
Provided that $h <   f_l/(c_0 l_0)$, $\bbeta^\dagger_h$ falls in the interior of $\Theta_h$, i.e., $ \| \bbeta^\dagger_h - \bbeta^* \|_{\bSigma}<  h$. By the definition of $\bbeta^\dagger_h$ in the beginning of the proof, we must have $\bbeta^*_h \in \Theta_h$; otherwise if  $\bbeta^*_h \notin \Theta_h$, $\bbeta^\dagger_h$ lies on the boundary of $\Theta_h$, which leads to contradiction.
Consequently, $\bbeta^*_h = \bbeta^\dagger_h$ satisfies the claimed bound \eqref{bias.ubd}. Moreover, by \eqref{pop.foc.lbd}, $Q_h(\cdot)$ is strictly convex in a neighborhood of $\bbeta^*_h$, thus verifying the uniqueness claim.

Next, to investigate the leading term in the bias, define the remainder
\#
	\Delta_h =  \bSigma^{-1/2} \bigl\{ \nabla Q_h(\bbeta^*_h) - \nabla Q_h(\bbeta^*)  -  \Jb ( \bbeta^*_h -\bbeta^* ) \big\} =    \bSigma^{-1/2}   \Jb ( \bbeta^*_h -\bbeta^* )   -  \bSigma^{-1/2}   \nabla Q_h(\bbeta^*) .  \nn
\#
Once again, using the mean value theorem for vector-valued functions, we find that
\#
	\Delta_h &  =  \Bigg\{     \bSigma^{-1/2}  \int_0^1  \nabla^2 Q_h \big( (1-t) \bbeta^* + t \bbeta^*_h \big) \, {\rm d}t  \, \bSigma^{-1/2}  -  \Jb_0 \Bigg\}  \bSigma^{1/2} ( \bbeta^*_h - \bbeta^* )   , \label{residual} 
\#
where $\Jb_0 =  \bSigma^{-1/2} \Jb  \bSigma^{-1/2} = \EE \{ f_{\varepsilon | \bx}(0) \bz \bz^\T\}$ and  $\bz = \bSigma^{-1/2}\bx$.
Under Conditions (B1) and (B2),  we derive that
\#
	&  \Bigg\|   \bSigma^{-1/2}  \int_0^1  \nabla^2 Q_h \big((1-t) \bbeta^* + t \bbeta^*_h  \big) \, {\rm d}t  \,  \bSigma^{-1/2}   - \Jb_0   \Bigg\|_2  \nn \\
&=\Bigg\|   \EE   \int_0^1  \int_{-\infty}^\infty K(u)  \bigl\{ f_{\varepsilon | \bx}(   t \langle  \bx,  \bbeta^*_h -\bbeta^* \rangle - hu ) - f_{\varepsilon | \bx}(0) \bigr\} \, {\rm d} u \, {\rm d} t \,  \bz \bz^\T \Bigg\|_2 \nn \\
& \leq  l_0 \sup_{\bu \in \mathbb{S}^{p-1}} \EE    \int_0^1  \int_{-\infty}^\infty K(u) \bigl(  | t \langle \bx, \bbeta^*_h - \bbeta^* \rangle | + h |u |   \bigr)  \, {\rm d}u \, {\rm d} t  \, (\bz^\T \bu)^2 \nn \\
& \leq \frac{l_0}{2 } \sup_{\bu \in \mathbb{S}^{p-1}}  \EE \bigl\{ |\langle \bx, \bbeta^*_h - \bbeta^* \rangle | (\bz^\T \bu)^2 \bigr\}  +  l_0 \kappa_1 h \nn \\
& \leq \frac{l_0}{2 } \mu_3 \| \bbeta_h^* - \bbeta^* \|_{\bSigma} +  l_0  \kappa_1 h. \nn
\#
This bound, together with \eqref{residual}, implies
\#
	\| \Delta_h \|_2 \leq   l_0    \bigl( 0.5  \mu_3 \| \bbeta_h^* - \bbeta^* \|_{\bSigma}   + \kappa_1 h \bigr) \| \bbeta_h^* - \bbeta^* \|_{\bSigma}  . \label{residual.bound}
\#
From the earlier bound \eqref{bias.ubd}, we see that $\| \Delta_h \|_2 \lesssim h^3$.

Turning to the gradient $\nabla Q_h(\bbeta^*) = \EE \{  \bar K(-\varepsilon/h) -\tau \} \bx$,  we apply a second-order Taylor series expansion to $F_{\varepsilon|\bx}$ to conclude that
\#
 \EE \bigl\{  \bar K(-\varepsilon /h) | \bx \bigr\}  - \tau  & =  \int_{-\infty}^\infty  K(u) \int_{0}^{-hu} \bigl\{  f_{\varepsilon | \bx} (t) - f_{\varepsilon | \bx} (0) \bigr\} \, {\rm d} t  \, {\rm d}u \nn \\
& =  \frac{1 }{2}  \kappa_2  h^2 \cdot  f_{\varepsilon | \bx}'(0) +   \int_{-\infty}^\infty  \int_{0}^{-hu}   \int_0^t   K(u) \bigl\{  f_{\varepsilon | \bx}'(v) - f_{\varepsilon | \bx}'(0) \bigr\}   \, {\rm d} v \, {\rm d} t  \, {\rm d}u  . \nn
\#
Under the Lipschitz continuity assumption on $f'_{\varepsilon| \bx}$, this further implies
\#
	 \bigg\| \bSigma^{-1/2} \nabla Q_h(\bbeta^*)     - \frac{1}{2} \kappa_2 h^2 \cdot  \bSigma^{-1/2}  \EE   \bigl\{  f_{\varepsilon | \bx}'(0)\bx \bigr\}  \bigg\|_2 \leq \frac{l_1}{6 } \kappa_3 h^3 . \label{gradient.approxi}
\#
Finally, combining \eqref{residual.bound} and \eqref{gradient.approxi} proves \eqref{bias.leading}.   \qed

\section{Proofs of Results in Section~\ref{sec:theory:lasso}}
\label{appendix:lasso}
Recall from ~\eqref{def:wb} that $\bw_h^* =  \bw_h(\bbeta^*)$ and $\bw_h(\bbeta) = \nabla \hat Q_h(  \bbeta ) - \nabla Q_h(\bbeta)$, $\bbeta \in \RR^p$.
We start with the following two lemmas that will be needed in proving Proposition~\ref{prop:RSC} and Theorem~\ref{thm:lasso-qr}. 

\begin{lemma} \label{covariate.moment}
Let $\bz = \bSigma^{-1/2} \bx \in \RR^p$ be the standardized feature vector which is isotropic, i.e., $\EE(\bz \bz^\T) = \Ib_p$. Under Condition~(B3), the $k$-th ($k\geq 3$) absolute moments of all the one-dimensional marginals of $\bz$ are uniformly bounded: $\mu_k := \sup_{\bu \in \mathbb{S}^{p-1}} \EE |\bz^\T \bu |^k  \leq  k! \upsilon_0^k$. In particular, $\mu_1 \leq \mu_2^{1/2} =1$.
\end{lemma}
 
\begin{lemma} \label{lem:subgradient}
Assume Conditions~(B1)--(B3) hold. Then, for any $t> 0$,  
\#
	  \|  \bw_h^*  \|_\infty \leq   \upsilon_0\sigma_{\bx}  \Biggl[   \sqrt{  \bigl\{\tau(1-\tau) + C h^2 \bigr\}\frac{2 t}{n}} + \max(1-\tau, \tau ) \frac{2t}{n} \Biggr]      \label{score.ubd}
\#
holds with probability at least $1-2pe^{-t}$, where $C=(\tau+1)l_0 \kappa_2 $.
\end{lemma}

\subsection{Proof of Proposition~\ref{prop:RSC}}

Given $r, l >0$, define the local cone-neighborhood of $\bbeta^*$ 
\#
	\Theta = \Theta(r, l) =\bigl\{ \bbeta \in \RR^p : \bbeta -\bbeta^*  \in \BB_{\bSigma}(r ) \cap  \CC_{\bSigma}( l) \bigr\}, \label{local.set}
\#
where $\BB_{\bSigma}(r) = \{ \bdelta \in \RR^p : \|    \bdelta  \|_{\bSigma} \leq  r \}$ and  $\CC_{\bSigma}( l) = \{ \bdelta \in \RR^p : \| \bdelta \|_1 \leq l \|\bdelta \|_{\bSigma} \}$.
Since the smoothed quantile objective \eqref{smooth.qloss} is convex and twice continuously differentiable, it follows from \eqref{smooth.Hessian} that
\#
   D(\bbeta) & := \bigl\langle  \nabla \hat Q_h(\bbeta) - \nabla \hat Q_h(\bbeta^*)  ,  \bbeta - \bbeta^* \bigr\rangle \nn \\
  &   =  \frac{1}{n} \sn\Bigg\{ \bar{K} \Bigg(  \frac{  \bx_i^\T \bbeta  - y_i}{h}\Bigg) - \bar{K}\Bigg( \frac{    -\varepsilon_i}{h} \Bigg) \Bigg\} \langle \bx_i, \bbeta - \bbeta^* \rangle . \label{def.D}
\#
For $i=1,\ldots, n$, define the events $E_i = \{ |   \varepsilon_i | \leq h/2 \} \cap \{  |\langle \bx_i, \bbeta - \bbeta^* \rangle | /  \| \bbeta - \bbeta^* \|_{\bSigma}  \leq  h/(2r)\}$, on which $|y_i -  \bx_i^\T \bbeta  | \leq h$ for any $\bbeta \in \bbeta^* + \BB_{\bSigma}(r)$. 
Since $\kappa_l = \min_{|u|\leq 1} K(u)>0$, $D(\bbeta)$ can be lower bounded as 
\#
 D(\bbeta)  \geq  \frac{\kappa_l}{n h} \sn \langle \bx_i , \bbeta - \bbeta^* \rangle^2 \mathbbm{1}_{E_i } , \label{eq:dblower1}
\#
where $ \mathbbm{1}_{E_i }$ is the indicator function of $E_i $. Thus, it suffices to bound the right-hand side of the above inequality from below uniformly over $\bbeta \in \Theta$. 

To deal with the discontinuity, we use the following smoothing technique from \cite{L2017}, which turns the objective into a Lipschitz continuous empirical process.
For $R>0$, define the function 
\#
	\varphi_R(u) =\begin{cases}
		u^2,   &  \mbox{ if }  |u | \leq  \frac{R}{2},   \\
		\{ u-R \sign(u)  \}^2,   &  \mbox{ if }  \frac{R}{2} \leq |u | \leq R, \\
		0,  & \mbox{ if } |u| >R,  
	\end{cases} 
 \nn
\#
which is $R$-Lipschitz continuous, and satisfies
\#
  \varphi_{c R}(c u) = c^2 \varphi_R(u) ~\mbox{ for any } c>0 , \quad    u^2 \mathbbm{1}(|u| \leq R / 2)	\leq \varphi_R(u) \leq  u^2  \mathbbm{1}(|u| \leq R ) . \label{phi.bound}
\# 
Together,  \eqref{eq:dblower1} and \eqref{phi.bound} imply
\#
		\frac{  D(\bbeta ) }{ \| \bbeta - \bbeta^* \|_{\bSigma}^2 } \geq     \kappa_l    \cdot \underbrace{    \frac{1}{n h  } \sn  \varphi_{  h/(2r) } \big( \langle \bx_i, \bbeta  - \bbeta^* \rangle/\| \bbeta - \bbeta^* \|_{\bSigma} \big) \cdot \chi_i  }_{ =:  D_0(\bbeta)} , \label{def.D0}
\#
where $\chi_i = \mathbbm{1} (|\varepsilon_i | \leq h/2)$.

In the following, we bound the expectation $\EE \{   D_0(\bbeta ) \}$ and the random fluctuation $  D_0(\bbeta ) - \EE \{   D_0(\bbeta )\}$ over $\bbeta \in \Theta$, respectively. 
For the binary variable $\chi_i $, using Condition~(B1) we have
\#
	  | \EE( \chi_i  | \bx_i ) -  h     f_{\varepsilon | \bx}(0)   |  & \leq   \int_{-h/2}^{h/2}   |  f_{\varepsilon | \bx } (t   ) - f_{\varepsilon | \bx} (0)  | \, {\rm d} t   \leq  l_0 h^2/4.   \label{indicator.bound}
\#
Moreover, write $\bdelta = \bbeta - \bbeta^*$ for $\bbeta\in \Theta $, and define the one-dimensional marginal $\xi_{\bdelta} = \bx^\T \bdelta  / \| \bdelta \|_{\bSigma}$ such that $\EE (\xi_{\bdelta}^2) =1$.  
Provided $h\leq f_l/(2l_0)$, it follows from \eqref{phi.bound} and \eqref{indicator.bound} that
\#
 &	\EE \big\{\varphi_{  h/(2r)  } (   \bx_i^\T  \bdelta / \| \bdelta \|_{\bSigma}  ) \cdot \chi_i   \big\}    =    \EE \big\{\varphi_{  h/(2r)    } ( \xi_{\bdelta} ) \cdot \chi_i    \big\}    \geq  \frac{7}{8} f_l h \cdot  \bigl( 1 -  \EE \bigl\{ \xi_{\bdelta}^2   \mathbbm{1}_{ | \xi_{\bdelta} | >  h/(4r) }   \bigr\} \bigr)    .\nn
\# 
For any $u>0$, by the sub-exponential condition on $\bx \in \RR^p$, we have
\#
  \EE \bigl\{ \xi_{\bdelta}^2  \mathbbm{1}(|\xi_{\bdelta}| >   u  )  \bigr\}  &=2 \EE \left\{ \int_{0}^{\infty} t \cdot    \mathbbm{1}(|\xi_{\bdelta}| >t) \cdot   \mathbbm{1}(|\xi_{\bdelta}| >   u  )  \,{\rm d}  t    \right\}    \nn\\
  &=2  \int_{0}^{u} t\cdot   \EE\bigl\{    \mathbbm{1}(|\xi_{\bdelta}| >t) \cdot \mathbbm{1}(|\xi_{\bdelta}| > u  )   \bigr\}   {\rm d}  t  +2  \int_{u}^{\infty} t \cdot \PP(   |\xi_{\bdelta}| >t  ) \,  {\rm d}  t    \nn\\
  &=   u^2\PP(|\xi_{\bdelta }  | >    u )+2 \upsilon_0^2  \int_{    u/ \upsilon_0 }^\infty t\cdot \PP(|\xi_{\bdelta}/ \upsilon_0  | \geq t)   \,{\rm d}  t    \nn \\
 & \leq     u^2  e^{- u/ \upsilon_0 }  + 2 \upsilon_0^2 \int_{   u/\upsilon_0    }^\infty  t e^{-t}   \,{\rm d}  t \nn \\
 & =  \bigl(   u^2 + 2 \upsilon_0 u + 2 \upsilon_0^2   \bigr) e^{-u/\upsilon_0  } \nn, 
\#
where  the third equality follows from a change of variable.
As long as  $r \leq h/(20\upsilon_0^2)$, taking $u= h/(4r) \geq 5 \upsilon_0^2$  in the above bound yields $\EE \{ \xi_{\bdelta}^2    \mathbbm{1}(|\xi_{\bdelta}| >  h/(4r) )  \} <1/4$. Consequently,
\#
 \inf_{\bbeta \in \Theta } \EE \{ D_0(\bbeta) \}  \geq   \frac{21}{32} f_l     \mbox{~~as long as~~}   20\upsilon_0^2 r \leq  h\leq f_l/(2 l_0) . \label{D0.mean.lbd1}
\#

Next we evaluate the random fluctuation term
\#
	\Omega : = \sup_{  \bdelta \in   \CC_{\bSigma}( l)    }   |   D_0(\bdelta) - \EE \{   D_0(\bdelta) \}   |  . \label{Deltah.def}
\#
Write $ \omega_{\bdelta}(\bx_i, \varepsilon_i ) =    \varphi_{ h/(2r) } ( \bx_i^\T \bdelta / \| \bdelta \|_{\bSigma} ) \cdot \chi_i / h$, so that $ D_0(\bdelta)  = (1/n) \sn  \omega_{\bdelta}(\bx_i, \varepsilon_i )   $. 
 By \eqref{def.D0} and \eqref{indicator.bound}, and the fact that $\varphi_R(u) \leq  (R/2)|u|$, we have 
 $$
  0 \leq   \omega_{\bdelta}(\bx_i, \varepsilon_i )   \leq    (4r)^{-2} h  ~~\mbox{ and }~~ \EE \omega^2_{\bdelta}(\bx_i, \varepsilon_i )  \leq     (4r)^{-2} \cdot 9f_u h/ 8  .
 $$ 
With the above preparations, we apply Theorem~7.3 in \cite{B2003} (a refined Talagrand's inequality) to conclude that, for any $t>0$,
\#
	\Omega & \leq \EE \Omega  + (\EE\Omega )^{1/2}  \frac{1}{2r}\sqrt{\frac{ h t}{  n }} +    \frac{3}{2} f_u^{1/2}  (4r)^{-1}\sqrt{\frac{ h t}{n }} +  \frac{h}{(4r)^2}\frac{   t}{ 3 n   }   \nn \\
	& \leq   \frac{5}{4} \EE\Omega  +   \frac{3}{2} f_u^{1/2}  (4r)^{-1} \sqrt{\frac{h t}{n }} + (4+1/3)  \frac{h t}{(4r)^2 n}  \label{Deltah.concentration}
\#
with probability at least $1-e^{-t}$, where the second step follows from the inequality that $ab\leq a^2/4 + b^2$ for all $a,b\in\RR$.

It then remains to bound the expectation $\EE \Omega$. 
By Rademacher symmetrization,
\#
 \EE \Omega  \leq 2 \EE \Bigg\{  \sup_{ \bdelta  \in  \CC_{\bSigma}(l) }    \frac{1}{n} \sn  e_i    \omega_{\bdelta}(\bx_i, \varepsilon_i )   \Bigg\}  , \nn
\# 
where $e_1,\ldots, e_n$ are independent Rademacher random variables. Since $\chi_i = \mathbbm{1}(|\varepsilon_i| \leq h/2) \in \{ 0 , 1\}$, $ \omega_{\bdelta}(\bx_i, \varepsilon_i )$ can be written as $ \omega_{\bdelta}(\bx_i, \varepsilon_i )= h^{-1} \varphi_{h  / (2r)} ( \chi_i \bx_i^\T \bdelta / \| \bdelta \|_{\bSigma})$.
By the Lipschitz continuity of $\varphi_R(\cdot)$, $ \omega_{\bdelta}(\bx_i, \varepsilon_i )$ is a $(2r)^{-1}$-Lipschitz function in $\chi_i \bx_i^\T \bdelta / \| \bdelta \|_{\bSigma}$, i.e.,  for any sample $(\bx_i, \varepsilon_i)$ and parameters $\bdelta , \bdelta'\in \RR^p$, 
\#
	|   \omega_{\bdelta}(\bx_i, \varepsilon_i )  -  \omega_{\bdelta'}(\bx_i, \varepsilon_i ) |  \leq  \frac{1}{2r}  |  \chi_i  \bx_i ^\T \bdelta / \| \bdelta \|_{\bSigma} -  \chi_i  \bx_i^\T \bdelta' / \| \bdelta' \|_{\bSigma}  | \label{eq:lips}.
\#
Moreover,  $\omega_{\bdelta}(\bx_i, \varepsilon_i )  = 0$ for any $\bdelta$ such that $\chi_i  \bx_i ^\T \bdelta / \| \bdelta \|_{\bSigma} = 0$. To use Talagrand's contraction principle to bound the Rademacher complexity,  define the subset $T\subseteq \RR^n$
$$
	T = \big\{ \bt = (t_1, \ldots , t_n)^\T : t_i = \langle \chi_i \bx_i , \bdelta / \| \bdelta \|_{\bSigma}  \rangle,  i=1,\ldots, n, \, \bdelta \in \CC_{\bSigma}(l) \big\},
$$
and contractions $\phi_i:\RR \to \RR$ as $\phi_i(t) = (2r/h) \cdot   \varphi_{ h  /(2r)} (t)$. By \eqref{eq:lips}, $|\phi(t)-\phi(s)|\leq |t-s|$ for all $t, s \in \RR$. 
Applying Talagrand's contraction principle (see, e.g., Theorem~4.12 and (4.20) in \cite{LT1991}), we have
\#
 \EE  \Omega  & \leq 2 \EE \Bigg\{  \sup_{ \bdelta  \in  \CC_{\bSigma}(l) }    \frac{1}{n} \sn  e_i      \omega_{\bdelta}(\bx_i, \varepsilon_i )    \Bigg\} \nn \\
&  = \frac{1}{r} \EE \Bigg\{ \sup_{\bt \in T} \frac{1}{n} \sn e_i \phi_i(t_i) \Bigg\} \nn \\
&   \leq  \frac{1}{r} \EE \Bigg( \sup_{\bt \in T} \frac{1}{n} \sn e_i  t_i  \Bigg)  \nn \\
&=   \frac{1}{r}\EE \Bigg\{  \sup_{ \bdelta  \in  \CC_{\bSigma}(l) }    \frac{1}{n} \sn  e_i      \langle  \chi_i \bx_i   , \bdelta / \| \bdelta \|_{\bSigma} \rangle  \Bigg\} \nn \\
& \leq    \frac{l}{   r } \cdot  \EE  \bigg\|  \frac{1}{n} \sn e_i  \chi_i \bx_i   \bigg\|_\infty  , \label{Delta.meanbound}
\# 
where  the last inequality follows from the cone constraint that $\|\bdelta \|_1 \leq   l \| \bdelta \|_{\bSigma}$. The problem is then boiled down to bounding the expectation on the right-hand side of \eqref{Delta.meanbound}.
For each $1\leq j\leq p$, define the partial sum  $S_j =  \sn  e_i   \chi_i     x_{ij}$, of which each summand satisfies  $\EE ( e_i  \chi_i x_{ij} ) =0$ and $\EE( e_i    \chi_i  x_{ij} )^2 \leq \sigma_{jj}  c_h :=  \sigma_{jj} \bigl(  f_u  h  + l_0 h^2/4   \bigr)$ due to \eqref{indicator.bound}. In addition, for $k=3,4,\ldots$,
\#
	\EE |e_i   \chi_i  x_{ij}|^k  &\leq  \upsilon_0^k \sigma_{jj}^{k/2} c_h  \cdot k \int_0^\infty t^{k-1} \PP(|x_{ij} | \geq \upsilon_0 \sigma^{1/2}_{jj}t) \, {\rm d} t \nn \\
& \leq  \upsilon_0^k \sigma_{jj}^{k/2}   c_h \cdot k \int_0^\infty t^{k-1} e^{-t} \, {\rm d} t \nn \\ 
& =  k! \upsilon_0^k \sigma_{jj}^{k/2} c_h  \leq \frac{k!}{2} \cdot \upsilon_0^2 \sigma_{jj} c_h \cdot (2 \upsilon_0\sigma_{jj}^{1/2})^{k-2} . \nn
\#
Following the proof of Theorems~2.10 and 2.5 in \cite{BLM2013}, it can be shown that for all $\lambda \in (0,1/c)$, $ \log \EE e^{\lambda S_j} \leq \psi(\lambda) := \frac{v\lambda^2}{2(1-c\lambda)}$ and
$$
 \EE \max_{1\leq j\leq p} |S_j |   \leq   \inf_{\lambda \in (0,1/c)}  \biggl\{  \frac{\log(2p) + \psi(\lambda)}{\lambda} \biggr\} = \sqrt{2v \log(2p)} +   c \log(2p)  ,
$$ 
where $v= \upsilon_0^2 \sigma_{\bx}^2 c_h  \cdot  n$ and $c=2\upsilon_0 \sigma_{\bx}$.
Re-arranging terms and using \eqref{Delta.meanbound} yield
\#
 \EE  \Omega  \leq       \upsilon_0 \sigma_{\bx}  \frac{l}{r} \Biggl\{  \frac{3}{2} f_u^{1/2}\sqrt{  \frac{   h \log(2p)}{n }}  +   \frac{2\log(2p)}{n  } \Biggr\}.  \label{Deltah.mean}
\#
Consequently, it follows from \eqref{Deltah.def}, \eqref{Deltah.concentration} with $t=\log(2p)$ and \eqref{Deltah.mean} that
\#
  & \Omega    = \sup_{  \bdelta \in   \CC_{\bSigma}(l)    } |   D_0(\bdelta) - \EE \{   D_0(\bdelta) \}  | \nn \\
& \leq  \upsilon_0 \sigma_{\bx}  \frac{l}{r} \Biggl\{  \frac{15}{8} f_u^{1/2} \sqrt{  \frac{ h \log(2p)}{n }}  + \frac{5}{2}  \frac{ \log(2p)}{n  } \Biggr\}    +  \frac{3}{2} f_u^{1/2}  (4r)^{-1} \sqrt{  \frac{h \log (2p)}{n }} +   (4+1/3) \frac{h \log(2p)}{(4r)^2 n }     \label{Deltah.ubd}
\#
with probability at least $1-(2p)^{-1}$.

Finally, from the bounds  \eqref{def.D}, \eqref{def.D0}, \eqref{D0.mean.lbd1} and \eqref{Deltah.ubd}  we conclude that as long as $n \geq C  f_u f_l^{-2} (l/r)^2 h \log (2p)$ for a sufficiently large $C$ depending only on $(\upsilon_0, \sigma_{\bx})$,  
$$
\inf_{\bbeta \in \bbeta^* + \BB_{\bSigma}(r) \cap \CC_{\bSigma}(l)}\frac{ \langle  \nabla \hat Q_h(\bbeta) - \nabla \hat Q_h(\bbeta^*)  ,  \bbeta - \bbeta^* \rangle}{\| \bbeta - \bbeta^* \|_{\bSigma}^2 } \geq \frac{1}{2} \kappa_l f_l
$$
holds with probability at least $1-(2p)^{-1}$, as claimed.
\qed

\subsection{Proof of Theorem~\ref{thm:lasso-qr}}
Let $\cS\subseteq [p] $ be the active set of $\bbeta^*$ with cardinality $s=|\cS|$.
The symmetric Bregman divergence between $\hat \bbeta =\hat \bbeta_h$ and $\bbeta^*$ is defined as 
\begin{equation}
\label{eq:sbd}
\bigl \langle   \nabla \hat Q_h(\hat \bbeta) -  \nabla \hat Q_h(\bbeta^*) , \hat{\bbeta}  - \bbeta^*  \bigr\rangle\ge 0.
 \end{equation}
The proof of Theorem~\ref{thm:lasso-qr} involves establishing upper and lower bounds for the symmetric Bregman divergence~\eqref{eq:sbd}.  

\medskip
\noindent
{\sc Step 1: Upper Bound}.
By the first-order optimality condition of~\eqref{lasso.sqr}, there exists  a subgradient $\hat \bg  \in \partial \|\hat{\bbeta}\|_1$ such that $\nabla \hat Q_h(\hat \bbeta) +  \lambda \hat \bg = \mathbf{0}$. 
Set $\hat{\bdelta}=\hat{\bbeta}-\bbeta^*$.  By the definition of subgradient, we have
\#
\langle  \hat \bg  ,\bbeta^*  - \hat \bbeta \rangle &\le \|\bbeta^* \|_1- \|\hat \bbeta \|_1  =  \|\bbeta^*_{\cS}\|_1- \|\hat{\bdelta}  +  \bbeta^* \|_1\nn \\
&= \|\bbeta^*_{\cS}\|_1- \|\hat{\bdelta}_{  \cS^{\cc}}\|_1  - \|\hat{\bdelta}_{\cS} +  \bbeta^*_{\cS}\|_1 \le \|\hat{\bdelta}_{\cS} \|_1- \|\hat{\bdelta}_{  \cS^{\cc}}\|_1,
\label{thm:lasso-lad:proof1}
\#
where the last inequality holds by the reverse triangle inequality. 
Substituting the first-order optimality condition into \eqref{eq:sbd} yields
\#
 &  \bigl\langle   \nabla \hat Q_h(\hat \bbeta) -  \nabla \hat Q_h(\bbeta^*)   , \hat{\bbeta}  - \bbeta^*  \bigr\rangle \nn \\
& =  \lambda  \langle  \hat \bg  ,  \bbeta^*  - \hat \bbeta   \rangle +  \bigl\langle  \nabla \hat Q_h(\bbeta^*) - \nabla  Q_h(\bbeta^*)    , \bbeta^* - \hat \bbeta   \bigr\rangle + \bigl\langle \nabla Q_h(\bbeta^*)  , \bbeta^* - \hat \bbeta \bigr\rangle   \nn \\
& \leq  \lambda \bigl(  \|\hat{\bdelta}_{\cS} \|_1- \|\hat{\bdelta}_{  \cS^{\cc}}\|_1\bigr) + \underbrace{  \| \nabla \hat Q_h(\bbeta^*) -  \nabla  Q_h(\bbeta^*)   \|_{\infty} }_{ =:   \| \bw^*_h \|_\infty }  \|    \hat \bdelta  \|_1  +   \underbrace{   \| \bSigma^{-1/2}  \nabla Q_h(\bbeta^*) \|_2 }_{ =:  b_h^*}\| \hat \bdelta  \|_{\bSigma} ,\nn
\#
where $Q_h(\cdot)$ is the population smoothed quantile objective defined in \eqref{pseudo.parameter}. 
Here $  \| \bw^*_h \|_\infty$ is a stochastic term that determines the statistical error, and $b_h^*$ is the (deterministic) smoothing bias satisfying $b_h^*  \leq l_0 \kappa_2 h^2/2$ due to \eqref{score.mean.ubd}. Conditioned on the event $\{ \lambda \geq  2 \| \bw^*_h \|_\infty \}$, we have
\#
	&   \bigl\langle   \nabla \hat Q_h(\hat \bbeta) -  \nabla \hat Q_h(\bbeta^*)   , \hat{\bbeta}  - \bbeta^*  \bigr\rangle     \nn \\
	  	  &\leq \lambda \bigl(  \|\hat{\bdelta}_{\cS} \|_1- \|\hat{\bdelta}_{ \cS^{\cc}}\|_1 +   \|   \hat \bdelta \|_1/2 \bigr) +   b_h^*  \|   \hat \bdelta \|_{\bSigma} \nn \\ 
	& \leq   \frac{\lambda}{2}  \bigl(   3 \| \hat \bdelta_{\cS}  \|_1 - \|  \hat \bdelta_{  \cS^{\cc}} \|_1  \bigr) + b_h^*   \|   \hat \bdelta  \|_{\bSigma}\nn \\
	& \leq \frac{3}{2}    s^{1/2}   \lambda \| \hat \bdelta \|_2 +  b_h^* \|   \hat \bdelta  \|_{\bSigma} .   \label{foc.ubd}
\# 
Since $\langle   \nabla \hat Q_h(\hat \bbeta) -  \nabla \hat Q_h(\bbeta^*) , \hat{\bbeta}  - \bbeta^*  \rangle \ge 0$, as a byproduct, we see from~\eqref{foc.ubd} that $\hat  \bdelta$ satisfies the cone-type constraint $\| \hat \bdelta_{ {\cS}^{\cc }}  \|_1  \leq 3 \| \hat \bdelta_{\cS} \|_1   +2 \lambda^{-1} b_h^* \|   \hat \bdelta  \|_{\bSigma}$, from which it follows that
\#
		\| \hat \bdelta \|_1 \leq   4  s^{1/2} \| \hat \bdelta \|_2 + 2 \lambda^{-1} b_h^* \|   \hat \bdelta  \|_{\bSigma}   \leq   \bigl(   4 \gamma_p^{-1/2}   s^{1/2} +   l_0 \kappa_2 \lambda^{-1} h^2   \bigr)  \|   \hat \bdelta  \|_{\bSigma} .  \label{l1.cone.bound}
\#
We then let $h^2 \leq s^{1/2} \lambda$ hereinafter, so that conditioned on $\{ \lambda \geq  2 \| \bw^*_h \|_\infty \}$, $\hat \bbeta \in \bbeta^* + \CC_{\bSigma}(l)$ with $l= (4 \gamma_p^{-1/2}  + l_0 \kappa_2)  s^{1/2}$.


\medskip
\noindent
{\sc Step 2: Lower Bound}.
Set $r=h/(20\upsilon_0^2)$. Recall from Proposition~\ref{prop:RSC} that the RSC property only holds (with high probability) in a local neighborhood $\bbeta^*+ \BB_{\bSigma}(r) \cap  \CC_{\bSigma}(l) $, to which $\hat \bbeta$ does not necessarily belong.
Similarly to the proof of Proposition~\ref{prop:bias}, we define $\eta = \sup\{ u\in [0,1] :  \bbeta^* + u ( \hat \bbeta - \bbeta^* ) \in \BB_{\bSigma}(r) \}$, and an intermediate vector $\tilde{\bbeta} = (1-\eta) \bbeta^* + \eta \hat{\bbeta}$ that falls in $\bbeta^*+ \BB_{\bSigma}(r)$.
By this definition, we have $\eta=1$ if $\hat \bbeta \in \bbeta^* + \BB_{\bSigma}(r)$, and $\eta \in (0,1)$  if $\hat \bbeta   \notin \bbeta^* + \BB_{\bSigma}(r)$. In the latter case, $\wt{\bbeta}$ lies at the boundary of $ \bbeta^* +\BB_{\bSigma}(r)$.
Since $\wt \bbeta - \bbeta^* = \eta( \hat \bbeta - \bbeta^*)$, by \eqref{l1.cone.bound} we also have  $\wt \bbeta \in \bbeta^* + \CC_{\bSigma}(l)$ conditioned  on $\{ \lambda \geq  2 \| \bw^*_h \|_\infty \}$.
Consequently, conditioned on $\{  \lambda \geq  2 \| \bw^*_h \|_\infty \} \cap \cE_{{\rm rsc}}(r, l, \kappa)$ with $\kappa = (\kappa_l f_l)/2$,
\#
  \bigl\langle   \nabla \hat Q_h(\wt \bbeta ) - \nabla \hat Q_h(\bbeta^*) ,  \wt \bbeta - \bbeta^*  \bigr\rangle  \geq  \frac{1 }{2} \kappa_l f_l \, \|  \wt \bbeta - \bbeta^* \|_{\bSigma}^2  . \label{foc.lbd}
\#

\medskip
\noindent
{\sc Step 3: Combining Lower and Upper Bounds}. 
To bridge the upper and lower bounds obtained above, we apply \eqref{pop.foc} with $( Q_h, \bbeta^*_h, \bbeta^{\dagger}_h)$ replaced by $(\hat Q_h, \hat \bbeta, \wt \bbeta)$ to conclude that
\begin{align}
	    \bigl\langle   \nabla \hat Q_h(\wt \bbeta ) - \nabla \hat Q_h(\bbeta^*) ,  \wt \bbeta - \bbeta^*  \bigr\rangle   \leq \eta \bigl\langle   \nabla \hat Q_h(\hat \bbeta ) - \nabla \hat Q_h(\bbeta^*) ,  \hat{\bbeta}  - \bbeta^*  \bigr\rangle.   \nn
\end{align}
This, combined with \eqref{foc.ubd} and \eqref{foc.lbd}, implies that conditioned on $\{  \lambda \geq  2 \| \bw^*_h \|_\infty \} \cap \cE_{{\rm rsc}}(r, l, \kappa)$,  
\#
	  \frac{1}{2} \kappa_l   f_l\,  \| \wt \bdelta \|_{\bSigma}^2  \leq   \frac{3}{2}   s^{1/2} \lambda \| \wt \bdelta \|_2  +  b_h^*  \|   \wt \bdelta  \|_{\bSigma}   \leq   \frac{3}{2}   \gamma_p^{-1/2}  s^{1/2}   \lambda   \| \wt \bdelta \|_{\bSigma} +   \frac{1}{2}  l_0 \kappa_2 h^2    \|   \wt \bdelta  \|_{\bSigma} , \nn
\#
where $\tilde\bdelta = \tilde{\bbeta} - \bbeta^*$. Canceling $\| \wt \bdelta \|_{\bSigma}$ and re-arranging the terms yield
\#
   \| \wt \bdelta \|_{\bSigma}  \leq      \frac{1}{\kappa_l f_l}   \bigl(  3 \gamma_p^{-1/2}   s^{1/2} \lambda + l_0 \kappa_2 h^2 \bigr) \leq   \frac{1}{\kappa_l f_l}   \bigl(  3 \gamma_p^{-1/2}    + l_0 \kappa_2 \bigr) s^{1/2} \lambda     .   \label{thm:lasso-lad:proof3}
\#

It remains to control the probability of the event $\{  \lambda \geq  2 \| \bw^*_h \|_\infty \} \cap \cE_{{\rm rsc}}(r, l, \kappa)$, where $l=( 4 \gamma_p^{-1/2}    +   l_0 \kappa_2)s^{1/2}$ and $\kappa = (\kappa_l f_l)/2$.
By Proposition~\ref{prop:RSC} and Lemma~\ref{lem:subgradient}, we take  
\#
\lambda = 2 \upsilon_0\sigma_{\bx}  \Biggl\{    \sqrt{  \bigl\{ \tau(1-\tau) + (1+\tau) l_0 \kappa_2  h^2 \bigr\} \frac{ \log (2p)}{n}} + \max(1-\tau, \tau ) \frac{2\log (2p)}{n} \Biggr\} , \label{lambda.choice}
\#
so that $\{  \lambda \geq  2 \| \bw^*_h \|_\infty \} \cap \cE_{{\rm rsc}}(r, l, \kappa)$ occurs with probability at least $1-p^{-1}$ as long as $$
   \frac{ \sigma_{\bx}^2 f_u}{(\kappa_l f_l)^2}  \frac{s\log(p)}{n}  \lesssim  h \leq  f_l/(2l_0)  .
$$ 
This certifies the error bound \eqref{thm:lasso-lad:proof3} for $\wt \bbeta$.
Assume further that
 $$
   \frac{1}{\kappa_l f_l}   \bigl(  3 \gamma_p^{-1/2}    + l_0 \kappa_2  \bigr) s^{1/2} \lambda < r =  \frac{h}{20\upsilon_0^2},
 $$
 then $\wt \bbeta$ falls in  the interior of $\BB_{\bSigma}(r)$ with high probability.  Via proof by contradiction, we must have $\eta=1$ and thus $\hat \bbeta = \wt \bbeta$ also satisfies \eqref{thm:lasso-lad:proof3}. This proves the claimed bounds.  \qed

\bigskip
\section{Proofs of Results in Section~\ref{sec:theory:concave}}
\label{appendix:concave}
We first provide two high-level results on the cone property and $\ell_2$-error bound of the weighted $\ell_1$-penalized smoothed QR estimator that solves \eqref{general.lasso}. Recall that $\bw_h^* = \bw_h(\bbeta^*)\in \RR^p$ and $b_h^* = b_h(\bbeta^*)$, where $\bw_h(\bbeta) = \nabla \hat Q_h(  \bbeta ) - \nabla Q_h(\bbeta) $ and  $b_h(\bbeta) =  \| \bSigma^{-1/2}  \nabla  Q_h(\bbeta) \|_2$.
Lemma~\ref{lem:cone.property} provides conditions under which the optimal solution to the convex problem \eqref{general.lasso} falls in an $\ell_1$-cone. 

\begin{lemma}   \label{lem:cone.property}
Let $\cT$  be a subset of $[p]$ satisfying $\cS \subseteq \cT$, and let $\bbeta \in \RR^{p}$ be such that $\bbeta_{  {\cT}^{\cc}} = \textbf{0}$. 
Conditioned on $\{ \| \blambda_{ {\cT}^{\cc}} \|_{\min}  >    \|  \bw_h(\bbeta) \|_{\infty} \}$, any optimal solution $\hat  \bbeta $ to \eqref{general.lasso} satisfies
\#
 \|  (\hat\bbeta  -\bbeta )_{ {\cT}^{{\rm c}}}  \|_1 \leq \frac{  \{ \| \blambda \|_\infty +  \| \bw_h(\bbeta) \|_\infty \}  \| (\hat \bbeta - \bbeta)_{\cT} \|_1    +  b_h(\bbeta)  \| \hat \bbeta -\bbeta \|_{\bSigma} }{\| \blambda_{ {\cT}^{{\rm c}}}  \|_{\min} -  \| \bw_h(\bbeta) \|_\infty } .     \nn
\# 
\end{lemma}

\begin{lemma} \label{lem:general.error}
Let $\cT$ be a subset of $[p]$ such that $\cS \subseteq \cT$ and $k=|\cT|$, and let $\blambda = ( \lambda_1,\ldots, \lambda_p)^\T $ satisfy $\| \blambda\|_\infty \leq \lambda$, $\| \blambda_{  \cT^\cc} \|_{\min} \geq a\lambda$ for some $0<a\leq 1$ and $\lambda \geq (s/\gamma_p)^{-1/2} b^*_h$.
Conditioned on the event $\{ \lambda \geq (2/a)   \|  \bw^*_h \|_\infty \}$,  any optimal solution $\hat \bbeta $ to \eqref{general.lasso} satisfies $\hat \bbeta    \in \bbeta^* + \CC_{\bSigma}(l)$, where $l=l(a,k):=(2+2/a)  (k/\gamma_p)^{1/2} +  (2/a) (s/\gamma_p)^{1/2}$.
Moreover, let $r, \kappa >0$ satisfy
\#
       \gamma_p^{-1/2}   \big(   0.5 a k^{1/2}  +2  s^{1/2}   \big) \lambda  <   r  \cdot  \kappa  . \label{RSC.scaling1}
\#	
Then, conditioned on $\{ \lambda \geq (2/a)   \|  \bw^*_h \|_\infty \}\cap \cE_{{\rm rsc}}(r, l, \kappa)$, 
\#
 \| \hat \bbeta -\bbeta^* \|_{\bSigma}  & \leq    \kappa^{-1}   \bigl \{ \gamma_p^{-1/2}   \bigl( \|  \bw^*_{h,  \cT} \|_2 +   \|\blambda_{\cS} \|_2 \bigr)  + b_h^*  \bigr\}   \nn \\
 &  \leq      \kappa^{-1}  \gamma_p^{-1/2}   \big(   0.5 a k^{1/2}  + 2 s^{1/2} \big)  \lambda  . \label{general.l2error}
\#
\end{lemma}

Lemma~\ref{lem:subgradient:lowd} provides a probabilistic bound for the stochastic term $ \|  \bw^*_{h, \cS} \|_2$, which determines the oracle rate of convergence.

\begin{lemma} 
\label{lem:subgradient:lowd}
Assume that Conditions~(B1)--(B3) hold. Then, for any $t> 0$,  
\#
	  \|\Sb^{-1/2} \bw_{h,  \cS}^* \|_2 \leq   3  \upsilon_0  \Biggl[   \sqrt{  \bigl\{\tau(1-\tau) + C h^2 \bigr\}\frac{2 s + t}{n}} + \max(1-\tau, \tau ) \frac{2s+ t }{n} \Biggr]      \label{score.ubd}
\#
holds with probability at least $1-e^{-t}$, where $C=(\tau+1)l_0 \kappa_2 $ and $ \Sb = \EE(\bx_{ \cS} \bx_{  \cS}^\T) \in \RR^{ s \times s}$.
\end{lemma}

\subsection{Proof of Theorem~\ref{thm:LLA}} 
The proof is based a deterministic analysis conditioning on the event $\{ \| \bw_h^* \|_\infty \leq 0.5 q'(\alpha_0) \lambda \}$ for some $\lambda \geq (s/\gamma_p)^{-1/2} b^*_h$. We extend the argument used in the proof of Theorem~4.2 in \cite{FLSZ2018} with a more delicate treatment of the local RSC property and smoothing bias. 
With an initial estimator $\hat \bbeta^{(0)} = \textbf{0}$, we have $\blambda^{(0)}  = (\lambda, \ldots, \lambda)^\T \in \RR^{p}$. Applying Lemma~\ref{lem:general.error} with $\cT = \cS$ and $a=1$  yields that, conditioned further on the event $\cE_{{\rm rsc}}(r, l(1,s), \kappa)$, 
\#
	 \| \hat \bbeta^{(1)} - \bbeta^* \|_{\bSigma} &  \leq   \kappa^{-1}   \bigl \{ \gamma_p^{-1/2}   \bigl( \| \bw^*_{h,  \cS} \|_2 +   \|\blambda_{\cS} \|_2 \bigr)  + b_h^*  \bigr\}    \nn \\
	 & \leq   \kappa^{-1}     \bigl \{ \gamma_p^{-1/2}   \bigl( 1+  0.5 q'(\alpha_0)    \bigr)s^{1/2} \lambda  + b_h^*  \bigr\} \nn \\
	 & \leq    \kappa^{-1}     \bigl \{   0.5 q'(\alpha_0)     + 2 \bigr\} (s/\gamma_p)^{1/2} \lambda ,\label{bound.step1}   
\#
where $l(1,s)= 6(s/\gamma_p)^{1/2}$.

To improve the statistical rate of $\hat \bbeta^{(\ell)}   = (\hat \beta^{(\ell)}_{1} ,\ldots, \hat \beta^{(\ell)}_{p})^\T$ at step $\ell\geq 2$, we need to control the magnitude of the false discoveries of the solution obtained from the previous step, that is, $\max_{j \in  \cS^{\cc}} | \hat \beta_{j}^{(\ell-1)}|$. 
Recall that $\blambda^{(\ell-1)}   = ( \lambda_1^{(\ell-1)} , \ldots, \lambda_{p}^{(\ell-1)} )^\T =  (   q_\lambda'( |\hat \beta_{1}^{(\ell-1)}| )  ,  \ldots  , q_\lambda'( |\hat \beta_{p}^{(\ell-1)}| ) )^\T$, where $q_\lambda(t) = \lambda^2 q(t/\lambda)$ for $t \geq 0$. Since $q'(\cdot)$ is monotone on $\RR^+$, large magnitudes of $ | \hat \beta_{j}^{(\ell-1)}|$ indicate small values of  $\lambda^{(\ell-1)}_j$.
Motivated by this observation, we construct an augmented index set $\cT_\ell$, satisfying $\cS \subseteq \cT_\ell \subseteq [p]$, in each step and control the magnitude of $\| \blambda^{(\ell-1)}_{\cT_\ell^{{\cc}}} \|_{\min}$.

For $\ell =  1, 2, \ldots$, define the index set
\#
	\cT_\ell = \cS \cup  \bigl\{ 1  \leq   j \leq  p :  \lambda_j^{(\ell-1)}  < q'(\alpha_0 ) \lambda  \bigr\} ,  \label{def:El}
\#
which depends on $\hat \bbeta^{(\ell-1)}$. Let $c>0$ be determined by equation \eqref{eqn.c}.
We claim that
\#
	 | \cT_\ell | < (c^2 +1 ) s ~~\mbox{ and }~~  \| \blambda^{(\ell-1)}_{\cT_\ell^{\cc}} \|_{\min} \geq  q'(\alpha_0) \lambda   . \label{scaling.stepl}
\#
Indeed, if these two inequalities hold, applying Lemma~\ref{lem:general.error} with $a=q'(\alpha_0)$, $k= (c^2 +1 ) s$ and $l  =  \{ ( 2+ \frac{2}{q'(\alpha_0)} )  (c^2 +1 )^{1/2} + \frac{2}{q'(\alpha_0)} \} (s/\gamma_p)^{1/2}$ implies that, conditioned on $\cE_{{\rm rsc}}(r, l , \kappa)$,
\#
	\| \hat \bbeta^{(\ell)} - \bbeta^* \|_{\bSigma}  & \leq   \kappa^{-1}  \bigl\{   \gamma_p^{-1/2} \bigl(  \| \bw^*_{h, \cT_\ell}\|_2 + \| \blambda_{\cS}^{(\ell-1)} \|_2   \bigr)  + b_h^* \bigr\}    \label{bound.stepl} \\ 
	& <       \kappa^{-1} \bigl\{   0.5 q'(\alpha_0)  (c^2+1)^{1/2} + 2  \bigr\}  ( s/\gamma_p)^{1/2} \lambda \nn \\
	& \leq  \gamma_p^{1/2}   \alpha_0  c s^{1/2} \lambda = r_{{\rm opt}} ,  \label{crude.bound.stepl}
\#
where we have used \eqref{alpha0.constraint} and \eqref{eqn.c} in the second and third inequalities, respectively.

We now verify the  claim \eqref{scaling.stepl}  by induction on $\ell$. The claim is trivial if $\ell=1$, in which case $\blambda^{(0)} = ( \lambda, \ldots, \lambda)^\T$ and $\cT_1 =\cS$. Next, assume that for some integer $\ell \geq 1$, \eqref{scaling.stepl} holds and so does \eqref{crude.bound.stepl}. First we show that $|\cT_{\ell+1}| < (c^2+1)s$. For any $j\in \cT_{\ell+1} \setminus \cS$, $q'_\lambda( | \hat \beta_{j}^{(\ell)}|  )  = \lambda_{j}^{(\ell)}   <  q'(\alpha_0 ) \lambda = q_\lambda'(\alpha_0 \lambda ) $, implying $| \hat \beta_{ j}^{(\ell)}|  > \alpha_0 \lambda$ by the monotonicity of $q'_\lambda$ on $\RR^+$.
Recalling that $\beta^*_{ j} = 0$ for $j\in  \cT_{\ell+1} \setminus \cS$ and that the bound \eqref{crude.bound.stepl} holds for $\hat \bbeta^{(\ell)}$ by induction, we have
\#
	 | \cT_{\ell+1} \setminus \cS |^{1/2}  & <   ( \alpha_0 \lambda )^{-1}   \| \hat \bbeta^{(\ell )}_{  \cT_{\ell+1} \setminus \cS} \|_2   =  ( \alpha_0 \lambda )^{-1}   \|  ( \hat \bbeta^{(\ell )}  - \bbeta^* )_{ \cT_{\ell+1} \setminus \cS} \|_2  \nn \\
	 & \leq  ( \alpha_0 \lambda )^{-1} \gamma_p^{-1/2}  \|   \hat \bbeta^{(\ell )}  - \bbeta^*  \|_{\bSigma}    \leq c s^{1/2} . \label{set.bound}
\#
Consequently, $|\cT_{\ell+1}| =| \cS | + |\cT_{\ell+1} \setminus \cS| < (c^2+1)s$, as claimed. Turning to $  \| \blambda^{(\ell )}_{\cT_{\ell+1}^{\cc}} \|_{\min} $, it follows from \eqref{def:El} that $\lambda^{(\ell)}_j  \geq q'(\alpha_0) \lambda$  for each $ j \in \cT_{\ell+1}^{\cc}$. This completes the proof of  \eqref{scaling.stepl}.

Thus far, we have shown that the bounds \eqref{bound.stepl} and \eqref{crude.bound.stepl} hold for every $\ell\geq 1$. Specifically, the latter implies $\hat \bbeta^{(\ell)} \in \bbeta^* + \BB_{\bSigma}( r_{{\rm opt}})$ for every $\ell \geq 1$, where $r_{{\rm opt}} \asymp s^{1/2} \lambda$. Next we will show that, when the signal is sufficiently strong and if a concave penality $q_\lambda$ is used, the error bound $r_{{\rm opt}}$ can be refined at each iteration.
By \eqref{bound.stepl}, a key step is to derive sharper bounds on 
\#
	\| \blambda_{\cS}^{(\ell-1)}\|_2 = \sqrt{ \sum\nolimits_{j\in \cS} \{ \lambda_j^{(\ell -1)} \}^2 } = \sqrt{ \sum\nolimits_{j\in \cS} \{  q'_\lambda( | \hat \beta_{ j}^{(\ell-1)} |) \}^2} ~~\mbox{ and }~~ 
	\|   \bw_{h, \cT_\ell }^*  \|_2. \nn
\#	
For each $j$, note that if $|  \hat \beta_{ j}^{(\ell-1)}  - \beta_{ j}^*  | \geq   \alpha_0 \lambda$,  $\lambda_j^{(\ell-1)} \leq \lambda \leq  \alpha_0^{-1}|  \hat \beta_{ j}^{(\ell-1)}  - \beta_{ j}^*  |$; otherwise if $|  \hat \beta_{ j}^{(\ell-1)}  - \beta_{ j}^*  | \leq   \alpha_0 \lambda$, $\lambda_j^{(\ell-1)} \leq q_\lambda'( ( |\beta_{ j}^*| -  \alpha_0 \lambda )_+ )$ due to the monotonicity of $q_\lambda'$. Therefore, we have
\#
	\| \blambda_{\cS}^{(\ell-1)}\|_2 \leq  \|    q_\lambda'( ( |\bbeta^*_{ \cS}| -   \alpha_0 \lambda )_+ ) \|_2 +  \alpha_0^{-1}  \| ( \hat \bbeta^{(\ell-1)} - \bbeta^* )_\cS \|_2 . \nn
\#
For $	\|   \bw_{h, \cT_\ell }^*  \|_2$, it follows from the triangle inequality and \eqref{set.bound} that
\#
 	\|   \bw_{h, \cT_\ell }^*  \|_2    & \leq   	\|   \bw_{h,  \cS  }^*  \|_2 +    |\cT_\ell \setminus \cS |^{1/2} \|  \bw_{ h,  \cS^{{\rm c}}}^* \|_\infty  \nn \\
& <  \|   \bw_{h,   \cS  }^*  \|_2   + \frac{  1  }{  \alpha_0 \lambda}  \|  \bw_{h,  \cS^{{\rm c}}}^* \|_\infty   \|  ( \hat \bbeta^{(\ell  -1)}  - \bbeta^* )_{ \cT_{\ell } \setminus \cS} \|_2  \nn \\
&  \leq   \|   \bw_{h,   \cS  }^*  \|_2   +  0.5 q'(\alpha_0)  \alpha_0^{-1}  \|  ( \hat \bbeta^{(\ell  -1)} - \bbeta^* )_{ \cT_{\ell } \setminus \cS} \|_2   . \nn
\#
Using the elementary inequality $a + b \cdot c /2 \leq \sqrt{ (1+c^2/4)(a^2 +b^2)}$ for $a,b, c\geq 0$, we obtain
\$
 \| \blambda_{\cS}^{(\ell-1)}\|_2 + 	\|   \bw_{h, \cT_\ell }^*  \|_2 
\leq   \|  q_\lambda'(   ( |\bbeta^*_{ \cS}| -   \alpha_0 \lambda )_+ ) \|_2 +  \|   \bw_{h,   \cS  }^*  \|_2 + \frac{\sqrt{ 1+\{q'(\alpha_0)/2\}^2}}{\alpha_0}   \|  ( \hat \bbeta^{(\ell  -1)} - \bbeta^* )_{ \cT_{\ell } } \|_2 .
\$
Substituting this bound into \eqref{bound.stepl} yields
\#
 & \| \hat \bbeta^{(\ell)} - \bbeta^* \|_{\bSigma}  \nn \\
 & \leq     \kappa^{-1}  \gamma_p^{-1/2}  \big\{     \|  q_\lambda'(  ( |\bbeta^*_{ \cS}| -   \alpha_0 \lambda )_+ ) \|_2  +   \|   \bw_{h,   \cS  }^*  \|_2   \big\}    +   \kappa^{-1}  b_h^*   +  \delta \cdot   \| \hat \bbeta^{(\ell-1)} - \bbeta^* \|_{\bSigma}  , \nn
\# 
where $\delta = \sqrt{ 1+\{q'(\alpha_0)/2\}^2} / (\alpha_0 \kappa \gamma_p) \in (0,1)$ by \eqref{alpha0.constraint}.
This proves \eqref{contraction.inequality}. In conjunction with \eqref{bound.step1}, the second bound \eqref{sequential.bound} follows immediately. \qed

 \subsection{Proof of Theorem~\ref{thm:ncvx}}

The proof is based on  Theorem~\ref{thm:LLA}, in conjunction with Proposition~\ref{prop:RSC} and Lemma~\ref{lem:subgradient}.
With the stated choice of regularization parameter $\lambda \asymp  \sigma_{\bx} \sqrt{\tau(1-\tau)\log(p)/n}$ and bandwidth constraint, applying Lemma~\ref{lem:subgradient} with $t=2\log(2p)$ implies that $\| \bw_h^* \|_\infty \leq 0.5 q'(\alpha_0) \lambda$ holds with probability at least $1- (2p)^{-1}$ provided that $h\leq \sqrt{ \tau(1-\tau)/\{(1+\tau) l_0 \kappa_2\}}$ and $n\geq \frac{\max(\tau, 1-\tau)^2}{\tau(1-\tau)} \log(p)$.

Next, we apply Proposition~\ref{prop:RSC} to control the probability of the event $\cE_{{\rm rsc}}(r, l, \kappa)$ from Theorem~\ref{thm:LLA},  where
\$
	 r = \frac{h}{20\upsilon_0^2} , \quad   l =  \Bigg\{ \bigg(2+\frac{2}{q'(\alpha_0) } \bigg)  (c^2+1)^{1/2} + \frac{2}{q'(\alpha_0)}  \Bigg\}  (s/\gamma_p)^{1/2}  , \quad \kappa = \frac{\kappa_l f_l}{2}, 
\$
and the constant $c>0$ is determined by equation \eqref{eqn.c}. Provided that $n h \gtrsim f_u f_l^{-2} s\log(p)$,  Proposition~\ref{prop:RSC} guarantees that event $\cE_{{\rm rsc}}(r, l, \kappa)$ holds with probability at least $1-(2p)^{-1}$.

Moreover, recall from \eqref{crude.bound.stepl} that $r_{{\rm opt}} = \gamma_p^{1/2} \alpha_0 c s^{1/2} \lambda$ and   $b^*_h \leq l_0 \kappa_2 h^2/2$.  As long as the bandwidth $h$ is such that $r_{{\rm opt}} \leq r$ and $b_h^*\leq (s/\gamma_p)^{1/2} \lambda$, we can apply Theorem~\ref{thm:LLA} to conclude that, conditioned on $\cE_{{\rm rsc}}(r, l, \kappa) \cap \{ \| \bw_h^* \|_\infty \leq 0.5 q'(\alpha_0) \lambda \}$,
\#
	 \| \hat \bbeta^{(\ell)} - \bbeta^* \|_{\bSigma} \leq  \delta^{\ell-1} r_{{\rm opt}} + (1-\delta)^{-1}  
	 \kappa^{-1}   \big[ \gamma_p^{-1/2} \big\{  \| q'_\lambda ( (  \bbeta^*_{\cS} - \alpha_0 \lambda)_+ ) \|_2 + \| \bw^*_{h, \cS} \|_2 \big\} +   0.5 l_0 \kappa_2 h^2 \big]  \label{multi-step.bound}
\#
for any $\ell\geq 2$, where $\delta = \sqrt{5}/(\alpha_0 \kappa_l f_l \gamma_p ) \in (0,1)$ and $ \| q'_\lambda ( ( \bbeta^*_{\cS} - \alpha_0 \lambda)_+ ) \|_2  = 0$ under the stated beta-min condition.
Applying  Lemma~\ref{lem:subgradient:lowd} to the oracle term $\| \bw^*_{h, \cS} \|_2$, we obtain that with probability at least $1-e^{-t}$,
\#
  \| \Sb^{-1/2}\bw_{h,    \cS  }^*  \|_2 \lesssim   \upsilon_0   \Bigg\{  \sqrt{\tau(1-\tau )\frac{s+t}{n}} + \max(\tau, 1-\tau ) \frac{s+t}{n} \Bigg\}. \nn
\#

Finally, turning to the first term on the right-hand of \eqref{multi-step.bound}, noting that $r_{{\rm opt}} \asymp (\kappa_l f_l)^{-1} s^{1/2} \lambda$, we have
\#
 \delta^{\ell-1}  s^{1/2} \lambda   \lesssim   (\kappa_l f_l)^{-1} \delta^{\ell-1} \sqrt{\frac{s\log(p)}{n}} \lesssim (\kappa_l f_l)^{-1}\sqrt{\frac{s}{n}} , \nn
\# 
provided that  $\ell \geq \lceil \log(\log p) / \log(1/\delta) \rceil$. Putting together the pieces yields the claimed bounds in \eqref{weak.oracle.rate}.   \qed

\section{Proofs of Results in Section~\ref{sec:oracle}} 
\label{appendix:oracle}

Since the strong oracle property concerns the closeness between the estimator and the oracle, we modify Lemma~\ref{lem:general.error} to obtain the following result.  Recall that $\gamma_p=\gamma_p(\bSigma) \in (0, 1]$ is the minimum eigenvalue of $\bSigma$.
 
\begin{lemma} \label{lem:general.error2}
Let $\cT$ be a subset of $[p]$ such that $\cS \subseteq \cT$ and $k=| \cT|$, and let $\blambda = ( \lambda_1,\ldots, \lambda_p)^\T $ satisfy $\| \blambda\|_\infty \leq \lambda$, $\| \blambda_{  \cT^\cc} \|_{\min} \geq a\lambda $ for some $0<a\leq 1$ and $\lambda>0$.
Conditioned on $\{ \|  \bw^\ora_h \|_\infty \leq 0.5 a \lambda \}$,  any optimal solution $\hat \bbeta $ of \eqref{weighted.lasso-lad} falls in the $\ell_1$-cone $\hat \bbeta^{\ora} + \CC_{\bSigma}(l)$, where $l= (2+2/a)(k/\gamma_p)^{1/2}$.
Moreover, let $r, \kappa>0$ satisfy
\#
	\gamma_p^{-1/2}  \big( 0.5 a k^{1/2} + s^{1/2}     \big) \lambda < r \cdot \kappa . \label{RSC.scaling2}
\#
Then, conditioned on $\{ \|  \bw^\ora_h \|_\infty \leq 0.5 a \lambda \} \cap \cG_{{\rm rsc}}(r, l, \kappa)$, 
\#
 \| \hat \bbeta - \hat \bbeta^\ora \|_{\bSigma}  & \leq    \kappa^{-1} \gamma_p^{-1/2}   \bigl( \|  \bw^\ora_{h,\cT} \|_2 +   \|\blambda_{\cS} \|_2 \bigr)      \leq        \kappa^{-1}  \gamma_p^{-1/2}  \bigl(   0.5 a k^{1/2}   +   s^{1/2} \bigr) \lambda      . \label{general.l2error2}
\#
\end{lemma}

\subsection{Proof of Theorem~\ref{thm:strong.oracle1}}

For $\ell = 1, 2, \ldots$, let $\cT_\ell = \cS \cup   \{ j \in [p] :  \lambda_j^{(\ell-1)}  < q'( \alpha_0 ) \lambda  \}$ be the index sets given in \eqref{def:El}.
Recall that $l =  \{2+2/q'( \alpha_0 )\} (c_1^2+1)^{1/2}  (s/\gamma_p)^{1/2}$, where $c_1>0$ is determined by equation \eqref{eqn.c1}.
Conditioned on $\{ \| \bw^{\ora}_h \|_\infty \leq  0.5 q'( \alpha_0 ) \lambda  \}  \cap \{ \| \hat \bbeta^{\ora} - \bbeta^* \|_{\bSigma} \leq r \} \cap \cG_{{\rm rsc}}(r, l, \kappa)$, applying Lemma~\ref{lem:general.error2} with $a=q'( \alpha_0 )/2$ and following the same argument as in the proof of Theorem~\ref{thm:LLA}, we obtain that $|\cT_\ell| < (c_1^2+1)s$ and
\#
 & 	\| \hat \bbeta^{(\ell)} - \hat \bbeta^{{\rm ora}} \|_{\bSigma}   \leq \kappa^{-1} \gamma_p^{-1/2} \bigl\{ \| \blambda^{(\ell-1)}_{\cS} \|_2  +    \| \bw^{\ora} _{h, \cT_\ell} \|_2  \bigr\}   \nn \\
 & <     \kappa^{-1} \gamma_p^{-1/2}   \big\{    0.5 q'( \alpha_0 ) (c_1^2+1)^{1/2} + 1 \big\}  s^{1/2} \lambda   = \gamma_p^{1/2}  \alpha_0  c_1 s^{1/2} \lambda   \leq r   . \label{oracle.bound1}
\# 
Furthermore, define a sequence of index sets 
$$
	\cS_\ell  = \bigl\{ j \in [p] :  | \hat \beta_j^{(\ell)} - \beta_j^* |  > \alpha_0  \lambda \bigr\}, \ \ \ell =0, 1, 2, \ldots .
$$
Given the initialization $\hat  \bbeta^{(0)} = \textbf{0}$, the stated beta-min condition ensures $\cS_0 = \cS$.

In order to establish the equivalence between $\hat \bbeta^{(\ell)}$ and $\hat \bbeta^\ora$, we need to derive sharper bounds on 
$ \| \blambda^{(\ell-1)}_{\cS} \|_2$ and $ \| \bw^{\ora} _{h, \cT_\ell} \|_2$ in \eqref{oracle.bound1}.
For $\| \blambda^{(\ell-1)}_{\cS} \|_2 $,  from the monotonicity of $q'_\lambda$ we see that
$\lambda_j^{(\ell-1)}  = q_\lambda' ( | \hat \beta^{(\ell-1)}_j  | )   \leq q_\lambda'(|\beta_j^* | -    \alpha_0 \lambda  )$  if $j\in  \cS \cap  \cS_{\ell -1}^{\cc}  $, and $\lambda_j^{(\ell-1)}  \leq \lambda$ for all the remaining $j$. Combined with the beta-min condition $\| \bbeta_{\cS}^* \|_{\min} \geq (\alpha_0 + \alpha_1) \lambda$, we obtain
\#
	\| \blambda^{(\ell-1)}_{\cS} \|_2 \leq  \| q_\lambda'(|\bbeta^*_{\cS}|  - \alpha_0 \lambda  ) \|_2 + \lambda |\cS \cap \cS_{\ell - 1}|^{1/2 } =\lambda |\cS \cap \cS_{\ell - 1}|^{1/2}  .  \nn
\#
Turning to $ \| \bw^\ora_{h,\cT_\ell }  \|_2$, recall that  $\bw^\ora_{h, \cS} = \textbf{0}$ and hence
$$
 \| \bw^\ora_{h,\cT_\ell } \|_2  = \|\bw^\ora_{h, \cT_\ell \setminus \cS} \|_2 \leq \| \bw_h^\ora  \|_\infty |\cT_\ell \setminus \cS |^{1/2} .
$$
For each $j\in  \cT_\ell \setminus \cS$, $\beta^*_j = 0 $ and $\lambda_j^{(\ell-1)} = q_\lambda'(|\hat \beta_j^{(\ell-1)}| ) < q'(\alpha_0) \lambda =  q_\lambda'(\alpha_0 \lambda )$. Hence, $| \hat \beta_j^{(\ell-1)} - \beta^*_j| =| \hat \beta_j^{(\ell-1)} |  >  \alpha_0 \lambda$, indicating $j\in   \cS_{\ell-1} \setminus \cS$. Therefore, we have $\cT_\ell \setminus \cS \subseteq \cS_{\ell -1} \setminus \cS$, from which it follows that
\#
	 \| \bw^\ora_{h,\cT_\ell} \|_2  \leq   \|  \bw_h^\ora \|_\infty |  \cS_{\ell -1} \setminus \cS |^{1/2}   . \nn
 \#
Since $\| \hat \bbeta^{(\ell)} - \hat \bbeta^{{\rm ora}} \|_{\bSigma} \geq \gamma_p^{1/2} \| \hat \bbeta^{(\ell)} - \hat \bbeta^{{\rm ora}} \|_2$, substituting the above bounds into \eqref{oracle.bound1} yields 
 \#
	\| \hat \bbeta^{(\ell)} - \hat \bbeta^{{\rm ora}} \|_2  &  \leq   \big\{  |\cS \cap \cS_{\ell - 1}|^{1/2} +   0.5 q'( \alpha_0 )    | \cS_{\ell -1} \setminus \cS |^{1/2}   \big\}  (\kappa \gamma_p)^{-1}   \lambda   \nn \\
	&  \leq   \frac{   \sqrt{1+ \{ q'( \alpha_0 )/2 \}^2}  }{\kappa \gamma_p }    | \cS_{\ell - 1} |^{1/2}    \lambda,  ~\mbox{ for every } \ell \geq 1 .   \label{oracle.bound2}
\#

By \eqref{oracle.bound2}, in order to prove $ \hat \bbeta^{(\ell)}= \hat \bbeta^{\ora}$ for some sufficiently large $\ell$, it suffices to show that the set $\cS_{\ell-1}$ is empty. By the definition of $\cS_\ell$, $\min_{j \in \cS_\ell} | \hat \beta_j^{(\ell)} - \hat \beta^{{\rm ora}}_{ j} |   >   \alpha_0  \lambda - \| \hat \bbeta^{{\rm ora}} -\bbeta^*\|_\infty$.  Provided  that
$$ 
	\| \hat \bbeta^{\ora} - \bbeta^* \|_\infty \leq    \left[   \alpha_0 -  \frac{  \sqrt{1+\{ q'(\alpha_0)/2\}^2 }   }{\delta \kappa  \gamma_p}   \right] \lambda ,
$$
we have
\#
	| \cS_\ell |^{1/2}  <  \frac{  \|  ( \hat \bbeta^{(\ell)} - \hat \bbeta^{{\rm ora}} )_{\cS} \|_2 }{  \alpha_0  \lambda - \| \hat \bbeta^{{\rm ora}} -\bbeta^*\|_\infty }   \leq    \delta | \cS_{\ell -1} |^{1/2}      , \ \ \ell \geq 1.
\#
Since $\cS_0 = \cS$ with $|\cS_0|=s$, we have  $| \cS_\ell |^{1/2}  <  \delta^\ell s^{1/2}$ for all $\ell\geq 1$. When $\ell\geq  \lceil \log (s^{1/2})/\log(1/\delta) \rceil$, $| \cS_\ell |<1$ and hence $\cS_\ell$ must be empty. Returning to the error bound \eqref{oracle.bound2}, we conclude that $ \hat \bbeta^{(\ell)} = \hat \bbeta^{{\rm ora}} $ for all $\ell \geq \lceil \log (s^{1/2})/\log(1/\delta) \rceil +1$. This completes the proof.  \qed

\subsection{Proof of Theorem~\ref{thm:strong.oracle2}}

To apply the deterministic result in Theorem~\ref{thm:strong.oracle1}, we need the following two lemmas to control the probability of the events in \eqref{oracle.events}. Specifically, Lemma~\ref{lem:RSC2} ensures that the local RSC event $\cG_{{\rm rsc}}(r, l ,\kappa)$ holds with high probability, and Lemma~\ref{lem:oracle.score} characterizes all the stochastic quantities that involve the oracle estimator. 

\begin{lemma} \label{lem:RSC2}
Let $(r, l ,h)$ satisfy
\#
	   24 \upsilon_1^2  r =  h \leq f_l/(2 l_0) ~~\mbox{ and }~~ n h \geq C   f_u f_l^{-2}   \max\bigl\{ s, l^2 \log(p)\bigr\} \label{RSC2.scaling}
\#
for some sufficiently large constant $C$ depending only on $(\upsilon_1, \sigma_{\bx})$.
Then, the event $\cG_{{\rm rsc}}(r, l, \kappa)$ holds with probability at least $1- (2p)^{-1}$, where $\kappa=  \kappa_l f_l /2$ and $\kappa_l = \min_{|u|\leq 1} K(u)$.
\end{lemma}

\begin{lemma} \label{lem:oracle.score}
Let $A_0\geq 1$ be the constant in \eqref{irrepresentable.cond}.
For any $t\geq 0$, the oracle score $\bw_h^\ora = \nabla \hat Q_h(\hat \bbeta^{\ora}) \in \RR^p$ and oracle estimator $\hat \bbeta^{\ora}$ satisfy the bounds
\#
 \| \bw_h^\ora \|_\infty \lesssim  \sqrt{\frac{\log(2 p)}{n} } +    A_0\Bigg\{   \sqrt{\frac{\log(s) +t}{n}} +  \sqrt{\frac{s+\log(p)}{n h }} \sqrt{\frac{s+t}{n}}  + h^2 \Bigg\}  \label{oracle.score.bound}
\#
and 
\#
 \|  ( \hat \bbeta^\ora - \bbeta^*)_{  \cS} \|_\infty \lesssim   \frac{s +t }{  h^{1/2} n}   + h^2  + \sqrt{\frac{\log(s) + t}{n}} \label{oracle.infty.bound}
\#
with probability at least $1-p^{-1}- 5e^{-t}$, provided that the sample size $n$ and bandwidth $h$ are subject to  $\max\{ \sqrt{(s+ t )/n }  , \sqrt{\log(p)/n} \} \lesssim h \lesssim 1$.
\end{lemma}

Returning to the main thread, we are now ready to use Theorem~\ref{thm:strong.oracle1} to establish the strong oracle property.
Set $r= h/( 24 \upsilon_1^2) $, $l= \{ 2 + \frac{2}{q'(\alpha_0)} \} (c_1^2+1)^{1/2}( s/\gamma_p)^{1/2}$, $\kappa = \kappa_l f_l /2$, and choose the bandwidth $h\asymp \{ \log(p) / n \}^{1/4}$ so that $r\asymp \{ \log(p) / n \}^{1/4}$.
Together, Lemma~\ref{lem:RSC2} and Lemma~\ref{lem:oracle.score} with $t=\log(n)$ imply that, with probability at least $1- 2 p^{-1} - 5 n^{-1}$,  the following bounds
\$
   \| \bw_h^\ora \|_\infty \lesssim  \sqrt{\frac{\log(p)}{n}}   , ~~ \|  \hat \bbeta^{\ora} - \bbeta^* \|_{\bSigma} \lesssim \sqrt{\frac{s+\log(n)}{n}}   ~\mbox{ and }~  \| \hat \bbeta^{\ora} - \bbeta^* \|_\infty \lesssim  \sqrt{\frac{\min\{s, \log(p)\}}{n}} 
\$
hold provided the sample size obeys $n\gtrsim \max\{ s^{8/3}/(\log p)^{5/3}, \log(p) \}$.

Finally, as required by \eqref{oracle.events} in Theorem~\ref{thm:strong.oracle1}, if we choose the regularization parameter $\lambda = C \sqrt{\log(p)/n}$ for a sufficiently large $C$, then the events in \eqref{oracle.events} hold with probability at least $1- 2 p^{-1} - 5 n^{-1}$ under the scaling $n\gtrsim \max\{ s^{8/3}/(\log p)^{5/3} ,  s^2 \log (p)\}$. We have thus verified all the requirements in Theorem~\ref{thm:strong.oracle1}, hence certifying  the strong oracle property.
\qed

\section{Proofs of Results in Section~\ref{sec:zero.bias}}
\label{appendix:zerobias}

\subsection{Proof of Proposition~\ref{prop:unique}}

By \eqref{smooth.Hessian} and the uniqueness of $b_h$, $m''(\alpha ) =   \int_{-\infty}^\infty K(u)  f_{\varepsilon }(\alpha- hu ) \, {\rm d} u $ satisfies that $m''(b_h )>0$.
Recall that $\beta^*_1$ denotes the intercept and $\bx= (1, \bx_-^\T)^\T$ with $\bx_- \in \RR^{p-1}$.  For any $\bbeta= (\beta_1 , \ldots, \beta_p)^\T \in \RR^p$,
\#
	Q_h(\bbeta) & = \EE  \ell_h(\varepsilon - (\beta_1 - \beta^*_1)  -   \bx_-^\T ( \bbeta_-  - \bbeta_-^* ) ) \nn \\
& = \EE_{\bx}   \EE \bigl \{  \ell_h(\varepsilon - (\beta_1 - \beta^*_1)  -  \bx_-^\T ( \bbeta_-  - \bbeta_-^* )  ) | \bx \bigr\} \nn \\
& = \EE  \bigl\{  m( \beta_1 - \beta^*_1   +  \bx_-^\T ( \bbeta_-  - \bbeta_-^* )  ) \bigr\} \nn \\
& \geq  m(b_h )  =  \EE \ell_h( \varepsilon - b_h) = Q_h( \bbeta^\star )  ,
\#
where $\bbeta^\star = (\beta_1^* + b_h, \bbeta^{* \T}_-)^\T \in \RR^p$. This implies that $Q_h(\bbeta^\star) = \min_{\bbeta \in \RR^p}Q_h(\bbeta)$.
Furthermore, compute the Hessian matrix $\nabla^2 Q_h(\bbeta) = \EE \{ K_h(  \bx^\T ( \bbeta - \bbeta^* ) - \varepsilon ) \bx \bx^\T \}$. In particular, $\nabla^2 Q_h(\bbeta^\star) =  \EE \bigl\{ K_h( b_h - \varepsilon ) \bx \bx^\T \bigr\} = m''(b_h )\EE (   \bx\bx^\T )$ is positive definite, so that $\bbeta^\star$ is the unique minimizer of $\bbeta \mapsto Q_h(\bbeta)$. This ensures  that $\bbeta^*_h = \bbeta^\star$, as claimed. \qed

Next we characterize the order of $b_h$ as a function of $h$. 
Similarly to the proof of Proposition~\ref{prop:bias}, we define $\wt b$ as follows: if $| b_h | \leq  \kappa_2^{1/2} h$, set $\wt b = b_h$; otherwise if $| b_h | >   h$, set $\wt b =  \eta b_h$ for some $\eta \in (0,1)$ so that $|\wt b | =  \kappa_2^{1/2}  h$.
By \eqref{pop.foc},
\#
	 0\leq  \{ m'( \wt b ) - m'(0)  \}  \wt b \leq   \{   m'(b_h) - m'(0) \}  \wt b =  -  m'(0)  \wt b. \nn
\#
For the left-hand side,
\#
	m'(  \wt b ) - m'(0) & = \int_0^{  \wt b } m''(t) \, {\rm d}t  = \int_0^{  \wt b }  \int_{-\infty}^\infty K(u) f_{\varepsilon}(t-hu) \, {\rm d} u  \, {\rm d}t \nn \\
& =  f_{\varepsilon}(0)   \wt b  + \int_0^{  \wt b }  \int_{-\infty}^\infty K(u)  \bigl\{ f_{\varepsilon}(t-hu)  - f_{\varepsilon}(0) \bigr\} \, {\rm d} u  \, {\rm d}t , \nn
\#
implying
\#
\{ m'(  \wt b  ) - m'(0)  \}   \wt b  \geq  f_{\varepsilon}(0)  \wt b^2 - \frac{l_0}{2} |  \wt b |^3 - l_0 \kappa_1 h \cdot    \wt b^2 .\nn
\#
For the right-hand side, we have $|m'(0)| = | \int_{-\infty}^\infty  K(u)  \{ F_{\varepsilon}(-hu) -F_{\varepsilon}(0) \} \, {\rm d} u  | \leq \frac{l_0}{2} \kappa_2 h^2$. Combining the above upper and lower bounds, we find that
\#
    f_\varepsilon(0)    \wt b^2 \leq  \frac{l_0}{2} \kappa_2 h^2 | \wt b | + \frac{l_0}{2} | \wt b |^3 + l_0 \kappa_1  h    \wt b^2 \leq  \bigl(  \kappa_2 + \kappa_1 \kappa_2^{1/2}  \bigr) l_0 h^2 | \wt b |, \nn
\#
where the first inequality uses the fact that $| \wt b| \leq \kappa_2^{1/2} h$. Canceling $|\wt b|$ gives
\#
	 |\wt b | \leq  \underbrace{   \bigl(\kappa_2^{1/2} + \kappa_1\bigr)  }_{= c_1} \frac{l_0  h }{ f_\varepsilon(0)  }  \cdot    \kappa_2^{1/2} h .\nn
\#
As long as  $c_1 l_0 h < f_{\varepsilon}(0)$,  the above inequality implies $|\wt b| < \kappa_2^{1/2} h$.
By the definition of $\wt b$, we must have $\wt b = b_h$; otherwise $|\wt b| = \kappa_2^{1/2}h$ which leads to contradiction. This completes the proof of  \eqref{inter.bias.ubd1}. \qed

\subsection{Proof of Theorem~\ref{thm:new.weak.oracle}}

By the definition of $\bbeta^*_h$, we have $\nabla Q_h(\bbeta^*_h) = \textbf{0}$. Replacing $\bbeta^*$ by $\bbeta^*_h$ in \eqref{def:wb},  the smoothing error term $b_h^*$ now becomes zero. Modifying the proof of Theorem~\ref{thm:LLA}  accordingly, the conclusions therein remain valid, but now with $b_h^* = 0$ and
$$
	\bw_h^* = \bw_h(\bbeta^*_h) = \frac{1}{n} \sn  \bigl\{  \bar{K}( (b_h - \varepsilon_i)/h)  - \tau \bigr\} \bx_i  .
$$
Once again, the key is to show that event $\cE_{{\rm rsc}}(r, l ,\kappa) \cap \{ \| \bw^*_h \|_\infty \leq 0.5 q'(\alpha_0) \lambda \}$ holds with high probability, where $\cE_{{\rm rsc}}$ is given in \eqref{event.rsc} with $\bbeta^*$ replaced by $\bbeta^*_h$.

Proceed similarly to the proof of Lemma~\ref{lem:subgradient} and Lemma~\ref{lem:subgradient:lowd}, we obtain that with probability at least $1-(2p)^{-1}$,
\#
 \| \bw_h^* \|_\infty \lesssim \sigma_{\bx} \Bigg\{ \sqrt{ \frac{\log(2p)}{n}} + \frac{\log(2p)}{n}  \Bigg\}, \label{new.score.max.bound}
\#
and for any $t>0$, 
\#
	  \| \Sb^{-1/2} \bw_{h,   \cS }^* \|_2 \lesssim   \sqrt{\frac{s+ t}{n}} + \frac{s+t}{n}   \label{new.score.l2.bound}
\#
holds with probability at least $1-e^{-t}$,

Next, in order to show that Proposition~\ref{prop:RSC} remains valid if $\bbeta^*$ is replaced by $\bbeta_h^*$, it suffices to change the definition of the event $E_i$ in \eqref{eq:dblower1} to
\#
	 E_i = \bigl\{  | \varepsilon_i - b_h | \leq h/2 \bigr\} \cap  \left\{ \frac{|\langle \bx_i, \bbeta - \bbeta_h^* \rangle |}{ \| \bbeta - \bbeta_h^* \|_{\bSigma}}  \leq  \frac{h}{2r} \right\} . \nn
\#
Moreover, note that
\#
   | \EE  \mathbbm{1}(  | \varepsilon_i  - b_h | \leq h/2 )  -  h     f_0   |  & \leq   \int_{-h/2}^{h/2}   |  f_{\varepsilon  } (t  + b_h  ) - f_\varepsilon(0) | \, {\rm d} t   \leq  l_0 h^2/4 + l_0 b_h h. \nn
\#
Keep all other statements the same, we obtain that with probability at least $1-(2p)^{-1}$, the event $\cE_{{\rm rsc}}(r, l, \kappa)$ with $r = h/(20\upsilon_0^2)$, $l=   \{  (2 + \frac{2}{q'(\alpha_0)}) (c^2+1)^{1/2}  + \frac{2}{q'(\alpha_0)}  \} (s/\gamma_p)^{1/2}$ and $\kappa = \kappa_l f_0/2$ holds as long as $\sigma_{\bx}^2 f_0^{-1}s\log(p)/n \lesssim h \lesssim f_0$.

With a penalty level $\lambda \asymp \sigma_{\bx} \sqrt{\log(p)/n}$, we conclude from Theorem~\ref{thm:LLA} that with probability at least $1-p^{-1} - e^{-t}$,
\# 
	 \| \hat \bbeta^{(\ell)} - \bbeta^*_h \|_{\bSigma} & \lesssim \delta^{\ell-1} f_0^{-1}\sqrt{\frac{s\log(p)}{n}} +(1-\delta)^{-1} f_0^{-1}\sqrt{\frac{s+t}{n}}   \nn
\#
holds for every $\ell \geq 2$, provided that $n\gtrsim s\log(p) + t$ and $\sigma_{\bx} f_0^{-1}\sqrt{s\log(p)/n} \lesssim h \lesssim f_0 $, where $\delta = \sqrt{4+ \{ q'(\alpha_0)\}^2}/(\alpha_0 \kappa_l f_0 \gamma_p)$. This completes the proof by letting $\ell \geq \lceil  \log(\log p) / \log(1/\delta)\rceil$. \qed

\subsection{Proof of Proposition~\ref{thm:new.oracle}}

Define the oracle smoothed quantile loss and its population counterpart as
$$
	\hat Q^\ora_h(\bbeta) = \frac{1}{n} \sn \ell_h(y_i -  \bx_{i, \cS}^\T  \bbeta   )  ~~\mbox{ and }~~  Q^\ora_h(\bbeta)  = \EE \hat Q^\ora_h(\bbeta)  , \ \ \bbeta \in \RR^{ s}. 
$$
With some abuse of notation, we write $\bbeta^*_h = \bbeta^*_{h,  \cS} \in \RR^s$ and $\hat \bbeta^\ora \in \argmin_{\bbeta \in \RR^{ s}} 	\hat Q^\ora_h(\bbeta)$. The concentration bound \eqref{new.oracle.concentration} follows from the same argument that was used to prove \eqref{oracle.concentration}. In particular, the restricted strong convexity of $\hat Q^\ora_h$ around $\bbeta^*_h$ is established similarly as in the proof of Theorem~\ref{thm:new.weak.oracle}, and the $h^2$-term vanishes because $\nabla Q^{\ora}_h(\bbeta^*_h) = \textbf{0}$.

To prove \eqref{new.oracle.bahadur}, define the stochastic process
\#
	\Delta(\bbeta ) = \Sb^{-1/2}    \bigl\{ \nabla  \hat Q^\ora_h(\bbeta) -\nabla \hat Q^\ora_h(\bbeta_h^*) -\Db_h (\bbeta - \bbeta_h^*)  \big\} ,     \ \  \bbeta \in \RR^s , \label{def:Delta}
\#
where by the independence of $\bx$ and $\varepsilon$, $\Db_h = \nabla^2 Q^\ora_h(\bbeta^*_h) =  m''(b_h) \cdot  \Sb$. We will bound the supremum $\sup_{\bbeta \in \bbeta_h^* + \BB_{\Sb}(r)} \| \Delta(\bbeta )  - \EE \Delta(\bbeta )  \|_2$ using the same argument as in the proof of Theorem~4.2 in \cite{HPTZ2020}. It then suffices to evaluate $\EE \Delta(\bbeta)$. By the mean value theorem for vector-valued functions,
\#
\EE \Delta(\bbeta) & = \Sb^{-1/2}  \int_0^1 \nabla^2   Q^\ora_h( (1-t) \bbeta_h^* + t \bbeta )  \, \mathrm{d}t   \,  \bigl(  \bbeta - \bbeta_h^* \bigr)  -  \Sb^{-1/2}\Db_h  (\bbeta - \bbeta_h^*) \nn \\
& =  \Biggl\{   \Sb^{-1/2}    \int_0^1 \nabla^2   Q^\ora_h((1-t) \bbeta_h^* + t \bbeta  ) \,  \mathrm{d}t \,   \Sb^{-1/2}    - m''(b_h) \Ib_s \Biggr\}  \Sb^{1/2}(\bbeta - \bbeta_h^*)  . \label{exp.Delta.oracle}
\#
Note that, for every $\bbeta \in \RR^s$,
\#
  \nabla^2   Q_h^\ora(\bbeta  ) & = \EE \bigl\{ K_h(   \bx_{  \cS}^\T \bbeta - y ) \bx_{ \cS} \bx_{  \cS}^\T \bigr\}   =   \EE   \Biggl\{  \int_{-\infty}^\infty K(u)  f_\varepsilon( \bx_{  \cS}^\T( \bbeta  - \bbeta^*_h) + b_h - h u ) \, {\rm d} u \cdot   \bx_{   \cS} \bx_{  \cS}^\T \Biggr\}. \nn
\#
Moreover, write $\bdelta = \bbeta - \bbeta_h^*$ for $\bbeta \in \bbeta_h^* + \BB_{\Sb}(r)$ so that
$$
	 \nabla^2   Q_h^\ora((1-t) \bbeta_h^* + t \bbeta  ) =  \EE \Biggl\{ \int_{-\infty}^\infty K(u)  f_\varepsilon(   \bx_{  \cS}^\T \bdelta \cdot t + b_h - h u ) \, {\rm d} u \cdot   \bx_{ \cS} \bx_{  \cS}^\T  \Biggr\} . 
$$
Consequently, for any $t\in [0,1]$,
\#
	& \Bigl\|  \Sb^{-1/2} \nabla^2   Q^\ora_h( (1-t) \bbeta_h^* + t \bbeta ) \, \Sb^{-1/2}   -  m''(b_h) \Ib_s  \Bigr\|_2 \nn \\
	& =  \Biggl\|  \Sb^{-1/2}    \EE  \biggl[  \int K(u) \bigl\{ f_{\varepsilon } (    \bx_{  \cS}^\T  \bdelta  \cdot t + b_h - h u)    - f_{\varepsilon  } (b_h - h u ) \bigr\} \, {\rm d} u  \, \bx_{  \cS} \bx_{  \cS}^\T  \biggr] \Sb^{-1/2}   \Biggr\|_2   \nn \\
& \leq  l_0 t\cdot  \sup_{ \| \bu \|_2 = 1 }  \EE \bigl\{  \langle\bx_{  \cS} ,  \Sb^{-1/2} \bu \rangle^2  |  \bx_{  \cS}^\T \bdelta  |  \bigr\}      \nn \\
& \leq  l_0 t \cdot   \Bigg(\sup_{ \| \bu \|_2 = 1 }  \EE  \langle\bx_{  \cS},  \Sb^{-1/2} \bu \rangle^4 \Bigg)^{1/2}  \Bigg( \EE \langle \bx_{ \cS} , \bdelta  \rangle ^2 \Bigg)^{1/2}  \nn \\
& \leq      \mu_4^{1/2}  l_0  r t   , \nn
\#
where $\mu_4 = \sup_{  \bu \in \mathbb{S}^{p-1} }  \EE\langle \bSigma^{-1/2} \bx , \bu \rangle^4$. Together with \eqref{exp.Delta.oracle}, this leads to
\#
	\sup_{\bbeta \in \bbeta_h^* +  \BB_{\Sb} (r) }  \|   \EE \Delta(\bbeta)  \|_2  \leq    \frac{l_0}{2   } \mu_4^{1/2}    r^2 . \label{br.remainder.ubd1}
\#

Furthermore, observe that
\#
  m''(b_h) = \EE \{ K_h (b_h-\varepsilon ) \}  =   \int_{-\infty}^\infty  K( u ) f_\varepsilon( b_h - h u)  \, {\rm d} u  \geq f_{\varepsilon}(0) - l_0 ( b_h + \kappa_1 h   ) \geq \frac{1}{2} f_{\varepsilon}(0) , \nn
\#
where the last inequality holds provided that $h$ is sufficiently small.  
Combining this with \eqref{br.remainder.ubd1} and (B.31) of \cite{HPTZ2020}, we conclude that for any $r, t>0$,
\#
		\sup_{\bbeta \in \bbeta_h^* + \BB_{\Sb}(r) }\bigl \|  \Sb^{-1/2} \bigl\{ \nabla  \hat Q^\ora_h(\bbeta) -\nabla \hat Q^\ora_h(\bbeta_h^*)  - \Db_h (\bbeta - \bbeta_h^*) \big\} \bigr\|_2 \lesssim  \Biggl( \sqrt{\frac{s+t}{n h}}+  r \Biggr) r 
\# 
with probability at least $1-e^{-t}$ as long as $\sqrt{(s+t)/n} \lesssim h\lesssim 1$.  Taking $\bbeta = \hat \bbeta^\ora$ and $r\asymp \sqrt{(s+t)/n}$, \eqref{new.oracle.bahadur} follows from \eqref{new.oracle.concentration} and the fact that $\nabla  \hat Q^\ora_h(\hat \bbeta^\ora) = \textbf{0}$. \qed

\subsection{Proof of Theorem~\ref{thm:new.strong.oracle}}

Similarly to the proof of Theorem~\ref{thm:strong.oracle2}, the proof of Theorem~\ref{thm:new.strong.oracle} is based on Lemmas~\ref{lem:RSC2} and \ref{lem:oracle.score} with slight modifications.
In the proof of Lemma~\ref{lem:RSC2}, change the event $E_i$ used in \eqref{def:D12} to 
\#
   E_i = \bigl\{ |\varepsilon_i  - b_h | \leq h /2 \bigr\}  \cap   \bigl\{    |\langle  \bx_i, \bbeta_2 - \bbeta_h^* \rangle |  \leq  h/4 \bigr\} \cap \bigl\{    |\langle  \bx_i, \bbeta_1 - \bbeta_2 \rangle |  \leq  \|   \bbeta_1 - \bbeta_2  \|_{\bSigma} \cdot h / (4r)    \bigr\}  , \nn
\#
and keep all other arguments, the conclusions of  Lemma~\ref{lem:RSC2} remain valid.

Recall that $\bw^\ora_h = \nabla \hat Q_h(\hat \bbeta^\ora)$.
The following result refines Lemma~\ref{lem:oracle.score} under the additional independence assumption.

\begin{lemma} \label{lem:new.oracle.score}
Let $A_1 \geq 1$ be the constant in \eqref{irrepresentable.cond2}. For any $t\geq 0$, the oracle score $\bw_h^\ora = \nabla \hat Q_h(\hat \bbeta^{\ora}) \in \RR^p$ and oracle estimator $\hat \bbeta^{\ora}$ satisfy the bounds
\#
 \| \bw_h^\ora \|_\infty \lesssim  \sqrt{\frac{\log(  p)}{n} } +    A_1\Bigg\{   \sqrt{\frac{\log(s) +t}{n}} + \frac{(s+t)^{1/2}(s+\log p)^{1/2} }{h^{1/2} n}   \Bigg\}  \label{new.oracle.score.bound}
\#
and 
\#
 \|    \hat \bbeta^\ora - \bbeta^*  \|_\infty \lesssim   \frac{s +t }{  h^{1/2} n}   + \sqrt{\frac{\log(s) + t}{n}} \label{new.oracle.infty.bound}
\#
with probability at least $1-p^{-1}- 5e^{-t}$, provided that the sample size $n$ and bandwidth $h$ are subject to  $\sqrt{(s+ t )/n }\lesssim h \lesssim 1$ and $h\gtrsim \sqrt{(s+\log p ) /n}$.
\end{lemma}

The rest of the proof then proceeds similarly to the proof of Theorem~\ref{thm:strong.oracle2}, and thus is omitted.
\qed


\section{Proof of Auxiliary Lemmas}
\label{appendix:auxiliary}

\subsection{Proof of Lemma~\ref{covariate.moment}}

Condition~(B3) ensures that $\PP(|\bz^\T \bu | \geq  \upsilon_0  t) \leq e^{-t}$ for any $t \geq 0$ and $\bu \in \mathbb{S}^{p-1}$. For any $k\geq 1$, this implies
\$
 \EE |\bz^\T \bu |^k = \upsilon_0^k  k \int_0^\infty  t^{k-1}\PP(|\bz^\T \bu | \geq  \upsilon_0   t ) \, {\rm d}t \leq  \upsilon_0^k  k \int_0^\infty  t^{k-1} e^{-t} \, {\rm d}t = k! \upsilon_0^k .
\$
Taking the supremum over $\bu \in \mathbb{S}^{p-1}$ proves the claimed bound. \qed

\subsection{Proof of Lemma~\ref{lem:subgradient}}
To facilitate the proof, let $\xi_i = \bar K(-\varepsilon_i  /h) -   \tau$.  Taking $\bbeta = \bbeta^*$ in the gradient function \eqref{smooth.Hessian} yields
\#
\nabla \hat Q_h(\bbeta^*)  =  \frac{1}{n}  \sn    \bigl\{ \bar K(-\varepsilon_i/h) - \tau \bigr\} \bx_i  = \frac{1}{n} \sn \xi_i \bx_i. \nn
\#
The upper bound for $\| \nabla \hat Q_h(\bbeta^*) - \nabla  Q_h(\bbeta^*)  \|_\infty = \| (1/n) \sn \{\xi_i \bx_i - \EE (\xi_i \bx_i) \} \|_\infty$ involves two quantities that are related to the kernel function: $\EE \{\bar K^2(  - \varepsilon /h  ) | \bx  \}$ and   $\EE\{  \bar K(  - \varepsilon /h) | \bx\}$.  We start with obtaining an upper bound for  $\EE \{\bar K^2(  - \varepsilon /h  ) | \bx  \}$. 
By a change of variable and integration by parts, we obtain
\#
\EE \bigl\{ \bar K^2( - \varepsilon/h) | \bx \bigr\}     &=  \int_{-\infty}^{\infty} \bar K^2(- u/h)  f_{\varepsilon\mid \bx} (u) \,  \mathrm{d} u  \nn \\
 &=  h \int_{-\infty}^{\infty}  \bar K^2(v)  f_{\varepsilon\mid \bx} (    -vh) \, \mathrm{d} v  \nn \\
&=  2 \int_{-\infty}^{\infty}   K(v) \bar K(v)  F_{\varepsilon\mid \bx} (   -vh) \, \mathrm{d} v \label{eq:maxnorm1}.
\#
By the fundamental theorem of calculus and the fact that $F_{\varepsilon \mid \bx} (0) = \tau$, we have
\#  
  F_{\varepsilon \mid \bx} (   -vh)  & = F_{\varepsilon \mid \bx} (0) + \int_0^{ - vh}   f_{\varepsilon |\bx}(t) \, {\rm d} t \nn \\
& = \tau  + (  - hv) f_{\varepsilon \mid\bx} (0) + 
 \int_0^{    - vh}   \bigl\{  f_{\varepsilon |\bx}(t) - f_{\varepsilon | \bx}(0) \bigr\} \, {\rm d} t   . \label{eq:maxnorm2}
\#
Moreover,  it can be shown that 
\#
  a_K :=  \int_{-\infty}^{\infty} vK(v) \bar K(v) \, \mathrm{d} v =  \int_0^{\infty} K(v) \{1-K(v)\} \, \mathrm{d}v > 0 ~~\mbox{ and } ~~ a_K \leq \kappa_1, \label{eq:maxnorm2-2}
\#
where $\kappa_1=\int |u| K(u) \,{\rm d} u$.

Substituting~\eqref{eq:maxnorm2} into~\eqref{eq:maxnorm1}, and by \eqref{eq:maxnorm2-2}, we obtain
\#
& \EE \bigl\{ \bar K^2( -\varepsilon/h) | \bx \bigr\}  \nn \\
  &= 2    \tau   \int_{-\infty}^{\infty} K(v) \bar K(v) \, \mathrm{d} v    - 2hf_{\varepsilon\mid \bx} (0) \int_{-\infty}^{\infty} vK(v) \bar K(v)\, \mathrm{d} v      \nn \\
  & ~~~~ +   2 \int_{-\infty }^\infty \int_0^{   - vh}    \bigl \{ f_{\varepsilon | \bx}(t) - f_{\varepsilon | \bx}(0) \bigr\}  K(v) \bar K(v) \, {\rm d} t \, {\rm d} v    \nn \\
  & \leq   \tau     -  2 a_K h  f_{\varepsilon\mid \bx} (0)       +  l_0 h^2  \cdot   \int_{-\infty}^{\infty} v^2 K(v) \bar K(v) \,  \mathrm{d} v     \nn \\
  & \leq  \tau   +   l_0 \kappa_2  h^2 ,\label{eq:maxnorm2-2.1}
 \#
 where the first inequality holds using the Lipschitz condition on $f_{\varepsilon \mid \bx}$ in Condition (B1), and the last inequality holds by Condition (B2) on the kernel function.
Through a similar calculation, $\EE \bigl\{ \bar K(-\varepsilon/h) | \bx \bigr\}  = \tau  + \int_{-\infty}^\infty \int_0^{ - hv}    \bigl \{ f_{\varepsilon | \bx}(t) - f_{\varepsilon | \bx}(0) \bigr\}  K(v)   \, {\rm d} t \, {\rm d} v$. The Lipschitz condition on $f_{\varepsilon |\bx}$ then ensures that
\#  
   \bigl|  \EE \bigl\{ \bar K( -\varepsilon /h) | \bx \bigr\}  -  \tau   \bigr| \leq  \frac{l_0}{2}   \kappa_2 h^2. \label{eq:maxnorm2-2.2}
\#
Together, \eqref{eq:maxnorm2-2.1} and \eqref{eq:maxnorm2-2.2} imply
\#
	\EE(\xi^2 | \bx ) & \leq \tau(1-\tau)    + (\tau+1)l_0 \kappa_2 h^2 = \tau(1-\tau ) + C h^2,  \label{cond.var.bound}
\#
where $\xi= \bar K(  -\varepsilon  /h) -  \tau $ and $C=(\tau+1)l_0 \kappa_2$.

With the above preparations, we are now ready to prove \eqref{score.ubd}. First, we use Bernstein's inequality to bound each $(1/n)\sum_{i=1}^n \{  \xi_i x_{ij}- \EE (\xi_i x_{ij})\} $, and then apply a union bound over $j=1,\ldots, p$.  Note that $\xi_1 x_{1j}- \EE (\xi_1 x_{1j}), \ldots, \xi_n x_{nj}- \EE (\xi_n x_{nj})$ are independent zero-mean random variables, and by \eqref{cond.var.bound},
\#
	\EE ( \xi_i x_{ij})^2 & = \EE_{\bx} \bigl\{ x_{ij}^2 \cdot \EE(\xi_i^2 | \bx_i ) \bigr\}  \leq \tau(1-\tau) \sigma_{jj} + C \sigma_{jj} h^2 . \nn
\#
Under Condition (B3), we have $\PP \bigl(|x_{ij}|\ge \sigma_{jj}^{1/2}\upsilon_0 t \bigr) \leq e^{-t}$ for all $t\geq 0$.
Noting that $|\xi_i| \leq \max(1-\tau, \tau) $, we have for  $k=2,3,\ldots$, 
\#
\EE \bigl( | \xi_i x_{ij} |^k\bigr) &\le \max( 1-\tau, \tau)^{k-2}\, \EE_{\bx}  \bigl\{  | x_{ij}|^k \cdot\EE(\xi_i^2 | \bx_i ) \bigr\}    \nn \\
&\le   \max( 1-\tau, \tau)^{k-2}  \bigl\{ \tau(1-\tau) + C  h^2 \bigr\}  \upsilon^k_0 \sigma_{jj}^{k/2} \int_0^{\infty} \PP (|\sigma_{jj}^{-1/2}x_{ij}|\ge \upsilon_0 t) kt^{k-1}\,\mathrm{d}t \nn\\
&\le  \max( 1-\tau, \tau)^{k-2}  \bigl\{ \tau(1-\tau) + C  h^2 \bigr\}    \upsilon^k_0 \sigma_{jj}^{k/2} k \int_0^{\infty} t^{k-1}e^{-t}\, \mathrm{d}t\nn\\
&=   k!   \max( 1-\tau, \tau)^{k-2} \bigl\{ \tau(1-\tau) + C  h^2 \bigr\}     \upsilon^k_0 \sigma_{jj}^{k/2} \nn \\
&\le \frac{k!}{2} \cdot    \bigl\{ \tau(1-\tau) + C  h^2 \bigr\}   \upsilon_0^2 \sigma_{jj}    \cdot  \bigl\{ 2 \max( 1-\tau, \tau) \upsilon_0 \sigma_{jj}^{1/2} \bigr\} ^{k-2}. \label{eq:bern:cond}
\#
Consequently, it follows from  Bernstein's inequality that for every $t\geq 0$,
\#
   \biggl| \frac{1}{n} \sn  \{\xi_i x_{ij} - \EE (\xi_i x_{ij})\}  \biggr| \leq   \upsilon_0\sigma_{jj}^{1/2} \left[   \sqrt{ \bigl\{ \tau(1-\tau) + C  h^2 \bigr\}  \frac{2t}{n}} + \max(1-\tau, \tau ) \frac{2t}{n} \right]  \nn
\#
with probability at least $1-2e^{-t}$. Finally, we apply a union bound to reach the conclusion \eqref{score.ubd}.      \qed

\subsection{Proof of Lemma~\ref{lem:cone.property}}
Since the objective function in \eqref{general.lasso} is convex, by the first-order optimality condition, 
there exists  a subgradient  $\hat \bg \in \partial \| \hat \bbeta  \|_1$ such that
$\nabla \hat Q_h(\hat \bbeta ) +   \blambda  \circ \hat \bg = \textbf{0}$. 
Using the fact that the subdifferential of a convex function is monotone increasing, for any $\bbeta \in \RR^{p}$, we have
\#
	     0 & = \langle \nabla \hat Q_h(\hat \bbeta )    +   \blambda  \circ \hat \bg , \hat \bbeta - \bbeta \rangle \nn  \\
	     & =  \underbrace{ \langle \nabla \hat Q_h(\hat \bbeta )  - \nabla \hat Q_h(  \bbeta ) , \hat \bbeta - \bbeta \rangle }_{\geq 0} + \langle  \nabla \hat Q_h( \bbeta ) , \hat \bbeta - \bbeta \rangle + \langle  \blambda  \circ \hat \bg , \hat \bbeta - \bbeta \rangle  \nn \\ 
	     &  \geq   \langle  \nabla \hat Q_h( \bbeta ) - \nabla Q_h(\bbeta) , \hat \bbeta - \bbeta \rangle + \langle \nabla Q_h(\bbeta), \hat \bbeta - \bbeta \rangle  + \langle  \blambda  \circ \hat \bg  , \hat \bbeta - \bbeta \rangle  \nn \\ 
	     & \geq -  \|\bw_h(\bbeta)\|_{\infty}   \| \hat \bbeta - \bbeta  \|_1 -   b_h(\bbeta)  \| \hat \bbeta - \bbeta \|_{\bSigma}    + \langle   \blambda  \circ \hat \bg  , \hat \bbeta - \bbeta \rangle .  \nn
\#

In addition, by the definition of the subgradient, $\langle   \blambda  \circ  \hat \bg , \hat \bbeta \rangle = \|  \blambda  \circ \hat \bbeta  \|_1$. 
Thus, for any $\bbeta\in \RR^p$ satisfying $\bbeta_{ \cT^{{\rm c}}}=\textbf{0}$, we can decompose $ \langle  \blambda  \circ  \hat \bg , \hat \bbeta - \bbeta \rangle$ according to the subset $\cT \subseteq [p]$ as 
\#
  \langle  \blambda  \circ \hat \bg  , \hat \bbeta - \bbeta \rangle & =  \langle  ( \blambda \circ \hat \bg)_{ \cT^{{\rm c}}} ,  (\hat \bbeta  - \bbeta )_{ \cT^{{\rm c}}}  \rangle +  \langle  ( \blambda \circ  \hat \bg )_{\cT}, ( \hat \bbeta - \bbeta )_{\cT} \rangle \nn \\
& \geq \| \blambda_{ \cT^{{\rm c}}} \|_{\min} \| ( \hat \bbeta  - \bbeta )_{ \cT^{{\rm c}}} \|_1 - \|\blambda_{\cT} \|_\infty \| (\hat \bbeta - \bbeta )_{\cT} \|_1. \nn
\#
Re-arranging the terms leads to the stated result. \qed

\subsection{Proof of Lemma~\ref{lem:general.error}}

The proof of Lemma~\ref{lem:general.error} is based on Lemma~\ref{lem:cone.property} and a similar localized analysis used in the proof of Theorem~\ref{thm:lasso-qr}. 
Define an intermediate vector $\tilde{\bbeta} = (1-\eta) \bbeta^* + \eta \hat{\bbeta}$, where $\eta = \sup\{ u\in [0,1] :  \bbeta^* + u ( \hat \bbeta - \bbeta^* ) \in \BB_{\bSigma}(r) \}$, and note that $\wt \bbeta \in \bbeta^*+ \BB_{\bSigma}(r)$.
If $\hat{\bbeta} \in  \bbeta^* + \BB_{\bSigma}(r)$, $\wt \bbeta $ coincides with $\hat \bbeta$; otherwise, $\wt \bbeta$ lies on the boundary of $\bbeta^* +\BB_{\bSigma}(r)$ with $\eta$ strictly less than 1.
 
We first show that $\tilde{\bbeta}  \in \bbeta^* +  \CC_{\bSigma}(l(a,k))$.
By a variant of \eqref{pop.grad.diff} and the optimality of $\hat \bbeta$, we have
\#
	 0 \leq   \langle \nabla \hat Q_h(\wt \bbeta)-    \nabla \hat Q_h(\bbeta^*)  ,  \wt \bbeta - \bbeta^*  \rangle    \leq \eta \langle   \nabla \hat Q_h (\hat \bbeta)  -   \nabla \hat Q_h(\bbeta^*)  , \hat{\bbeta}  - \bbeta^*  \rangle    \label{foc.v2}
\#
and $ \nabla \hat Q_h (\hat \bbeta)  + \blambda \circ  \hat \bg = \textbf{0}$ for some $ \hat \bg \in \partial \| \hat \bbeta  \|_1$.
Lemma~\ref{lem:cone.property} ensures that, conditioned on $\{ \lambda \geq  (2/a) \| \bw^*_h \|_\infty\}$, $\hat  \bbeta  -\bbeta^* $ obeys the following cone-type constraint: 
\#
 \|  (\hat  \bbeta  -\bbeta^* )_{\cT^{{\rm c}}} \|_1  &   \leq \frac{  ( \| \blambda \|_\infty +  \| \bw_h^*  \|_\infty ) \| (\hat \bbeta - \bbeta^*)_{\cT} \|_1     +  b_h^*\, \| \hat \bbeta -\bbeta^* \|_{\bSigma} }{\| \blambda_{{\cT}^{{\rm c}}}  \|_{\min} -   \|   \bw_h^* \|_\infty }      \nn\\
&\leq    \Bigg(  1 + \frac{2 \lambda}{ \| \blambda_{  \cT^{\cc} } \|_{\min}}  \Bigg) \| (\hat \bbeta - \bbeta^*)_{\cT} \|_1 +  \frac{2   }{ \| \blambda_{ \cT^{\cc} } \|_{\min}}   b^*_h  \,\| \hat \bbeta - \bbeta^* \|_{\bSigma} \nn \\
& \leq  (1+2/a) \| (\hat \bbeta - \bbeta^*)_{\cT} \|_1   + 2 b_h^*  \| \hat \bbeta - \bbeta^* \|_{\bSigma}  /(a \lambda)  , \nn
\#
where the second and third inequalities follow from the assumed condition on $\blambda$,  i.e., $\lambda \ge \|\blambda\|_{\infty}$ and $\| \blambda_{ \cT^{\cc} } \|_{\min} \ge a\lambda$. Since $\lambda \geq (s/\gamma_p)^{-1/2} b^*_h$, it follows that 
\#
\| \hat  \bbeta  -\bbeta^* \|_1 &\leq   (2+2/a) \| (\hat \bbeta - \bbeta^*)_{\cT} \|_{1}     + 2 b_h^*  \| \hat \bbeta - \bbeta^* \|_{\bSigma}  /(a \lambda)\nn\\
&\leq   (2+2/a)k^{1/2} \| (\hat \bbeta - \bbeta^*)_{ \cT} \|_{2}    + 2 b_h^*  \| \hat \bbeta - \bbeta^* \|_{\bSigma}  /(a \lambda)\nn\\
&\le  \big\{ (2+2/a) (k/\gamma_p)^{1/2} +  (2/a) (s/\gamma_p)^{1/2}  \big\}   \| \hat \bbeta - \bbeta^* \|_{\bSigma}\nn\\
& =: l(a,k) \| \hat \bbeta - \bbeta^* \|_{\bSigma} , \nn
\#
where $\gamma_p = \gamma_p(\bSigma)$ is the minimum eigenvalue of $\bSigma$.
Thus
$
	 \wt \bbeta - \bbeta^* \in  \BB_{\bSigma}(r) \cap  \CC_{\bSigma}(  l(a,k) )  .
$
Conditioned on the event $\cE_{{\rm rsc}}(r,l(a,k),\kappa)$,
\#
	\langle \nabla \hat Q_h(\wt \bbeta)-    \nabla \hat Q_h(\bbeta^*)  ,  \wt \bbeta - \bbeta^*  \rangle    \geq   \kappa \cdot   \|   \wt \bbeta - \bbeta^*  \|_{\bSigma}^2  . \label{FOC.lbd}
\#

Turning to the right-hand side of \eqref{foc.v2}, we have
\#
	\langle   \nabla \hat Q_h (\hat \bbeta)  -   \nabla \hat Q_h(\bbeta^*)  , \hat{\bbeta}  - \bbeta^*  \rangle   =   - \langle  \nabla \hat Q_h(\bbeta^*) , \hat \bbeta - \bbeta^* \rangle   - \langle \blambda \circ \hat \bg , \hat \bbeta   -  \bbeta^* \rangle =:  \Pi_1 + \Pi_2.   \label{Pi12}
\#
It suffices to obtain upper bounds for $\Pi_1$ and $\Pi_2$.
Recall that $ \bw_h^*  =\nabla \hat Q_h(\bbeta^*) - \nabla Q_h(\bbeta^*)$  and $\hat \bdelta = \hat \bbeta - \bbeta^*$. Then, $|\Pi_1|$ can be upper bounded by
\#
 |\Pi_1 |   =   \langle \bw_h^*, \hat \bdelta \rangle + \langle \nabla Q_h(\bbeta^*) , \hat \bdelta \rangle  
    \leq \|   \bw_{ h,  \cT}^* \|_2 \|  \hat \bdelta_{ \cT} \|_2 +  \| \bw^*_{h,  \cT^{{\rm c}}} \|_\infty \| \hat \bdelta_{ \cT^{{\rm c}}} \|_1 + b_h^*  \| \hat \bdelta \|_{\bSigma} .
 \label{Pi1}
\#
For $\Pi_2$,  consider the decomposition 
\#
  \langle \blambda \circ \hat \bg , \hat\bdelta  \rangle    =    \langle  ( \blambda \circ  \hat \bg  )_{ {\cS}} ,  \hat\bdelta_{{\cS}}   \rangle  +   \langle  ( \blambda \circ  \hat \bg  )_{ \cT \setminus \cS } ,   \hat\bdelta_{\cT \setminus \cS }    \rangle   +   \langle  ( \blambda \circ  \hat \bg )_{ {\cT}^{{\rm c}}} ,   \hat\bdelta_{{\cT}^{{\rm c}}}    \rangle    \nn  
\#
and note that 
$$
	   \langle  ( \blambda \circ   \hat \bg  )_{ {\cS}} ,  \hat\bdelta_{{\cS}}   \rangle \geq - \|  \blambda_{\cS} \|_2  \|    \hat\bdelta_{{\cS}}  \|_2 .
$$
Since $\bbeta^*_{ \cS^{\cc}} = \textbf{0}$ and $ \hat \bg  \in \partial \| \hat \bbeta  \|_1$, $ \langle  ( \blambda \circ  \hat \bg )_{ \cT \setminus \cS } ,   ( \hat \bbeta  - \bbeta^*)_{\cT \setminus \cS }    \rangle =  \langle  ( \blambda \circ   \hat \bg )_{ \cT \setminus \cS } ,    \hat \bbeta_{\cT \setminus \cS}   \rangle \geq 0$ and 
$$
	 \langle  ( \blambda \circ   \hat \bg  )_{  {\cT}^{{\rm c}}} ,  (  \hat \bbeta  -\bbeta^* )_{  {\cT}^{{\rm c}}}    \rangle =   	 \langle  ( \blambda \circ   \hat \bg )_{  {\cT}^{{\rm c}}} ,   \hat \bbeta_{  {\cT}^{{\rm c}}}    \rangle = \| ( \blambda \circ   \hat \bbeta  )_{  {\cT}^{{\rm c}}} \|_1 \geq \| \blambda_{   {\cT}^{{\rm c}}} \|_{\min}  \| (\hat \bbeta   - \bbeta^* )_{  {\cT}^{{\rm c}}} \|_1 .
$$
Combining the above equations, we conclude that 
\#
	\Pi_2 \leq   \|  \blambda_{\cS} \|_2  \|    \hat\bdelta_{{\cS}}  \|_2  -  \| \blambda_{  {\cT}^{{\rm c}}} \|_{\min}  \|  \hat\bdelta_{  {\cT}^{{\rm c}}} \|_1  . \label{Pi2:upper}
\#
Substituting \eqref{Pi2:upper} and \eqref{Pi1} into \eqref{Pi12} implies
\#
	 & \langle    \nabla \hat Q_h(\hat \bbeta) - \nabla \hat Q_h(\bbeta^*)  , \hat{\bbeta}  - \bbeta^*  \rangle  \nn \\
& \leq   \|   \bw_{h,  \cT}^* \|_2 \|  \hat \bdelta_{  \cT} \|_2  +    \|  \blambda_{\cS} \|_2  \|  \hat \bdelta_{ \cS} \|_2  + b_h^*  \| \hat \bdelta \|_{\bSigma}     -  \bigl(  \| \blambda_{  {\cT}^{{\rm c}}} \|_{\min}  - \| \bw^*_{h,   \cT^{{\rm c}}} \|_\infty\bigr ) \| \hat \bdelta_{   \cT^{{\rm c}}}  \|_1  . \label{FOC.ubd}
\#

Recall that $\eta \hat \bdelta = \wt \bdelta$.
Provided $\| \blambda \|_\infty \leq \lambda$ and $\| \blambda_{  {\cT}^{{\rm c}}} \|_{\min}  \geq a \lambda \geq  2\| \bw_h^*  \|_\infty$, substituting \eqref{FOC.lbd} and \eqref{FOC.ubd}  into \eqref{foc.v2} yields
\#
  \kappa \,  \| \wt \bdelta \|_{\bSigma}^2 & \leq     \|   \bw_{h,  \cT}^* \|_2 \|  \wt \bdelta_{  \cT} \|_2  +    \|  \blambda_{\cS} \|_2  \|  \wt \bdelta_{ \cS} \|_2  + b_h^*  \| \wt \bdelta \|_{\bSigma}     \nn \\
  & \leq   \gamma_p^{-1/2}  \big( k^{ 1/2 } \|  \bw_h^* \|_\infty + s^{1/2}  \|  \blambda \|_\infty    \big) \|\wt  \bdelta \|_{\bSigma}   +  b_h^*  \|\wt  \bdelta \|_{\bSigma}  \nn \\
  & \leq    \big\{\gamma_p^{-1/2}  \big( k^{1/2} a /2 + s^{1/2}     \big)  \lambda   + b_h^* \big\}  \|\wt  \bdelta \|_{\bSigma}, \nn
\#
from which it follows that $ \| \wt \bdelta \|_{\bSigma}  \leq   \kappa^{-1}  \{\gamma_p^{-1/2}  ( k^{1/2} a /2 + s^{1/2}       )  \lambda   + b_h^*\}$. The constraint~\eqref{RSC.scaling1} ensures that $\wt \bbeta$ falls in the interior of $\bbeta^* + \BB_{\bSigma}(r)$. Therefore, we must have $\eta =1$ and $\hat \bbeta = \wt \bbeta$.
\qed
 
 \subsection{Proof of Lemma~\ref{lem:subgradient:lowd}}
Recall that $\bw_h^* = \nabla \hat Q_h(\bbeta^*) - \nabla Q_h(\bbeta^*)$ is the centered score function evaluated at $\bbeta^*$ and that $\xi_i = \bar K(-\varepsilon_i  /h) -   \tau$. 
Thus,
$
\bw^*_{h,   \cS}= (1/n) \sn \{\xi_i \bx_{i, {\cS}}-  \EE(\xi_i \bx_{i,  {\cS}})\} \in \RR^{s}.
$ 
We first obtain an upper bound for $\|  \Sb^{-1/2}  \bw^*_{h,   \cS}\|_2 = \sup_{\|\bu\|_2=1} \langle \bu,  \Sb^{-1/2} \bw^*_{h,  \cS} \rangle $.  
Using a covering argument, for any $\epsilon\in (0,1)$, there exists an $\epsilon$-net $\cN_\epsilon$ of the unit sphere with cardinality $|\cN_{\epsilon}|\le (1+2/\epsilon)^{s}$ such that 
$\|  \Sb^{-1/2}  \bw^*_{h,  \cS}\|_2 \leq (1-\epsilon)^{-1} \max_{\bu\in \cN_{\epsilon}} \langle \bu, \Sb^{-1/2} \bw^*_{h, \cS} \rangle$.
Thus, it suffices to obtain an upper bound for $\langle \bu,  \Sb^{-1/2} \bw^*_{h,  \cS} \rangle$ for each direction $\bu \in \cN_{\epsilon}$.

For each direction $\bu\in\cN_{\epsilon}$, let 
$
\gamma_{\bu,i} = \langle \bu,  \Sb^{-1/2}  \{\xi_i \bx_{i, {\cS}} -\EE(\xi_i \bx_{i,  {\cS}} )\}\rangle.
$
We employ the Bernstein's inequality to bound $(1/n)\sum_{i=1}^n \gamma_{\bu,i}$.  
Note that $\gamma_{\bu,i}$ has mean zero, and by~\eqref{cond.var.bound}, the variance can be upper bounded as 
\[
\mathrm{var}(\gamma_{\bu,i}) \leq    \{\tau(1-\tau) + C h^2\} \cdot \EE  \langle\bu, \Sb^{-1/2} \bx_{i,\cS}\rangle^2.
\]
Under Condition~(B3), we have $\PP (| \langle \bu, \Sb^{-1/2} \bx_{i,\cS} \rangle|\ge \upsilon_0 t)    \le e^{-t}$ for all $t\geq 0$.
Noting that $|\xi_i| \leq \max(1-\tau, \tau) $, we have for  $k=2,3,\ldots$, 
\#
\EE \bigl( | \langle \bu, \Sb^{-1/2} \bx_{i,\cS} \xi_i \rangle|^k\bigr) &\le \max( 1-\tau, \tau)^{k-2}\, \EE_{\bx}  \bigl\{  | \langle \bu, \Sb^{-1/2} \bx_{i,\cS} \rangle|^k \cdot\EE(\xi_i^2 | \bx_i ) \bigr\}    \nn \\
&\le   \max( 1-\tau, \tau)^{k-2}  \bigl\{ \tau(1-\tau) + C  h^2 \bigr\}  \upsilon^k_0 \int_0^{\infty}\PP (| \langle \bu, \Sb^{-1/2} \bx_{i,\cS} \rangle|\ge \upsilon_0 t) kt^{k-1}\,\mathrm{d}t \nn\\
&\le  \max( 1-\tau, \tau)^{k-2}  \bigl\{ \tau(1-\tau) + C  h^2 \bigr\}    \upsilon^k_0  k \int_0^{\infty} t^{k-1}e^{-t}\, \mathrm{d}t\nn\\
&=   k!   \max( 1-\tau, \tau)^{k-2} \bigl\{ \tau(1-\tau) + C  h^2 \bigr\}     \upsilon^k_0 \nn \\
&\le \frac{k!}{2} \cdot    \bigl\{ \tau(1-\tau) + C  h^2 \bigr\}   \upsilon_0^2     \cdot  \left\{ 2 \max( 1-\tau, \tau) \upsilon_0\right\} ^{k-2}. \label{eq:bern:cond}
\#
Consequently, it follows from  Bernstein's inequality that for every $t\geq 0$,
\#
 \frac{1}{n} \sn \gamma_{\bu,i}  \leq   \upsilon_0 \left[   \sqrt{ \bigl\{ \tau(1-\tau) + C  h^2 \bigr\}  \frac{2t}{n}} + \max(1-\tau, \tau ) \frac{2t}{n} \right]  \nn
\#
with probability at least $1-e^{-t}$.  

Finally, we apply a union bound over all vectors $\bu \in \cN_{\epsilon}$ and obtain 
\#
\|  \Sb^{-1/2}\bw^*_{h,    \cS}\|_2 \leq \frac{\upsilon_0}{1-\epsilon}\left[   \sqrt{ \bigl\{ \tau(1-\tau) + C  h^2 \bigr\}  \frac{2t}{n}} + \max(1-\tau, \tau ) \frac{2t}{n} \right]
\#
with probability at least $1- e^{\log(1+2/\epsilon) s -t}$. 
Selecting $\epsilon = 0.313$  and taking $t=  2s + y$ yield the claimed bound. \qed

\subsection{Proof of Lemma~\ref{lem:general.error2}}

The proof is similar to that of Lemma~\ref{lem:general.error}. We therefore only provide the key steps. As before, construct an intermediate vector $\tilde \bbeta = (1-\eta ) \hat \bbeta^\ora + \eta \hat \bbeta$ satisfying $\tilde{\bbeta}  \in   \hat \bbeta^\ora +  \BB_{\bSigma}(r)$, where $\eta\in (0,1]$ is chosen such that (i) $\eta=1$ if $\hat  \bbeta   \in  \hat \bbeta^\ora +  \BB_{\bSigma}(r)$, and (ii) $\eta\in(0,1)$ if $\hat  \bbeta  \notin  \hat \bbeta^\ora +  \BB_{\bSigma}(r)$. In the latter case, $\wt \bbeta$ lies on the boundary of $\bbeta^* + \BB_{\bSigma}(r)$.

We first show that $\tilde{\bbeta} \in \hat \bbeta^\ora + \CC(l )$ conditioned on the event $\{ \| \bw^{\ora}_h  \|_\infty \leq 0.5 a \lambda \}$, where $l=(2+2/a) (k/\gamma_p)^{1/2}$. By \eqref{pop.foc} and the optimality of $\hat \bbeta$, we have
\#
   \langle \nabla \hat Q_h(\wt \bbeta)-    \nabla \hat Q_h(\hat \bbeta^\ora)  ,  \wt \bbeta - \hat \bbeta^\ora  \rangle    \leq \eta \langle   \nabla \hat Q_h (\hat \bbeta)  -   \nabla \hat Q_h(\hat \bbeta^\ora)  , \hat{\bbeta}  - \hat \bbeta^\ora  \rangle    \label{foc.oracle}
\#
and $\nabla \hat Q_h (\hat \bbeta)  + \blambda \circ \hat \bg = \textbf{0}$ for some $ \hat \bg  \in \partial \| \hat \bbeta  \|_1$.
Following the proof of Lemma~\ref{lem:cone.property} with $\bbeta = \hat \bbeta^\ora$, it can be similarly shown that conditioned on $\{ \| \bw^{\ora}_h  \|_\infty \leq 0.5 a \lambda \}$,
\#
 \|  (\hat  \bbeta  - \hat \bbeta^\ora  )_{  \cT^{{\rm c}}} \|_1  &   \leq \frac{  \{ \| \blambda \|_\infty +  \| \bw^\ora_h \|_\infty \}  \| (\hat \bbeta - \hat \bbeta^\ora )_{\cT} \|_1    }{\| \blambda_{ {\cT}^{{\rm c}}}  \|_{\min} -   \| \bw^\ora_h \|_\infty   }      \nn\\
&\leq    \Bigg(  1 + \frac{2 \lambda}{ \| \blambda_{ \cT^{\cc} } \|_{\min}}  \Bigg) \| (\hat \bbeta -   \hat \bbeta^\ora )_{\cT} \|_1     \leq  (1+2/a) \| (\hat \bbeta -  \hat \bbeta^\ora )_{\cT} \|_1   . \nn
\#
Here there is no bias term because $\bw^\ora_h$ is the score function evaluated at $\hat \bbeta^\ora$ without subtracting the mean. Consequently,  $\| \hat  \bbeta  - \hat \bbeta^\ora \|_1 \leq   (2+2/a) \| (\hat \bbeta - \hat \bbeta^\ora )_{\cT} \|_{1}    \leq   (2+2/a) (k/\gamma_p)^{1/2} \| \hat \bbeta - \hat \bbeta^\ora \|_{\bSigma}$, implying $\wt \bbeta  \in  \hat \bbeta^\ora +   \CC_{\bSigma}( l)$. Furthermore, if the event $\cG_{{\rm rsc}}(r, l, \kappa)$ occurs,
\#
	\langle \nabla \hat Q_h(\wt \bbeta)-    \nabla \hat Q_h(\hat \bbeta^\ora )  ,  \wt \bbeta - \hat \bbeta^\ora  \rangle    \geq   \kappa \,   \|   \wt \bbeta - \hat \bbeta^\ora  \|_{\bSigma}^2  . \label{foc.oracle.lbd}
\#

Let $\hat \bdelta = \hat \bbeta - \hat \bbeta^\ora$ and $\wt \bdelta = \wt \bbeta - \hat \bbeta^\ora = \eta \hat \bdelta$.
For the right-hand side of \eqref{foc.oracle}, by a similar argument to that leads to \eqref{FOC.ubd}, we obtain
\#
	 & \langle    \nabla \hat Q_h(\hat \bbeta) - \nabla \hat Q_h( \hat \bbeta^\ora)  , \hat{\bbeta}  -  \hat \bbeta^\ora  \rangle  \nn \\
& \leq   \|   \bw_{h,  \cT}^\ora \|_2 \| \hat \bdelta_{   \cT} \|_2  +    \|  \blambda_{\cS} \|_2  \| \hat \bdelta_{ \cS} \|_2      -  \bigl(  \| \blambda_{ \cT^{{\rm c}}} \|_{\min}  - \| \bw^\ora_{h, \cT^{{\rm c}}} \|_\infty\bigr ) \|  \hat \bdelta_{ \cT^{{\rm c}}}  \|_1  . \label{foc.oracle.ubd}
\#
Given the stated conditioning, it follows from \eqref{foc.oracle}, \eqref{foc.oracle.lbd} and \eqref{foc.oracle.ubd} that
\#
  \kappa \,  \| \wt \bdelta \|_{\bSigma}^2 & \leq     \|   \bw_{h, \cT}^\ora \|_2 \|  \wt \bdelta_{\cT} \|_2  +    \|  \blambda_{\cS} \|_2  \|  \wt \bdelta_{ \cS} \|_2   \leq   \gamma_p^{-1/2}  \big(  \|   \bw_{h, \cT}^\ora \|_2   +    \|  \blambda_{\cS} \|_2 \big) \|\wt  \bdelta \|_{\bSigma}    . \nn
\#
Canceling out a factor of $\| \wt \bdelta \|_{\bSigma}$ from both sides  yields
\[
 \| \wt \bdelta \|_{\bSigma}  \leq    \kappa^{-1} \gamma_p^{-1/2}  \big( 0.5 a k^{1/2} + s^{1/2}     \big)  \lambda  < r, 
\]
where the second inequality follows from \eqref{RSC.scaling2}.  Consequently, $\wt \bbeta$ falls in the interior of $ \hat \bbeta^\ora + \BB_{\bSigma}(r)$, thus enforcing $\eta =1$ and $\hat \bbeta = \wt \bbeta$.  This proves the claimed bound \eqref{general.l2error2}. \qed

\subsection{Proof of Lemma~\ref{lem:RSC2}}

The proof is based on an argument similar to that in the proof of Lemma~\ref{prop:RSC} and also Proposition~2 in \cite{L2017}.
Since the bandwidth $h$ plays a critical role in subsequent analysis, we provide details of the proof that highlight its connection with the sample size.

For each pair $(\bbeta_1, \bbeta_2)$, write $\bdelta = \bbeta_1 - \bbeta_2$, and similarly to \eqref{def.D} and \eqref{eq:dblower1},
\#
	D(\bbeta_1 , \bbeta_2  ) & :=   \langle \nabla  \hat Q_h( \bbeta_1 ) -  \nabla \hat Q_h( \bbeta_2 ), \bbeta_1 - \bbeta_2 \rangle   \geq \frac{\kappa_l }{n h } \sn (\bx_i^\T \bdelta )^2  \mathbbm{1}_{E_i}  , \label{def:D12}
\#
where with slight abuse of notation, $\mathbbm{1}_{E_i}$ is the indicator function of the event
\#
	E_i   = \bigl\{ |\varepsilon_i | \leq h /2 \bigr\}  \cap   \bigl\{    |\langle  \bx_i, \bbeta_2 - \bbeta^* \rangle |  \leq  h/4 \bigr\} \cap \bigl\{    | \bx_i^\T \bdelta |  \leq  \| \bdelta  \|_{\bSigma} \cdot h / (4 r)    \bigr\}   , \nn
\#
on which $\max\bigl\{ | y_i -   \bx_i^\T \bbeta_1  | , | y_i - \bx_i^\T \bbeta_2 |  \bigr\} \leq h$ for all $\bbeta_1  \in  \bbeta_2 + \BB_{\bSigma}(r )$. In addition to the function $\varphi_R$ introduced in the proof of Lemma~\ref{prop:RSC},  we further define
\#  
  \phi_R(u)  =  \mathbbm{1}(|u|\leq R/2) + \{ 2-  (2/R) \sign(u)\} \mathbbm{1}(R/2\leq |u| \leq R) , \nn
\# 
which is a smoothed version  of the indicator function $u \mapsto \mathbbm{1}(|u| \leq R)$ and satisfies $\mathbbm{1}(|u| \leq R/2) \leq    \phi_R(u)  \leq \mathbbm{1}(|u| \leq R)$.
Consequently,
\#
	D(\bbeta_1 , \bbeta_2  ) & \geq \kappa_l \cdot  \| \bdelta \|_{\bSigma}^2 \cdot  \underbrace{    \frac{1}{nh} \sn  \chi_i   \cdot  \varphi_{h/(4r)}(   \bx_i^\T \bdelta / \| \bdelta \|_{\bSigma}  ) \phi_{ h/4 } (\langle \bx_i, \bbeta_2 -\bbeta^* \rangle )    }_{ = :  D_0(\bbeta_1, \bbeta_2)}  \nn \\
	& = \kappa_l \cdot \| \bdelta \|_{\bSigma}^2 \cdot \bigl\{   \EE  D_0(\bbeta_1, \bbeta_2) +  D_0(\bbeta_1, \bbeta_2) - \EE  D_0(\bbeta_1, \bbeta_2)  \bigr\} ,	 \label{def:D0}
\#
where $ \chi_i = \mathbbm{1}(|\varepsilon_i| \leq h/2)$. Provided $h\leq f_l/(2l_0)$, the earlier result \eqref{indicator.bound} implies
$$
     7 f_l h/ 8 \leq   \EE( \chi_i | \bx_i ) \leq   9  f_u h / 8 ~\mbox{ almost surely}.
$$

To bound the mean $\EE  D_0(\bbeta_1, \bbeta_2)$ from below, applying \eqref{indicator.bound} and inequalities $ \varphi_R(u) \geq  u^2 \mathbbm{1}(|u| \leq R/2) $ and $\phi_{R} (u) \geq \mathbbm{1} (|u|\leq R/2)$ yields
\#
	& \EE \bigl\{  \chi_i  \cdot  \varphi_{h/(4r )}(   \bx^\T  \bdelta/\|  \bdelta   \|_{\bSigma}  ) \phi_{h/4} (\langle \bx , \bbeta_2 -\bbeta^* \rangle )  \bigr\} \nn \\
&\geq  \frac{7}{8} f_l h  \,  \EE \bigl\{  \varphi_{  h / (4r ) }(  \bx^\T \bdelta / \|  \bdelta   \|_{\bSigma}   ) \phi_{ h/4 } (\langle \bx , \bbeta_2 -\bbeta^* \rangle )   \bigr\} \nn \\
& \geq   \frac{7}{8} f_l h  \,   \Bigl(  1  - \EE \bigl\{ \xi_{\bdelta}^2 \mathbbm{1}_{ |\xi_{\bdelta}| > h/(8r) } \bigr\} -  \EE  \bigl\{ \xi_{\bdelta}^2 \mathbbm{1}_{ | \langle \bx, \bbeta_2 - \bbeta^* \rangle | > h/8 } \bigr\}     \Bigr)   , \nn
\#
where $\xi_{\bdelta} =  \bx^\T \bdelta / \|  \bdelta   \|_{\bSigma}$ is such that $\EE \xi^2_{\bdelta} = 1$.
Under Condition~(B3$'$) with $\upsilon_1\geq 1$, for any $\bdelta \in \RR^p$ and  $u >0$ we have
\#
  \EE \bigl\{ \xi_{\bdelta}^2 \mathbbm{1}(|\xi_{\bdelta}| >   u  )  \bigr\}  \leq    2  u^2  e^{- u^2/2 \upsilon_1^2 }  + 4 \upsilon_1^2 \int_{   u/\upsilon_1    }^\infty  t e^{-t^2/2}   \,{\rm d}  t   =  \bigl(   2 u^2 + 4  \upsilon_1^2   \bigr) e^{- u^2/2 \upsilon_1^2 } .\nn
\#
Moreover, for $\bbeta_2 \in \bbeta^* + \BB_{\bSigma}(r/2)$,
\#
 & \EE  \bigl\{ \xi_{\bdelta}^2 \mathbbm{1}_{ | \langle \bx, \bbeta_2 - \bbeta^* \rangle | > h/8 } \bigr\}    \leq   \bigl(  \EE \xi_{\bdelta}^4 \bigr)^{1/2} 
\PP\bigl(  |\langle \bx, \bbeta_2 - \bbeta^* \rangle | > h/8  \bigr)^{1/2}  \leq 4 \sqrt{2} \upsilon_1^2   e^{-  h^2 / (8 \upsilon_1 r )^2 } , \nn
\#
where we have used the fact that $\EE \xi_{\bdelta}^4 \leq 16 \upsilon_1^4$.
From the above three moment inequalities, we find that as long as  $24 \upsilon_1^2 r \leq h$, or equivalently, $h/(8 r) \geq 3 \upsilon_1^2$,
\#
 \EE D_0(\bbeta_1, \bbeta_2) > 0.66 f_l ~\mbox{ for all }~\bbeta_1 \in \bbeta_2 + \BB_{\bSigma}(r) ~\mbox{ and }~ \bbeta_2 \in \bbeta^* + \BB_{\bSigma}(r/2)  . \label{D0.mean.lbd}
\#

To bound $| D_0(\bbeta_1, \bbeta_2) - \EE  D_0(\bbeta_1, \bbeta_2)|$ uniformly over $(\bbeta_1, \bbeta_2) \in  \Lambda(r, l ) = \{ (\bbeta_1, \bbeta_2): \bbeta \in \bbeta_1 + \BB_{\bSigma}(r) \cap \CC_{\bSigma}(l), \bbeta_2 \in \bbeta^* + \BB_{\bSigma}(r/2) , {\rm supp}(\bbeta_2) \subseteq \cS \}$,   define
$$
 \Omega(r,l) = \sup_{(\bbeta_1, \bbeta_2 ) \in  \Lambda(r, l ) }   \big\{  -  D_0(\bbeta_1, \bbeta_2) +  \EE D_0(\bbeta_1, \bbeta_2)  \big\}  .
$$
Write $D_0(\bbeta_1, \bbeta_2)  = (1/n) \sn \omega_{\bbeta_1, \bbeta_2}(\bx_i, \varepsilon_i)$, where 
$$
	\omega_{\bbeta_1, \bbeta_2}(\bx_i, \varepsilon_i) =  (\chi_i/h) \cdot \varphi_{h/(4r)}(  \bx_i^\T  \bdelta/ \| \bdelta\|_{\bSigma} )  \phi_{  h/4 } (\langle \bx_i, \bbeta_2 -\bbeta^* \rangle )   .
$$
Note that $\varphi_R(u)\leq (R/2) |u|$ and $\phi_R(u)\in [0,1]$. Then, for $h\leq f_l/(2l_0)$, 
\#
	  0 \leq  \omega_{\bbeta_1, \bbeta_2}(\bx_i, \varepsilon_i)  \leq (8r)^{-2} \cdot  h   ~~\mbox{ and }~~ \EE \omega^2_{\bbeta_1, \bbeta_2}(\bx_i, \varepsilon_i ) \leq   (8r)^{-2} \cdot  9f_u h / 8    . \nn
\# 
Again, using Bousquet's version of Talagrand's inequality yields that, for any $t>0$,
\#
	 \Omega(r,l) & \leq  \EE   \Omega(r,l)  + \sqrt{  \frac{ 9 f_u/8  + 2  \EE   \Omega(r,l) }{(8r)^2} \frac{2h t}{n} } +    \frac{h}{(8r)^2}  \frac{t}{3n}  \nn \\
	 &    \leq \frac{5}{4}\EE  \Omega(r,l) +  \frac{3}{16}\sqrt{    \frac{f_u h t}{ r^2 n}} +\frac{13}{3} \frac{ h t }{ (8r)^2 n}   \label{Bousquet.concentration2}
\#
holds with probability at least $1 - e^{-t}$. 
To bound $\EE   \Omega(r,l) $, we proceed with a different method to that in the proof of Lemma~\ref{prop:RSC}.
Using symmetrization with Rademacher random variables and by the connection between Gaussian and Rademacher
complexities (see, e.g. Lemma~4.5 in \cite{LT1991}), we obtain
\#
	\EE   \Omega(r,l)   \leq 2 \cdot \sqrt{\frac{\pi}{2}} \cdot   \EE \Biggl\{ \sup_{(\bbeta_1, \bbeta_2 ) \in  \Lambda(r, l) }   \GG_{\bbeta_1, \bbeta_2 }   \Biggr\}   , \label{exp.ubd1}
\#
where $\GG_{\bbeta_1, \bbeta_2 } := (nh)^{-1} \sn  g_i   \chi_i \cdot  \varphi_{ h/(4r)}(   \bx_i^\T \bdelta/\|  \bdelta  \|_{\bSigma} ) \phi_{ h/4} (\langle \bx_i, \bbeta_2 -\bbeta^* \rangle )$ with $\bdelta = \bbeta_1 -\bbeta_2$, and $g_i$ are independent standard normal random variables that are independent of the observations.
Let $\EE^*$ be the conditional expectation given $\{(y_i, \bx_i)\}_{i=1}^n$.
Note that $\{ \GG_{\bbeta_1, \bbeta_2}\}_{(\bbeta_1, \bbeta_2) \in \Lambda(r,l)}$ is a (conditional) Gaussian process and $\GG_{\bbeta^*, \bbeta^*} = 0$. 
We then apply the Gaussian comparison theorem to bound  $ \EE^* \{ \sup_{(\bbeta_1, \bbeta_2 ) \in \Lambda(r, l)}  \GG_{\bbeta_1, \bbeta_2 } \}$, from which an upper bound for $\EE  \{ \sup_{(\bbeta_1, \bbeta_2 ) \in \Lambda(r,l) }  \GG_{\bbeta_1, \bbeta_2 } \}$ follows immediately. 
For $(\bbeta_1, \bbeta_2), (\bbeta_1',\bbeta_2') \in \Lambda(r, l)$, write $\bdelta = \bbeta_1 - \bbeta_2$ and $\bdelta' = \bbeta_1'-\bbeta_2'$.
Consequently,
\#
  \GG_{\bbeta_1, \bbeta_2 } -  \GG_{\bbeta'_1, \bbeta'_2 }  
	& =  \GG_{\bbeta_1, \bbeta_2 } -  \GG_{\bbeta_2'+\bdelta, \bbeta'_2 } +  \GG_{\bbeta_2'+\bdelta , \bbeta'_2 }  - \GG_{\bbeta'_1, \bbeta'_2 }  \nn \\
	 & =   \frac{1}{n h} \sn   g_i  \chi_i \cdot  \varphi_{h/(4r)} ( \bx_i^\T \bdelta / \| \bdelta \|_{\bSigma} )  \bigl\{ \phi_{  h/4 } (\langle \bx_i, \bbeta_2 -\bbeta^* \rangle ) - \phi_{ h/4 } (\langle \bx_i, \bbeta_2' -\bbeta^* \rangle )  \bigr\}   \nn \\
	 & \quad ~+  \frac{1}{n h }  \sn  g_i  \chi_i \cdot  \phi_{  h/4 } (\langle \bx_i, \bbeta_2' -\bbeta^* \rangle )  \bigl \{ \varphi_{ h/(4r) } (  \bx_i^\T \bdelta / \| \bdelta \|_{\bSigma}  ) - \varphi_{ h/(4r) } (  \bx_i^\T  \bdelta' / \| \bdelta' \|_{\bSigma}  ) \bigr\}   . \nn
\#
Note that $\phi_R$ and $\varphi_R$ are, respectively, $(2/R)$- and $R$-Lipschitz continuous, and  $\varphi_R(u)\leq (R/2)^2$. Consequently,
\#
	& 	\EE^*( \GG_{\bbeta_1, \bbeta_2 } -  \GG_{\bbeta_2'+\bdelta, \bbeta'_2 } )^2 \nn \\
&    \leq \frac{1}{n^2} \sn   \frac{h^2}{(8r)^4} \bigg( \frac{8}{h} \bigg)^2   \langle \bx_i, \bbeta_2 - \bbeta_2' \rangle^2 \chi_i =    \bigg(\frac{1}{8 r^2 n } \bigg)^2   \sn     \langle \bx_i, \bbeta_2 - \bbeta_2' \rangle^2 \chi_i  \label{var.ubd1}
\#
and
\#
 \EE^*( \GG_{\bbeta_2'+ \bdelta, \bbeta_2' } -  \GG_{\bbeta_1' , \bbeta'_2 } )^2  
& \leq  \frac{1}{ (nh)^2 } \sn   \bigl\{    \varphi_{ h/(4r)} ( \bx_i^\T  \bdelta / \| \bdelta \|_{\bSigma}  ) - \varphi_{ h/(4r )} (  \bx_i^\T  \bdelta' / \| \bdelta' \|_{\bSigma}  ) \bigr\}^2 \chi_i  \nn \\
& \leq   \bigg(  \frac{1}{4 r n }    \bigg)^2  \sn \bigl(  \bx_i^\T  \bdelta / \| \bdelta \|_{\bSigma}  -  \bx_i^\T  \bdelta' / \| \bdelta' \|_{\bSigma}   \bigr)^2 \chi_i .  \label{var.ubd2}
\# 
Motivated by \eqref{var.ubd1}, \eqref{var.ubd2} and the inequality 
$$\EE^*( \GG_{\bbeta_1, \bbeta_2 } -  \GG_{\bbeta'_1, \bbeta'_2 } )^2 \leq 2\EE^*( \GG_{\bbeta_1, \bbeta_2 } -  \GG_{\bbeta_2'+\bdelta, \bbeta'_2 } )^2 + 2\EE^*( \GG_{\bbeta_2'+ \bdelta, \bbeta_2' } -  \GG_{\bbeta_1' , \bbeta'_2 } )^2, $$ 
we define another Gaussian process $\{ \ZZ_{\bbeta_1, \bbeta_2}\}_{(\bbeta_1, \bbeta_2) \in \Lambda(r,l)}$ as
\#
	\ZZ_{\bbeta_1, \bbeta_2} &  = \frac{\sqrt{2}}{8r^2 n }   \sn g_i' \langle \bx_i, \bbeta_2 - \bbeta^* \rangle \chi_i  + \frac{\sqrt{2}}{4r n }  \sn g_i'' \frac{\langle \bx_i, \bbeta_1 -\bbeta_2 \rangle}{ \| \bbeta_1 - \bbeta_2 \|_{\bSigma}}  \chi_i   \nn \\
	& =  \frac{\sqrt{2}}{8 r^2 n }  \sn \langle  g_i' \bx_{i, \cS}, ( \bbeta_2 - \bbeta^*  )_{ \cS}\rangle  \chi_i  + \frac{\sqrt{2}}{4rn }  \sn g_i'' \frac{\langle \bx_i, \bbeta_1 -\bbeta_2 \rangle}{ \| \bbeta_1 - \bbeta_2 \|_{\bSigma}} \chi_i , \nn
\#
such that $\EE^*(  \GG_{\bbeta_1, \bbeta_2 } -  \GG_{\bbeta'_1, \bbeta'_2 })^2 \leq \EE^*(  \ZZ_{\bbeta_1, \bbeta_2 } -  \ZZ_{\bbeta'_1, \bbeta'_2 })^2$, where $g_1', g_1'', \ldots, g_n', g_n''$ are i.i.d. standard normal random variables that are independent of all  the other variables. Applying Theorem~7.2.11 in \cite{V2018}---Sudakov-Fernique's Gaussian comparison inequality, we obtain
\#
 \EE^*  \Biggl\{ \sup_{(\bbeta_1, \bbeta_2 ) \in \Lambda(r,l) }  \GG_{\bbeta_1, \bbeta_2 } \Biggr\} \leq   \EE^*  \Biggl\{ \sup_{(\bbeta_1, \bbeta_2 ) \in   \Lambda(r,l) }  \ZZ_{\bbeta_1, \bbeta_2 } \Biggr\} ,  \label{Gaussian.sup.1}
\#
which remains valid if $\EE^*$ is replaced by $\EE$. To bound the supremum of $\ZZ_{\bbeta_1, \bbeta_2 }$ over $(\bbeta_1, \bbeta_2 ) \in  \Lambda(r,l)$, using the cone-like
constraint 
$\| \bbeta_1 -\bbeta_2 \|_1 \leq l\| \bbeta_1-\bbeta_2 \|_{\bSigma}$ and $\bbeta_2 \in \bbeta^*+ \BB_{\bSigma}(r/2)$, we deduce that
\#
	  \EE   \Biggl\{ \sup_{(\bbeta_1, \bbeta_2 ) \in \Lambda(r,l) }  \ZZ_{\bbeta_1, \bbeta_2 } \Biggr\}  
	&  \leq \frac{\sqrt{2}    }{16 r} \EE \bigg\| \frac{1}{n} \sn g_i'   \chi_i  \Sb^{-1/2} \bx_{i, \cS}   \bigg\|_2 +  \frac{\sqrt{2} l  }{ 4 r} \EE \bigg\| \frac{1}{n} \sn g''_i \chi_i  \bx_i    \bigg\|_\infty \nn \\
	& \leq    \frac{\sqrt{2}   }{ 16 r}  \sqrt{ \frac{9f_u h}{8}     \frac{   s }{n}} +  \frac{\sqrt{2} l  }{4 r} \EE \bigg\| \frac{1}{n} \sn g''_i  \chi_i \bx_i   \bigg\|_\infty .  \label{Gaussian.sup.2}
\#
This, combined with \eqref{exp.ubd1}, \eqref{Gaussian.sup.1} and \eqref{Gaussian.sup.2}, yields
\#
	\EE \Omega(r,l) \leq  \sqrt{ \pi } \,\Biggl\{ \frac{3}{16} \sqrt{  \frac{ h s }{2 r^2 n}}+  \frac{l}{ 2 r }  \EE  \Biggl( \max_{1\leq j\leq p} \Biggl|   \frac{1}{n} \sn g_i  \chi_i x_{ij}   \Biggr| \Biggr) \Biggr\}.  \label{exp.ubd3}
\#
It remains the bound the second term on the right-hand side of \eqref{exp.ubd3}. Write $S_j =  \sn g_i  \chi_i  x_{ij} $ for $j=1,\ldots, p$.
Under Condition~(B3$'$), for each $1\leq j\leq p$ and $k\geq 3$ we have
\#
	\EE |x_j |^k  \leq   2 \upsilon_1^k \sigma_{jj}^{k/2} k   \int_0^\infty t^{k-1} e^{-t^2/2} \, {\rm d} t = 2^{ k/2} \upsilon_1^k \sigma_{jj}^{k/2}  k \Gamma(k/2). \nn  
\#
Let $g\sim {\cal N}(0,1)$ be independent of $\bx$. By the Legendre duplication formula $\Gamma(k) \Gamma(k+1/2) = 2^{1-2k} \sqrt{\pi} \, \Gamma(2k)$, we have
\#
	\EE |g x_{j}|^k  & \leq  2^{k/2} \frac{\Gamma(\frac{k+1}{2})}{\sqrt{\pi}}   \cdot  2^{k/2} \upsilon_1^k \sigma_{jj}^{k/2} k \Gamma(k/2)
= 2 \upsilon_1^k \sigma_{jj}^{k/2}  k! . \nn
\#
Also recall that $\EE(\chi_i | \bx_i)\leq  c_h:= 9 f_u h/8$.
Hence, for any $0\leq \lambda < (2 \upsilon_1 \sigma_{jj}^{1/2})^{-1}$,
\#
 \EE e^{\lambda g \chi_i  x_j  } & = 1 + \frac{1}{2}  \lambda^2 \EE ( \chi_i x_j )^2 + \sum_{k=3}^\infty \frac{\lambda^k}{k!} \EE ( g  \chi_i x_j  )^k \nn \\
& \leq  1 + \frac{1}{2} c_h \sigma_{jj} \lambda^2 +2  c_h  \sum_{\ell=2}^\infty  \frac{\lambda^{2\ell}}{(2\ell)!} \upsilon_1^{2\ell} \sigma_{jj}^{\ell} (2\ell)! \nn \\
& =   1 + \frac{1}{2} c_h  \sigma_{jj} \lambda^2 +2  c_h  \sum_{\ell=2}^\infty \upsilon_1^{2\ell} \sigma_{jj}^{\ell}   \lambda^{2\ell}    \nn \\
& \leq 1 + \frac{1}{2}  \upsilon_1^2 c_h \sigma_{jj} \sum_{k=2}^\infty \lambda^k (2 \upsilon_1   \sigma_{jj}^{1/2})^{k-2} \nn \\
& \leq 1 +  \frac{1}{2}\frac{\upsilon_1^2 c_h  \sigma_{jj} \lambda^2}{1- 2 \upsilon_1 \sigma_{jj}^{1/2}  \lambda} . \nn
\#
It then follows that $\log \EE e^{\lambda S_j} \leq \frac{1}{2} \frac{\lambda^2 \cdot  \upsilon_1^2 c_h\sigma_{jj} n }{1 - 2\upsilon_1 \sigma_{jj}^{1/2} \lambda}$ for any $\lambda \in (0,  (2 \upsilon_1 \sigma_{jj}^{1/2})^{-1} )$. By symmetry, the same bound applies to $-S_j$. Applying Corollary~2.6 in \cite{BLM2013}, we obtain
\#
  \EE  \Biggl( \max_{1\leq j\leq p} \Biggl|   \frac{1}{n} \sn g_i \chi_i x_{ij}   \Biggr| \Biggr) =	\EE \max_{1\leq j\leq p} |S_j/n| 
  \leq   \upsilon_1 \sigma_{\bx}  \Biggl\{  \frac{3}{2}  \sqrt{\frac{ f_u h  \log(2p)}{n}} + \frac{2 \log(2p)}{n} \Biggr\}.   \label{exp.ubd4}
\#

Finally, take $r= h/(24\upsilon_1^2)$.
Combining \eqref{exp.ubd3}, \eqref{exp.ubd4} with the concentration bound  \eqref{Bousquet.concentration2}, we conclude that with probability at least $1- p^{-1}$, $\Omega(r,l)\leq 0.16 f_l$ as long as $n h \gtrsim    f_u f_l^{-2}   \max\{   s,  l^2 \log(p) \}$. This, together with \eqref{def:D12}, \eqref{def:D0} and \eqref{D0.mean.lbd}, proves the claim. \qed

\subsection{Proof of Lemma~\ref{lem:oracle.score}}

Let $\bw_h(\bbeta) = \nabla  \hat Q_h(\bbeta) -  \nabla   Q_h(\bbeta)$ be the centered score function as in \eqref{def:wb}. 
From the decomposition $\nabla  \hat Q_h(\hat \bbeta^{{\rm ora}})= \bw_h(\hat \bbeta^{{\rm ora}}) - \bw_h(\bbeta^*) +   \nabla  Q_h(\hat \bbeta^{{\rm ora}})  + \bw_h(\bbeta^*)$, we have
\#
 \|  \bw_h^\ora \|_\infty \leq  \|  \bw_h(\hat \bbeta^{{\rm ora}}) - \bw_h(\bbeta^*) \|_\infty +   \|  \nabla Q_h(\hat \bbeta^\ora) \|_\infty + \| \bw_h (\bbeta^*) \|_\infty.  \nn
\#
For $\| \bw_h (\bbeta^*) \|_\infty = \| \bw^*_h \|_\infty$, applying Lemma~\ref{lem:subgradient} to $\|  \bw^*_{h,\cS} \|_\infty$ and $\| \bw^*_{h, \cS^{\cc}}\|_\infty$ separately, we obtain that the following bounds
\#
\|  \bw^*_{h,\cS} \|_\infty \lesssim  \sigma_{\bx} \sqrt{\frac{\log(2s) + t}{n}} ~~\mbox{ and }~~
\|  \bw^*_{h,\cS^{\cc}} \|_\infty \lesssim  \sigma_{\bx} \sqrt{\frac{\log(2p) }{n}}     \label{score.diff.bound0}
\#
hold with probability at least $1-   e^{-t}$ and $1-(2p)^{-1}$, respectively, provided $n\gtrsim \log(2p) + t$.

In the following, we control the other two terms $  \|  \bw_h(\hat \bbeta^{{\rm ora}}) - \bw_h(\bbeta^*) \|_\infty$ and $ \|  \nabla Q_h(\hat \bbeta^\ora) \|_\infty$, separately, via empirical process arguments. The main difficulty is that the oracle $\hat \bbeta^{{\rm ora}}$ is also random and does not have a closed-form expression like the least squares estimator.

\medskip
\noindent
{\sc Step 1}: Bounding $\|  \bw_h(\hat \bbeta^{{\rm ora}}) - \bw_h(\bbeta^*) \|_\infty$. Define the oracle local neighborhood $\Theta^*_{\cS}(r) = \{ \bbeta \in \bbeta^* + \BB_{\bSigma}(r):  \bbeta_{ \cS^\cc} = \textbf{0} \}$.  Conditioned on the event $\{  \hat \bbeta^{{\rm ora}} \in \bbeta^* + \BB_{\bSigma}(r)\}$,  
\#
 \|  \bw_h(\hat \bbeta^{{\rm ora}}) - \bw_h(\bbeta^*)  \|_\infty \leq \sup_{\bbeta \in \Theta^*_{\cS}(r)} \|  \bw_h( \bbeta ) - \bw_h(\bbeta^*)  \|_\infty . 
 \label{local.fluct.max}
\#
For  $j \in [p]$,  let $\eb_j \in \RR^p$ be the canonical basis vectors in $\RR^p$.  For every $\bbeta \in \Theta^*_{\cS}(r)$, write $\bdelta = (\bbeta - \bbeta^*)_{\cS} \in \RR^s$, and note that $\| \bbeta - \bbeta^* \|_{\bSigma} = \| \bdelta \|_{\Sb}$, where $\Sb = \bSigma_{\cS \cS}$. Hence,
\#
   \sup_{\bbeta \in \Theta^*_{\cS}(r)} \|  \bw_h( \bbeta ) - \bw_h(\bbeta^*)  \|_\infty \leq  \sigma_{\bx}  \max_{1\leq j\leq p } \sup_{\| \bdelta \|_{\Sb}\leq r} | W_j(\bdelta) | ,\label{local.fluct.dec}
\#
where $W_{j} (\bdelta)= (1/n)\sn(  w_{ij}- \EE w_{ij} )$,  $w_{ij}    = \bar{x}_{ij}  \{  \bar K (  \langle  \bx_{i,  \cS } ,  \bdelta \rangle - \varepsilon_i  )/h) -  \bar K (  - \varepsilon_i  /h )  \}$ and $\bar x_{ij} = x_{ij} / \sigma_{jj}^{1/2}$.

 We will apply a concentration result for empirical processes in \cite{S2012} to bound the local fluctuation $\sup_{ \| \bdelta \|_{\Sb} \leq r} W_{j} (\bdelta)$. To this end, we need to control the exponential moments of $W_{j}(\bdelta)$. Note that
\#
\EE \Biggl\{ \bar K\biggl( \frac{   \langle   \bx_{i,  \cS },  \bdelta \rangle   - \varepsilon_i  }{h} \biggr) \,\bigg| \bx \Biggr\} &  = \int_{-\infty}^\infty   \bar K((   \langle  \bx_{i,  \cS } ,  \bdelta \rangle - t)/h) f_{\varepsilon_i | \bx_i } (t) \, {\rm d} t \nn  \\
& = h  \int_{-\infty}^\infty \bar K(u) f_{\varepsilon_i | \bx_i } (    \langle  \bx_{i,  \cS } ,  \bdelta \rangle  - uh ) \, {\rm d} u \nn \\
& = \int_{-\infty}^\infty F_{\varepsilon_i | \bx_i} (   \langle  \bx_{i,  \cS } ,  \bdelta \rangle - uh ) K(u) \, {\rm d} u . \nn
\# 
Similarly, $\EE \{ \bar K(-\varepsilon_i/h)  | \bx\} =\int_{-\infty}^\infty F_{\varepsilon_i | \bx_i} ( - uh ) K(u) \, {\rm d} u  $.  Under Conditions~(B1$'$) and (B2$'$), we have $|w_{ij}| \leq  h^{-1} | \bar{x}_{ij}   \langle  \bx_{i,  \cS } ,  \bdelta \rangle |$ and $| \EE (w_{ij}) | \leq  f_u  \EE | \bar{x}_{ij}  \langle  \bx_{i,  \cS } ,  \bdelta \rangle | \leq  f_u \| \bdelta \|_{\Sb}$. Moreover, by Minkowski's integral inequality, it can be shown that
\$
 \EE(w_{ij}^2 | \bx_i )  \leq  f_u  h^{-1} \bar x_{ij}^2   \langle  \bx_{i,  \cS } ,  \bdelta \rangle^2 .
\$
The above bounds together imply 
$$
\EE \bigl\{     (w_{ij} - \EE w_{ij})^2 | \bx_i \bigr\} \leq 2 \bigl( f_u^2  \| \bdelta \|_{\Sb}^2 + f_u h^{-1}   \bar{x}_{ij}^2  \langle  \bx_{i,  \cS } ,  \bdelta \rangle^2  \bigr). 
$$
For  every $\lambda \in \RR$ and $\bdelta \in \RR^s$, write $\lambda_* = \lambda /    \| \bdelta \|_{\Sb} $ and $\bdelta_* = \bdelta / \| \bdelta \|_{\Sb}$. Then, by the elementary inequality  $|e^u-1-u| \leq (u^2/2) e^{u\vee0}$, we obtain
\#
&  \EE e^{ \lambda  W_{j} (\bdelta) /    \| \bdelta \|_{\Sb}   } =  \prod_{i=1}^n \EE e^{ \frac{\lambda_*  }{n } (w_{ij} - \EE w_{ij}) }  \nn \\
& \leq \prod_{i=1}^n  \EE \Biggl\{ 1 + \frac{\lambda_*^2 }{  2 n^2 }  (w_{ij} - \EE w_{ij})^2  e^{ \frac{|\lambda_*  |}{  n} |w_{ij} - \EE w_{ij} |   } \Biggl\} \nn \\
& \leq   \prod_{i=1}^n  \Biggl\{ 1 + \frac{\lambda_*^2   }{  2 n^2 } e^{   \frac{|\lambda | f_u }{n}}\EE   (w_{ij} - \EE w_{ij})^2  e^{  \frac{  |\lambda  |}{  n h } | \bar{x}_{ij} \langle \bx_{i ,   \cS} , \bdelta_* \rangle  |   } \Biggl\} \nn \\
& \leq  \prod_{i=1}^n  \Biggl\{ 1 + \frac{\lambda^{ 2}  f_u^2  }{  n^2 } e^{     \frac{|\lambda |f_u }{n} }    \EE e^{ \frac{  |\lambda |}{   nh } |  \bar{x}_{ij}  \langle \bx_{i , \cS} , \bdelta_* \rangle |   }   +   \frac{\lambda^{ 2}  f_u   }{   n^2 h }   e^{    \frac{|\lambda | f_u  }{n} }    \EE  \bar{x}_{ij}^2 \langle \bx_{i , \cS} , \bdelta_* \rangle^2 e^{ \frac{  |\lambda |}{   n h } | \bar{x}_{ij}  \langle \bx_{i ,   \cS} , \bdelta_* \rangle |   } \Biggl\} .  \label{mgf.ubd1}
\#
Applying H\"older's inequality to the exponential moments on the right-hand side of \eqref{mgf.ubd1} yields that, for any $t>0$,
\#
 & \EE \bar{x}_{ij}^2 \langle \bx_{i ,    \cS} , \bdelta_*   \rangle^2 e^{ t |  \bar{x}_{ij}  \langle \bx_{i ,  \cS} , \bdelta_*  \rangle |   }   \leq    \big(  \EE \bar{x}_{ij}^4 e^{ t  \bar{x}_{ij}^2 } \big)^{1/2}   \cdot  \big( \EE \langle \bx_{i , \cS} , \bdelta_*   \rangle^4 e^{t  \langle \bx_{i ,   \cS} , \bdelta_* \rangle^2  }  \big)^{1/2} \nn
\#
and 
\#
 \EE e^{ t  | \bar{x}_{ij} \langle \bx_{i  , \cS} , \bdelta_* \rangle |   }  \leq \big(  \EE  e^{ t \bar{x}_{ij}^2 } \big)^{1/2} \cdot  \big( \EE e^{ t  \langle \bx_{i ,  \cS} , \bdelta_*  \rangle^2  } \big)^{1/2} . \nn
\#
For any unit vector $\bu \in \mathbb{S}^{p-1}$, let $Z_{\bu} =( \bz^\T \bu)^2/(4\upsilon_1^2 )$, where $\bz = \bSigma^{-1/2}\bx$. By Condition~(B3$'$), $\PP(Z_{\bu} \geq u) \leq 2 e^{-2u}$ for any $u\geq 0$. It can be shown that
$$
 \EE e^{Z_{\bu}}  = 1 +   \int_0^\infty e^u \PP(Z_{\bu} \geq u) {\rm d}u  \leq   3 ~~\mbox{ and }~~ \EE Z_{\bu}^2 e^{Z_{\bu}} = \int_0^\infty (u^2 +2u) e^u \PP(Z_{\bu} \geq u) {\rm d}u \leq 8.
$$
Substituting the above moment bounds into \eqref{mgf.ubd1}, we find that for any $|\lambda | \leq  \min  \{   n h /(4   \upsilon_1^2 ), n / f_u  \} $,
\#
   \EE e^{ \lambda   W_{j} (\bdelta) /   \| \bdelta \|_{\Sb}  }   \leq  \prod_{i=1}^n   \big\{  1 +  C  \upsilon_1^4 f_u  / (   n^2 h)  \big\}  \leq e^{ C \upsilon_1^4 f_u  / (   n h) } , \nn
\#
where $C>0$ is an absolute constant.
Similarly, it can be derived  that  for each pair $(\bdelta, \bdelta')$,
\#
  \EE e^{ \lambda    \{ W_{j} (\bdelta) - W_{j} (\bdelta')  \} /   \|  \bdelta  - \bdelta' \|_{\Sb}  }    \leq e^{C \upsilon_1^4 f_u  / (   n h)   }  ~\mbox{ for all }~ |\lambda | \leq  \min\bigl\{   n h /(4  \upsilon_1^2 ), n/ f_u\bigr\}    .\nn
\#
The above inequality certifies condition ($\mathcal{E}d$) in \cite{S2012} (see Section~2 in the supplement), so that Corollary~2.2 therein applies to the process $\{ W_{j} (\bdelta): \| \bdelta \|_{\Sb} \leq r \}$: with probability at least $1- e^{-u}$,
\#
 \sup_{\bbeta \in \Theta^*_{\cS}(r)}	\langle  \bw_h(\bbeta)-  \bw_h(\bbeta^*)  , \eb_j \rangle = \sup_{ \| \bdelta \|_{\Sb} \leq r} W_{j} (\bdelta)  \lesssim  \upsilon_1^2  f_u^{1/2}  \sigma_{\bx}    r \,  \sqrt{\frac{ s +  u}{n h}} , \nn
\#
provided $n h \gtrsim   (s+u)^{1/2}$. The same bound applies to $\sup_{ \| \bdelta \|_{\Sb} \leq r}  -W_{j} (\bdelta)$ by a similar argument. Taking  $u=2\log(2p)$ in  \eqref{local.fluct.dec}, and using the union bound, we obtain
\#
\sup_{\bbeta \in \Theta^*_{\cS}(r) }   \|  \bw_h(\bbeta)-  \bw_h(\bbeta^*)  \|_\infty  \lesssim  \sigma_{\bx} r \, \sqrt{\frac{\ s+ \log p }{ n h }}  \label{score.diff.bound1}
\#
with probability at least $1- (2p)^{-1}$ provided $n  h \gtrsim  ( s + \log p)^{1/2}$.

\medskip
\noindent
{\sc Step 2}: Bounding $\| \nabla Q_h(\hat \bbeta^\ora) \|_\infty$.  As before, we write $\bdelta = ( \bbeta -\bbeta^*)_{ \cS}$ for $\bbeta \in \Theta^*_{\cS}(r)$.   
In face, since the oracle score is such that $\bw^{\ora}_{h, \cS} = \textbf{0}$, it suffices to bound $\| \nabla Q_h(\hat \bbeta^\ora)_{\cS^{\cc}} \|_\infty$ instead. Similarly to \eqref{score.mean.ubd}, we have $ \| \nabla Q_h( \bbeta^* ) \|_\infty   \leq  0.5 l_0 \kappa_2 \sigma_{\bx}   h^2$. For any $\bbeta \in \Theta^*_{\cS}(r)$, note that
\#
  & \nabla Q_h( \bbeta )_{\cS^{\cc}} - \nabla Q_h( \bbeta^* )_{\cS^{\cc}}   =  \EE     \int_{-\infty}^\infty K(u) \bigl\{   F_{\varepsilon | \bx} (  \bx_{ \cS}^\T  \bdelta -  uh ) - F_{\varepsilon | \bx}(-uh) \bigr\}  \, {\rm d} u \cdot   \bx_{{\cS^{\cc}}} . \nn 
\# 
Using the Taylor series expansion twice, we get
\#
 F_{\varepsilon | \bx}( \bx_{  \cS}^\T  \bdelta -uh) -  F_{\varepsilon | \bx}(-uh)  =  f_{\varepsilon| \bx} (0 )  \cdot   \bx_{ \cS}^\T  \bdelta  + \int_{0}^{  \bx_{ \cS}^\T  \bdelta } \bigl\{  f_{\varepsilon | \bx}(t - hu) -  f_{\varepsilon | \bx}(0) \bigr\} \,  {\rm d} t .\nn
\#
Together, the last two displays imply
\#
 &  \bigl\|  \nabla Q_h( \bbeta )_{\cS^{\cc}} - \nabla Q_h( \bbeta^* )_{\cS^{\cc}}  -  \Jb_{\cS^{\cc} \cS}  \bdelta  \bigr\|_\infty \nn \\
 & \leq  0.5  l_0 \max_{j \in \cS^{\cc} } \EE |x_j |   (\bx_{\cS}^\T  \bdelta)^2 +   \kappa_1 h \max_{ j \in \cS^{\cc}} \EE |x_j \bx_{\cS}^\T  \bdelta |     \nn \\ 
 & \leq  0.5  l_0  \mu_4^{1/2} \sigma_{\bx} \| \bdelta \|_{\Sb}^2 +  l_0 \kappa_1  \sigma_{\bx}  h  \| \bdelta \|_{\Sb} ,  \nn
 \#
  where $ \Jb_{ \cS^{\cc}  \cS}  = \EE \{ f_{\varepsilon|\bx}(0) \bx_{\cS^{\cc}} \bx_{\cS} \} \in \RR^{(p-s)\times s}$. Putting together the pieces, we have shown that conditioned on $\{ \hat \bbeta^{\ora} \in \bbeta^* +\BB_{\bSigma}(r) \}$,
\#
\bigl\|  \nabla Q_h( \hat \bbeta^{{\rm ora}})_{\cS^{\cc}}  -  \Jb_{ \cS^{\cc} \cS}   (\hat \bbeta^{{\rm ora}}  -\bbeta^*)_{\cS} \bigr\|_\infty \leq  0.5  l_0 \sigma_{\bx}   \bigl(  \mu_4^{1/2}  r^2 +   2 \kappa_1    h r +   \kappa_2    h^2 \bigr) .  \label{score.diff.bound2}
\#
It remains to control  $ \| \Jb_{ \cS^{\cc} \cS}   (\hat \bbeta^{{\rm ora}}  -\bbeta^*)_{ \cS}  \|_\infty$, which is closely related to the $\ell_\infty$-error of the oracle estimator.
By \eqref{irrepresentable.cond},
\#
 & \bigl\|   \Jb_{ \cS^{\cc} \cS}   (\hat \bbeta^{{\rm ora}}  -\bbeta^*)_{  \cS}   \bigr\|_\infty 
  =  \bigl\|    \Jb_{ \cS^{\cc} \cS}    ( \Jb_{ \cS  \cS} )^{-1} \Jb_{ \cS  \cS}  (\hat \bbeta^{{\rm ora}}  -\bbeta^*)_{  \cS}   \bigr\|_\infty \nn \\
  & \leq \max_{j\in \cS^{\cc}}  \|   \Jb_{j \cS}( \Jb_{ \cS  \cS} )^{-1}   \|_1 \cdot  \| \Jb_{ \cS  \cS}   (\hat \bbeta^{{\rm ora}}  -\bbeta^*)_{  \cS}   \|_\infty  \leq A_0 \cdot  \| \Jb_{ \cS  \cS}   (\hat \bbeta^{{\rm ora}}  -\bbeta^*)_{  \cS}   \|_\infty .  \label{score.diff.bound3}
\# 
Next we derive a sharper bound for the oracle error $(\hat \bbeta^{{\rm ora}}  -\bbeta^*)_{  \cS}$ under $\ell_\infty$-norm, instead of using the trivial bound $\| \Jb_{ \cS  \cS}   (\hat \bbeta^{{\rm ora}}  -\bbeta^*)_{  \cS}   \|_\infty \leq \| \Jb_{ \cS  \cS}   (\hat \bbeta^{{\rm ora}}  -\bbeta^*)_{  \cS}   \|_2$. By Proposition~\ref{thm:oracle.smoothqr}, 
\$
 \| (\hat \bbeta^{\ora} - \bbeta^* )_{\cS} \|_{\Sb} \lesssim f_l^{-1} \Bigg(  \sqrt{\frac{s+t}{n}} + h^2 \Bigg)
\$ 
and
\$
 \Bigg\|  \Db  (\hat \bbeta^{\ora} - \bbeta^* )_{\cS}  +   \underbrace{ \frac{1}{n} \sn \big\{ \bar K(-\varepsilon_i/h) - \tau \} \bx_{i,\cS}   }_{= \nabla \hat Q_h(\bbeta^*)_{\cS} } \Bigg\|_{\Sb^{-1}} \lesssim \frac{s+t}{h^{1/2} n} + h \sqrt{\frac{s+t}{n}} + h^3
\$
hold with probability at least $1-3e^{-t}$, where $\Db = \Jb_{\cS \cS}$. The latter, combined with an earlier bound in \eqref{score.diff.bound0}, implies
\#
	 & \| \Db (\hat \bbeta^{\ora} - \bbeta^* )_{\cS} \|_\infty  \nn \\
	 & \leq   \| \Db (\hat \bbeta^{\ora} - \bbeta^* )_{\cS}  +  \nabla \hat Q_h(\bbeta^*)_{\cS} \|_\infty + \|   \nabla \hat Q_h(\bbeta^*)_{\cS}  \|_\infty \nn \\
	 & \leq   \| \Db (\hat \bbeta^{\ora} - \bbeta^* )_{\cS}  +  \nabla \hat Q_h(\bbeta^*)_{\cS} \|_2 +  \| \bw^*_{h, \cS} \|_\infty +  \|  \nabla Q_h(\bbeta^*)_{\cS} \|_{\infty} \nn \\
	 & \leq \gamma_1(\Sb)^{1/2} \cdot \| \Db (\hat \bbeta^{\ora} - \bbeta^* )_{\cS}  +  \nabla \hat Q_h(\bbeta^*)_{\cS} \|_{\Sb^{-1} } +   \| \bw^*_{h, \cS} \|_\infty +  \|  \nabla Q_h(\bbeta^*)_{\cS} \|_{\infty} \nn \\
	 & \lesssim \sqrt{\frac{\log(s) +t}{n}} + \frac{s+t}{h^{1/2} n} + h \sqrt{\frac{s+t}{n}} + h^2. \label{max.error.ubd1}
\#

Combining the bounds \eqref{score.diff.bound0}, \eqref{local.fluct.max}, \eqref{score.diff.bound1},  \eqref{score.diff.bound2}, \eqref{score.diff.bound3} and \eqref{max.error.ubd1}, we find that, with probability at least $1-p^{-1} - 4e^{-t}$,
\$
 \|  \nabla Q_h( \hat \bbeta^{{\rm ora}}) \|_\infty \lesssim \sqrt{\frac{\log(2 p)}{n} } +    A_0\Bigg\{   \sqrt{\frac{\log(s) +t}{n}} + \frac{s+t}{h^{1/2} n}  + h^2 \Bigg\}
\$
provided that $\sqrt{(s \vee \log p  +t )/n} \lesssim h \lesssim 1$. This proves \eqref{oracle.score.bound}.

Note that the $\ell_\infty$-error bound \eqref{max.error.ubd1} does not imply the desired bound on $\| (\hat \bbeta^{\ora} - \bbeta^* )_{\cS} \|_\infty $ directly. Using the same arguments, we obtain
\#
& \| (\hat \bbeta^{\ora} - \bbeta^* )_{\cS} \|_\infty  \nn \\
& \leq   \|  (\hat \bbeta^{\ora} - \bbeta^* )_{\cS}  + \Db^{-1}  \nabla \hat Q_h(\bbeta^*)_{\cS} \|_\infty + \|  \Db^{-1}   \nabla \hat Q_h(\bbeta^*)_{\cS}  \|_\infty \nn \\
& \leq   \|  (\hat \bbeta^{\ora} - \bbeta^* )_{\cS}  + \Db^{-1}  \nabla \hat Q_h(\bbeta^*)_{\cS} \|_2 + \|  \Db^{-1}    \bw^*_{h, \cS} \|_\infty +  \|  \Db^{-1}   \nabla Q_h(\bbeta^*)_{\cS} \|_\infty \nn \\
& \leq f_l^{-1} \gamma_s(\Sb)^{-1/2}  \cdot \| \Db (\hat \bbeta^{\ora} - \bbeta^* )_{\cS}  +  \nabla \hat Q_h(\bbeta^*)_{\cS} \|_{\Sb^{-1} } \nn \\
&~~~~~~~~~~~~~~ +    \|  \Db^{-1}      \bw^*_{h, \cS} \|_\infty + f_l^{-1}  \gamma_s(\Sb)^{-1/2}  \cdot  \| \Sb^{-1/2} \nabla Q_h(\bbeta^* )_{\Sb} \|_2 , \nn
\#
where we have used the fact $f_l \cdot \gamma_s(\Sb) \leq    \gamma_s(\Db) \leq \gamma_1(\Db) \leq f_u \cdot  \gamma_1(\Sb)$.
By \eqref{score.mean.ubd},  $\| \Sb^{-1/2} \nabla Q_h(\bbeta^* )_{\Sb} \|_2 \leq 0.5 l_0 \kappa_2 h^2$. For  $ \|  \Db^{-1}    \bw^*_{h, \cS} \|_\infty$, note that
$$
  \|  \Db^{-1}      \bw^*_{h, \cS} \|_\infty = \max_{1\leq j\leq s} \Bigg| \frac{1}{n} \sn (1-\EE) \{ \bar K(-\varepsilon_i/h) - \tau \} \langle  \Sb^{-1/2}\bx_{i, \cS},  \Sb^{1/2}  \Db^{-1}   \eb_j \rangle   \Bigg| ,
$$
where $\eb_j$'s are canonical basis vectors in $\RR^s$, and satisfy $\| \Sb^{ 1/2}  \Db^{-1}   \eb_j\|_2   \leq f_l^{-1}\gamma_s(\Sb)^{-1/2}$. Following the proof of Lemma~\ref{lem:subgradient}, it can be similarly shown that, with probability at least $1-e^{-t}$,
\#
  \| \Db^{-1}     \bw^*_{h, \cS} \|_\infty \lesssim  \sqrt{\frac{\log(s) + t}{n}}. \nn
\#
Putting together the pieces yields the stated result \eqref{oracle.infty.bound}.  \qed

\subsection{Proof of Lemma~\ref{lem:new.oracle.score}}

The proof parallels that of Lemma~\ref{lem:oracle.score}, and therefore we only provide an outline of the proof.
Recall that $\bw_h^\ora =  \nabla  \hat Q_h(\hat \bbeta^{{\rm ora}})$, $\bw_h(\bbeta) = \nabla  \hat Q_h(\bbeta) -  \nabla  Q_h(\bbeta)$ and $\nabla Q_h(\bbeta^*_h)=\textbf{0}$, where $Q_h(\bbeta) = \EE \hat Q_h(\bbeta)$. From the decomposition $\nabla  \hat Q_h(\hat \bbeta^{{\rm ora}})= \bw_h(\hat \bbeta^{{\rm ora}}) - \bw_h(\bbeta^*_h) +   \nabla  Q_h(\hat \bbeta^{{\rm ora}})  +  \bw_h(\bbeta^*_h) $ we see that
\#
 \|  \bw_h^\ora \|_\infty \leq  \|  \bw_h(\hat \bbeta^{{\rm ora}}) - \bw_h(\bbeta_h^*) \|_\infty +   \|  \nabla Q_h(\hat \bbeta^\ora) \|_\infty + \|  \bw_h(\bbeta^*_h)  \|_\infty.  \nn
\#
In fact, since $\nabla  \hat Q_h(\hat \bbeta^{{\rm ora}})_{\cS} = \textbf{0}$, essentially we only need to control $\| ( \bw_h^\ora)_{\cS^{\cc}} \|_\infty$.

To bound $ \|  \bw_h(\bbeta^*_h)  \|_\infty$,  treating $\bw_h(\bbeta^*_h)_{ \cS}\in \RR^s$ and $\bw_h(\bbeta^*_h)_{  \cS^{\cc}} \in \RR^{p-s}$ separately, it can be shown that as long as $n\gtrsim \log(2p)+t$,
\#
 \| \bw_h(\bbeta^*_h)_{ \cS} \|_\infty \lesssim \sigma_{\bx}\sqrt{\frac{\log(2s)+t}{n}} ~~\mbox{ and }~~  \|  \bw_h(\bbeta^*_h)_{ \cS^{\cc}}  \|_\infty \lesssim \sigma_{\bx} \sqrt{\frac{\log(2p)}{n}}    \label{new.score.diff.bound0}
\#
hold with probability at least $1-e^{-t}$ and $1-(2p)^{-1}$, respectively.
Turning to $\|  \bw_h(\hat \bbeta^{{\rm ora}}) - \bw_h(\bbeta_h^*) \|_\infty$, following the proof of \eqref{score.diff.bound1} it can be similarly shown that with probability at least $1- (2p)^{-1}$,
\#
\sup_{\bbeta \in \Theta^*_{h,\cS}(r) }   \|  \bw_h(\bbeta)-  \bw_h(\bbeta^*)  \|_\infty  \lesssim  \sigma_{\bx}  r \, \sqrt{\frac{s + \log(p) }{n h }}  \label{new.score.diff.bound1}
\#
provided $n h \gtrsim (s+\log p)^{1/2}$, where $\Theta^*_{h,\cS}(r) = \{ \bbeta \in \bbeta_h^* + \BB_{\bSigma}(r):  \bbeta_{  \cS^\cc} = \textbf{0} \}$.

It remains to bound $\| \nabla Q_h(\hat \bbeta^\ora)_{\cS^{\cc}}\|_\infty$.  Write $\bdelta = ( \bbeta -\bbeta_h^*)_{\cS}$ for $\bbeta \in \Theta^*_{h,\cS}(r)$, and note that
\#
  & \nabla Q_h( \bbeta ) - \underbrace{  \nabla Q_h( \bbeta_h^* ) }_{= \textbf{0}}   =  \EE     \int_{-\infty}^\infty K(u) \bigl\{   F_{\varepsilon } ( \bx_{  \cS}^\T \bdelta  + b_h -  uh ) - F_{\varepsilon }(b_h -uh) \bigr\}  \, {\rm d} u \cdot   \bx . \nn 
\# 
A Taylor expansion with integral remainder leads to
\#
& F_{\varepsilon } (  \bx_{  \cS}^\T \bdelta  + b_h -  uh )  \nn \\
& =  F_{\varepsilon }(b_h -  uh) + f_{\varepsilon}(b_h -  uh)  \bx_{ \cS}^\T \bdelta     + \int_{0}^{ \bx_{ \cS}^\T \bdelta }  \bigl\{  f_\varepsilon(t+b_h-uh) - f_\varepsilon(b_h-uh) \bigr\}  \, {\rm d} t .  \nn
\#
Noting that $\int K(u) f_\varepsilon(b_h-uh) \, {\rm d}u = \int  K_h( b_h-t )   f_\varepsilon(t) \, {\rm d} t = \EE K_h( b_h-\varepsilon) = m''(b_h)$, it follows that
\#
 &   \|  \nabla Q_h( \bbeta )_{\cS^{\cc}}  -  m''(b_h)\cdot  \bSigma_{\cS^{\cc} \cS }  \bdelta  \|_\infty \leq  \frac{ l_0  }{2}   \max_{ j \in \cS^{\cc}}   \EE\bigl\{   |x_j |   ( \bx_{ \cS}^\T \bdelta )^2 \bigr\}  \leq \frac{l_0}{2}   \mu_4^{1/2} \sigma_{\bx}   \| \bdelta \|_{\Sb}^2 . \nn
\#
Conditioned on the event $\{  \| \hat \bbeta^{{\rm ora}}  -\bbeta_h^* \|_{\bSigma} \leq r \}$, this implies 
\# \label{new.score.diff.bound2}
   \|  \nabla Q_h( \hat \bbeta^\ora )_{\cS^{\cc}}   \|_\infty \leq     m''(b_h)\cdot   \| \bSigma_{\cS^{\cc} \cS} (\hat \bbeta^{{\rm ora}}  -\bbeta_h^*)_{ \cS}   \|_\infty + \frac{l_0 }{2}  \mu_4^{1/2} \sigma_{\bx} r^2.
\#
Next we bound $\| \bSigma_{\cS^{\cc} \cS}  (\hat \bbeta^{{\rm ora}}  -\bbeta_h^*)_{ \cS}  \|_\infty$ using the Bahadur representation in Proposition~\ref{thm:new.oracle}. Recall that $\Db_h = m''(b_h) \Sb$ and $\Sb= \bSigma_{\cS \cS}$, we have
\$
 & m''(b_h) \| \bSigma_{\cS^{\cc} \cS} (\hat \bbeta^{{\rm ora}}  -\bbeta_h^*)_{  \cS}   \|_\infty    =     \| \bSigma_{\cS^{\cc} \cS} \Sb^{-1} \Db_h (\hat \bbeta^{{\rm ora}}  -\bbeta_h^*)_{  \cS}  \|_\infty \\ 
&  \leq \max_{j\in \cS^{\cc}} \| \bSigma_{j \cS} (\bSigma_{\cS\cS})^{-1} \|_1 \cdot \|\Db_h (\hat \bbeta^{{\rm ora}}  -\bbeta_h^*)_{  \cS } \|_\infty \leq A_1 \|\Db_h (\hat \bbeta^{{\rm ora}}  -\bbeta_h^*)_{  \cS } \|_\infty   ,
\$
where the last step is due to condition \eqref{irrepresentable.cond2}.
In view of Proposition~\ref{thm:new.oracle}, we write $\Db_h ( \hat \bbeta^{{\rm ora}} -  \bbeta_{h } ^* )_{ \cS} = -   \bw_h(\bbeta^*_h)_{ \cS} + \br_{h }$, where $\br_h \in \RR^{s}$ is the remainder of the Bahadur representation. Together, \eqref{new.score.diff.bound0} and Proposition~\ref{thm:new.oracle} imply that  with probability at least $1- 4 e^{-t}$,
\#
 & \|\Db_h (\hat \bbeta^{{\rm ora}}  -\bbeta_h^*)_{  \cS } \|_\infty \leq \| \bw_h(\bbeta^*_h)_{ \cS}  \|_\infty + \| \br_h \|_\infty  \nn \\
 & \leq  \| \bw_h(\bbeta^*_h)_{ \cS}  \|_\infty + \| \br_h \|_2 \leq  \| \bw_h(\bbeta^*_h)_{ \cS}  \|_\infty + \gamma_1(\Sb)^{1/2} \|  \Sb^{-1/2}\br_h \|_2    \lesssim   \sqrt{\frac{\log(s) + t}{n}} +  \frac{s+t}{h^{ 1/2} n}  \label{new.score.diff.bound3}
\#
and
\#
 \| (\hat \bbeta^{{\rm ora}}  -\bbeta_h^*)_{ \cS}   \|_{\Sb} \lesssim  \sqrt{\frac{s+t }{n}}   ,   \label{new.oracle.bh}
\#
provided that $\sqrt{(s+t)/n} \lesssim h \lesssim 1$.  Combining \eqref{new.score.diff.bound0}--\eqref{new.oracle.bh} proves \eqref{new.oracle.score.bound}.

Back to oracle estimator, from the previous decomposition we have
\#
	& \|     \hat \bbeta^\ora - \bbeta_h^*  \|_\infty =  \|  ( \hat \bbeta^\ora - \bbeta_h^*)_{  \cS} \|_\infty \nn \\ 
& \leq   \|  (  \hat \bbeta^\ora - \bbeta_h^*)_{  \cS}  +\Db_h^{-1} \bw_h(\bbeta_h^*)_{ \cS}\|_\infty +  \| \Db_h^{-1} \bw_h(\bbeta_h^*)_{  \cS} \|_\infty \nn \\
&     \leq      \frac{1}{m''(b_h) \sqrt{ \gamma_s(\Sb)}} \|\Sb^{-1/2}   \br_{h }    \|_2 + \frac{1}{m''(b_h)}  \|  \Sb^{-1} \bw_h(\bbeta_h^*)_{  \cS} \|_\infty . \nn
\#
Analogous to the first bound in \eqref{new.score.diff.bound0}, the $\ell_\infty$-norm of $   \Sb^{-1} \bw_h(\bbeta_h^*)_{ \cS} \in \RR^s$ can also be bounded as
\#
 \|  \Sb^{-1}  \bw_h(\bbeta_h^*)_{  \cS}  \|_{\infty } \lesssim   \sqrt{\frac{\log(s) + t}{n}}  \nn
\#
with probability at least $1- e^{-t}$.
Combining this with \eqref{new.oracle.bahadur}  proves \eqref{new.oracle.infty.bound}.  \qed


\end{document}